\documentclass[a4paper,11pt]{article}
\pdfoutput=1
\usepackage{jheppub}
\usepackage[mathscr]{eucal}
\usepackage{hhline}
\usepackage{psfrag}
\usepackage{mathrsfs} 
\usepackage{hyperref}
\usepackage{bm}
\usepackage{array}
\usepackage{tikz}
\usepackage{subcaption}
\usepackage{textcomp}
\usepackage{hyperref}
\usepackage{mathrsfs} 
\usepackage{enumitem}
\usepackage{caption}
\usepackage{epic}
\usepackage{eepic}
\usepackage{gensymb}
\usepackage{pifont}
\DeclareMathAlphabet{\mathpzc}{OT1}{pzc}{m}{it}

\makeatletter
\@addtoreset{equation}{section}
\makeatother

\def\bea{\begin{eqnarray}}
\def\eea{\end{eqnarray}}
\def\be{\begin{equation}}
\def\ee{\end{equation}}

\def\be{\begin{equation}}
\def\ee{\end{equation}}
\def\bdm{\begin{displaymath}}
\def\edm{\end{displaymath}}
\def\bea{\begin{eqnarray}}
\def\eea{\end{eqnarray}}

\def\ri{{\rm i}}
\def\half{\textstyle\frac{1}{2}}

\def\XXint#1#2#3{{\setbox0=\hbox{$#1{#2#3}{\int}$}
    \vcenter{\hbox{$#2#3$}}\kern-.5\wd0}}

\newcommand{\rd}{\mbox{d}}
\newcommand{\re}{\mbox{e}}

\DeclareMathAlphabet{\mathpzc}{OT1}{pzc}{m}{it}

\title{Equilibrium  density matrices for the 2D black hole sigma models
from an integrable spin chain}
\author[a]{Vladimir V. Bazhanov,}
\author[b]{Gleb A.  Kotousov,}
\author[c,d]{Sergei  L. Lukyanov}

\affiliation[a]{Department of Theoretical Physics,
         Research School of Physics and Engineering,\\
    Australian National University, Canberra, ACT 2601, Australia}
\affiliation[b]{DESY, Theory Group, Notkestrasse 85, Hamburg 22607, Germany}
\affiliation[c]{NHETC, Department of Physics and Astronomy,
     Rutgers University,\\
     Piscataway, NJ 08855-0849, USA}
\affiliation[d]{Kharkevich Institute for Information Transmission Problems,\\
Moscow, 127994, Russia}

\emailAdd{vladimir.bazhanov@anu.edu.au}
\emailAdd{gleb.kotousov@desy.de}
\emailAdd{sergei@physics.rutgers.edu}

\abstract{This work concerns the
quantum Lorentzian and Euclidean black hole non-linear sigma models.
For the Euclidean black hole sigma model an
 equilibrium density matrix is proposed, which reproduces 
 the modular invariant partition function from the
2001 paper of Maldacena, Ooguri and Son.
For the Lorentzian black hole sigma model, using its 
formulation as a gauged ${\rm SL}(2,\mathbb{R})$
WZW model, we describe the linear and Hermitian structure
of its space of states and also propose an expression 
for the  equilibrium density matrix.
Our analysis is guided by the results of the study of a certain
critical, integrable spin chain. In the scaling limit, the latter exhibits
the key features of the Lorentzian black hole sigma model including
the same global symmetries, the same algebra of
extended conformal symmetry and a continuous spectrum of conformal dimensions.
}

\arxivnumber{2010.10603}

\begin{document}
\captionsetup[figure]{labelfont={small},labelformat={default},labelsep=period,name={Fig.\!}}
\captionsetup[table]{labelfont={small},labelformat={default},labelsep=period,name={Tab.\!}}

\maketitle
\pagebreak

\section{Introduction}

\begin{figure}
\centering
\includegraphics[width=7.cm]{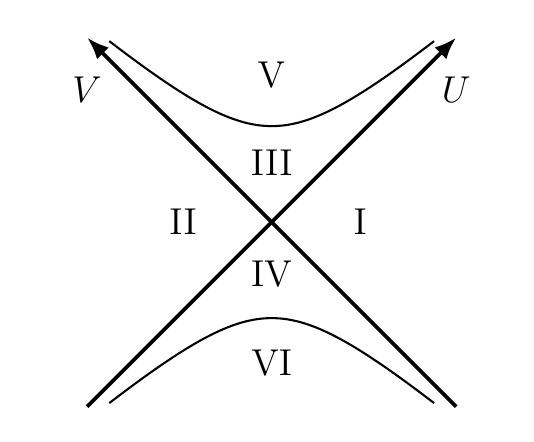}
\caption{\label{iaosi1212}\small Space-time diagram for the Lorentzian black hole \eqref{hsasaysaty}.
The cross 
defined by the equation $UV=0$ is a horizon, while the metric 
possesses a physical singularity on the hyperbola 
$UV=1$.
\label{fig20}}
 \end{figure}

It was observed by Witten in ref.\cite{Witten:1991yr}
  that  the two dimensional space equipped with the metric
\bea\label{hsasaysaty}
(\rd\sigma_{\rm\scriptscriptstyle LBH})^2=\frac{{\rm d} U{\rm d} V}{1-UV}
\eea
exhibits the characteristic features of a black hole geometry.
As  depicted in the space-time diagram in fig.\,\ref{fig20}, it possesses a horizon at $UV=0$ as well
as a curvature singularity at $UV=1$ just as the four dimensional Schwarzschild black hole
in terms of Kruskal coordinates. There is  no globally defined time coordinate
for the metric. Rather, there is a
Killing vector that is time-like only in regions I and II of
fig.\ref{fig20} and space-like in regions III and IV.
\bigskip

\begin{figure}[ht!]
\centering
\includegraphics[width=6.5cm]{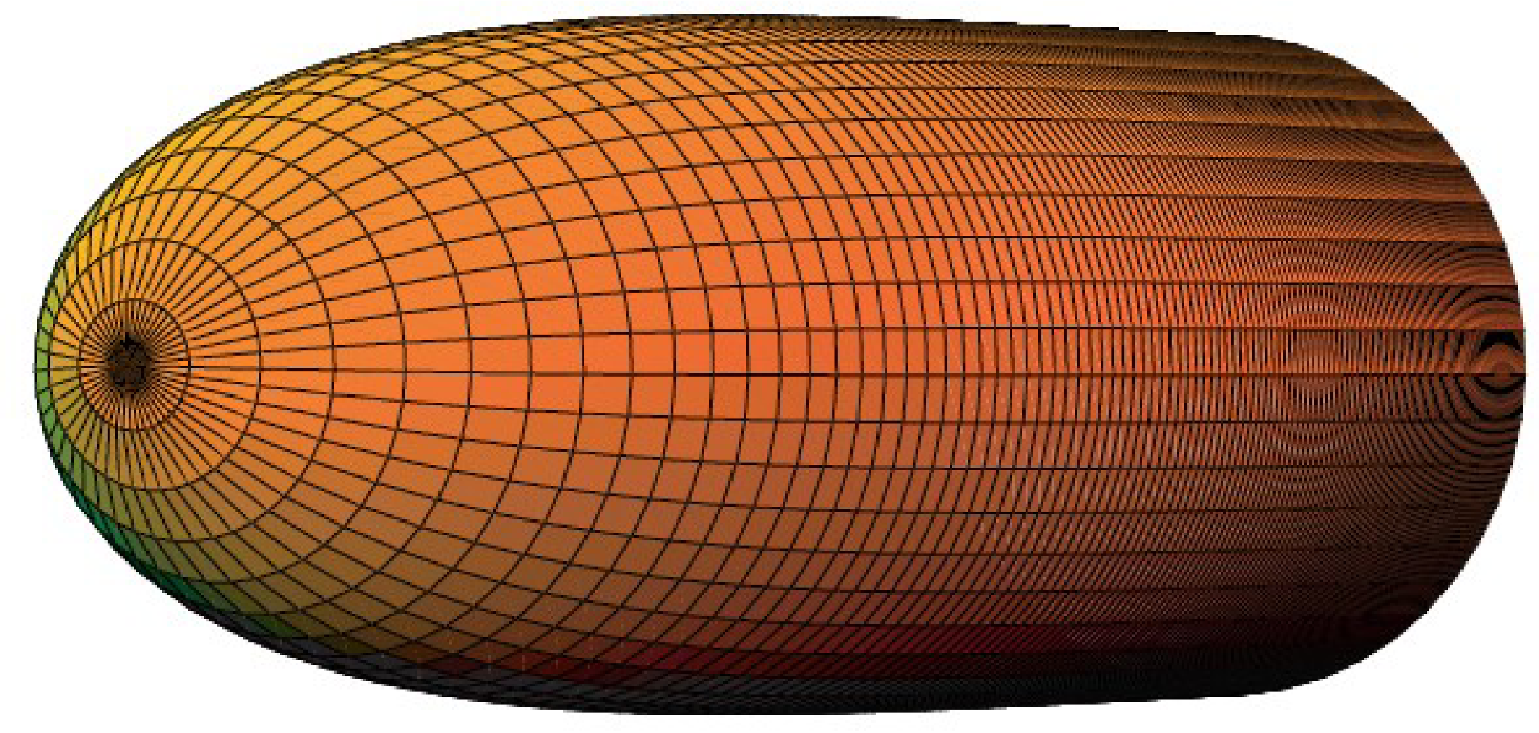}
\caption{Visualization of the $2D$ manifold with metric  \eqref{hsasaysaaatyA}.
\label{fig201}}
 \end{figure}

Similar to the $3+1$ dimensional black hole, the Euclidean version of  \eqref{hsasaysaty}
 is of prime interest.
The latter can be introduced in the spirit of the Hartle-Hawking    analytic continuation \cite{Hawking}.
It is performed  
by  a ``Wick rotation''  of the Killing coordinate $T=\half \,\log(-U/V)$ from region I to  pure imaginary values. This makes
$U$ and $V$ satisfy the
reality condition $V=-U^*$.
 Then, ignoring the overall negative sign, the Lorentzian metric  becomes  one  of  Euclidean signature:
 \bea\label{hsasaysaaatyA}
(\rd\sigma_{\rm\scriptscriptstyle EBH})^2=\frac{{\rm d} U{\rm d} U^*}{1+UU^*}\ .
\eea
The manifold equipped with   $(\rd\sigma_{\rm\scriptscriptstyle EBH})^2$  
may be embedded into three dimensional Euclidean space and visualized as a half-infinite cigar as shown in fig.\,\ref{fig201}.
The tip is located at $U=0$ while in the domain
 $|U|\gg1$, where the metric becomes flat, the 
target manifold resembles a half-infinite cylinder.

\bigskip

 For each of the metrics $(\rd\sigma_{\rm\scriptscriptstyle LBH})^2$ and $(\rd\sigma_{\rm\scriptscriptstyle EBH})^2$
 one can associate a Non-Linear Sigma Model (NLSM) whose classical dynamics is governed by the action\footnote{%
In this paper we 
use the following conventions.
We assume  Minkowski signature for the world-sheet $(x^0,x^1)\equiv (t,x)$
so that $\partial_0=\partial^0\equiv \frac{\partial}{\partial x^0}$, 
$\partial_1=-\partial^1\equiv \frac{\partial}{\partial x^1}$ with the space co-ordinate belonging to the segment 
$x\in[0,2\pi]$. The integration measure is taken to be $\rd^2 x=\rd x^0 \rd x^1$.
The  Plank constant is a positive dimensionless parameter. We'll also use 
 the Levi-Civita symbol  $\epsilon_{\mu\nu}=-\epsilon^{\mu\nu}$, below, which
  is defined as  $\epsilon_{01}=-\epsilon_{10}=1$.}
\begin{subequations}\label{hHYAYAYAYA}
\bea\label{iasdioi120143}
&&S_{\rm \scriptscriptstyle LBH}=\frac{1}{2\hbar}\int\rd^2 x \
\frac{\partial_\mu U{\partial}^\mu V}{1-UV}\qquad\quad
\qquad \\[-0.2cm]
&& \hspace{7cm}\big(\,\hbar\to  0^+\,\big) \nonumber 
\\[-0.2cm]\label{iasdioi1201435}
&&S_{\rm \scriptscriptstyle EBH}=\frac{1}{2\hbar}\int\rd^2 x \
\frac{\partial_\mu U{\partial}^\mu U^*}{1+UU^*}\ .
\eea
\end{subequations}
It is a challenging problem, in general, to assign a meaning to a quantum NLSM whose target
space metric is of Lorentzian signature. In particular the status of the quantum Lorentzian black hole NLSM is rather 
tentative. Contrary to this
the quantum Euclidean black hole NLSM is a  well studied CFT\cite{Elitzur:1991cb,Mandal:1991tz,Dixon:1989cg,Witten:1991yr,Dijkgraaf:1991ba,ZAM,Maldacena:2000hw,
Maldacena:2000kv,Hanany:2002ev,Ribault:2003ss,Schomerus:2005aq}.
\bigskip

In refs.\cite{Maldacena:2000kv,Hanany:2002ev}
 an explicit formula was presented for the  partition function  $Z_{\rm \scriptscriptstyle EBH}$. The latter is defined as a
Euclidean  path integral involving    the  Euclidean  action with  the world-sheet compactified on a  torus.
  The partition function contains a   divergence
   which  comes from the contribution of  the asymptotically flat domain of the target manifold:
\bea\label{oasid91020912}
Z_{\rm \scriptscriptstyle EBH}=Z^{\rm \scriptscriptstyle (sing)}_{\epsilon}
+Z^{\rm \scriptscriptstyle (reg)}_{\rm \scriptscriptstyle EBH}\ .
\eea
The singular part is somewhat universal.
It coincides with
the partition function of two free bosons, one of them being compactified, $\arg(U)\sim \arg(U)+2\pi$\,,
while the other,
 $\log|U|$\,, taking values in a segment  of 
length $\propto\log(1/\epsilon)$. Thus, as the regularization parameter $\epsilon\to 0$,
 \bea
 Z^{\rm \scriptscriptstyle (sing)}_{\epsilon}\propto\log(1/\epsilon)\to +\infty\ .
 \eea
On the other hand,
the finite part $Z^{\rm \scriptscriptstyle (reg)}_{\rm \scriptscriptstyle EBH}$ in \eqref{oasid91020912}
depends on the infra red regularization of the target manifold, which can be thought of
as being the result of interactions with an external thermostat.
In the works \cite{Maldacena:2000hw,Maldacena:2000kv,Hanany:2002ev} the Euclidean black hole NLSM occurs in the context
of bosonic string theory on ${\rm AdS}_3$.
This provides a particular  regularization. Its  major advantage, which makes it 
especially attractive from  the stringy point of view, is that  $Z_{\rm \scriptscriptstyle EBH}$
is  invariant w.r.t  modular transformations of the world-sheet torus.
Then  an immediate question arises concerning  the full equilibrium density matrix 
${\hat \rho}_{\rm \scriptscriptstyle EBH}$  corresponding to  such a regularization.
The latter can not be recovered from just the knowledge of  the partition function,
which is given by the  trace of ${\hat \rho}_{\rm \scriptscriptstyle EBH}$.\footnote{%
We use the non-standard convention that the trace of the (equilibrium) density matrix
rather than being one, coincides with the partition function.}
A conjecture for   ${\hat \rho}_{\rm \scriptscriptstyle EBH}$ was put forward
in the original work \cite{Maldacena:2000kv}.
However
 numerical analysis shows that the proposed expression   fails  to reproduce  the modular invariant
partition function.
\bigskip

In ref.\cite{Ikhlef:2011ay}
Ikhlef, Jacobsen and Saleur
made the interesting proposal that the
Euclidean black hole NLSM is
the  CFT governing the scaling limit of a certain integrable spin chain. 
This opened a potential way of obtaining  the
equilibrium density matrix ${\hat \rho}_{\rm \scriptscriptstyle EBH}$.
In the recent work \cite{Bazhanov:2019xvyA}  an extensive study of the spin chain
was performed and the original conjecture from \cite{Ikhlef:2011ay}
was re-examined.
Here, based on the results of that work,  we address the following questions:

\begin{enumerate}

\item[(i)]  What is the correct expression for ${\hat \rho}_{\rm \scriptscriptstyle EBH}$?

 \item[(ii)]  What is the space of states of the Lorentzian black hole NLSM
and how to assign to it a meaningful  equilibrium density matrix ${\hat \rho}_{\rm \scriptscriptstyle LBH}$?
In turn, how would  the latter be related to its Euclidean counterpart?

\end{enumerate}

 \section{Classical gauged ${\rm SL}(2,\mathbb{R})$ WZW model\label{sec231}}

\subsection{The classical action}
 
The Lorentzian black hole NLSM can be obtained
by gauging 
a  non-compact one dimensional subgroup
of the classical ${\rm SL}(2,\mathbb{R})$ 
WZW model.  Following the work  \cite{Witten:1991yr}
consider the classical action
\bea\label{ahasysysadd}
S&=&\frac{1}{\hbar}\int \rd^2 x\  \frac{1}{2}\ 
\Big[\,
\partial_\mu U\partial^\mu V+ \partial_\mu X\partial^\mu Y
+\log(X/Y)\, 
\epsilon^{\mu\nu}\partial_\mu U\partial_\nu V
\\[0.2cm]
&+&
a_\mu\,\big(Y\partial^\mu X-X\partial^\mu Y\big)+ \epsilon^{\mu\nu} a_\mu\,\big(U\partial_\nu V- V\partial_\nu U\big)
- a_\mu a^\mu\, XY
\,\Big] \ 
 .\nonumber
\eea
Here the integrand in the  first line  is just the classical Lagrangian density  of the  usual WZW model \cite{Witten:1983ar},
${\cal L}_{\scriptscriptstyle{\rm WZW}}[{\bf g}]$,
expressed via the
matrix entries of the  fundamental WZW   field
\bea\label{gaastsat}
{\bf g}=\begin{pmatrix}
X&U\\
-V&Y
\end{pmatrix}\in {\rm SL}(2,\mathbb{R})
\ .
\eea
Note that
the term involving $\log(X/Y)$ comes from the Wess-Zumino term and, up to a total derivative, can be rewritten in various
ways by employing the constraint  
\bea\label{usausau}
XY+UV=1\ .
\eea
The second line  in  \eqref{ahasysysadd} contains 
the gauge potential  $a_\mu$.  The action
is invariant w.r.t. the infinitesimal gauge  transformation of the form
\bea\label{aisiasi}
\delta X=\delta\omega\, X\ ,\ \ \  \delta Y=-\delta\omega\, Y\ ,\ \ \ \delta U=\delta V=0\ ;\ \ \ \delta a_\mu=\partial_\mu(\delta \omega)\ .
\eea
This can be seen by rewriting
the  Lagrangian density  corresponding to the action  \eqref{ahasysysadd} as
\bea\label{aoisid129812}
{\cal L}=\frac{1}{2}\
\bigg[\
\frac{\partial_\mu U\partial^\mu V}{1-UV}
- (1-UV)\  f_\mu f^\mu+\epsilon^{\mu\nu}\partial_\mu C_\nu
\,\bigg]
\eea
with
\bea\label{oasosaoas}
f_\mu=a_\mu-  \tfrac{1}{2}\  \partial_\mu  \log(X/Y)-\epsilon_{\mu\nu}\, J^\nu\ ,\ \  \ \ \ \ \ \ 
C_\mu=\tfrac{1}{2}\ \log(X/Y)\, (U\partial_\mu V-V\partial_\mu U)
\eea
and
\bea
J_\mu=\frac{1}{2}\ \frac{U\partial_\mu V-V\partial_\mu U}{1-UV}\ .
\eea
The extremum condition $\frac{\delta }{\delta a_\mu} S=0$ leads to the equation
\bea\label{isisai1a}
a_\mu= \tfrac{1}{2}\  \partial_\mu  \log(X/Y) +\epsilon_{\mu\nu}\, J^\nu\ .
\eea
The field strength corresponding to this vector potential is given by
\bea
\partial_\mu a_\nu-\partial_\nu a_\mu=\partial_\sigma J^\sigma\ \epsilon_{\mu\nu}\ .
\eea
It vanishes 
for any solution of the classical equations of motion, 
which includes  the continuity equation $\partial_\mu J^\mu=0$.

\bigskip
 In the orthodox formulation of the  gauged  ${\rm SL}(2,\mathbb{R})$ WZW model,
 the matrix valued field ${\bf g}$
 is  assumed to be  periodic:
 \bea\label{oisaiasias}
{\bf g}(t,x+2\pi)={\bf g}(t,x)\ .
 \eea
 If  we take $U$ and $V$  from the domain 
 \bea
 0\leq UV<1\ ,
 \eea
 it is natural to fix the
 gauge   by setting  \cite{Witten:1991yr}
 \bea\label{iaisaiassas}
 X=Y
 \eea
 which, in view of eq.\,\eqref{isisai1a},
results in
 $a_\mu=\epsilon_{\mu\nu}\, J^\nu$.
Then, after eliminating the auxiliary field $a_\mu$, the action
$S$ \eqref{ahasysysadd}  becomes   that of the Lorentzian  black hole NLSM \eqref{iasdioi120143}.
Note that, as was also pointed out in \cite{Witten:1991yr},
 if  we take the  ${\rm SL}(2,\mathbb{R})$ picture literally the 
full target space of the Lorentzian black hole NLSM would contain two copies
of the  regions III and IV in fig.\,\ref{iaosi1212} corresponding to the cases $X,Y>0$ and $X,Y<0$.
\bigskip

\bigskip
\subsection{Generalized boundary conditions\label{sec22}}

In what follows we'll consider the gauged  ${\rm SL}(2,\mathbb{R})$
WZW model, but
with more general boundary conditions than \eqref{oisaiasias}.
It is instructive to discuss the latter using 
another formulation of the classical field theory \cite{Gawedzki:1988nj,Dijkgraaf:1991ba}.
In this description the  Lagrange density  is just   the sum of that 
 of the WZW  model and  the massless Gaussian theory:
\bea
\label{kakaiaia}
{\tilde {\cal L}}={\cal L}_{\scriptscriptstyle{\rm WZW}}[{\bf G}]+2\, \partial \eta{\bar \partial}\eta\ .
\eea
The interaction   is introduced through the  constraints
\bea\label{aoasoasoas}
\Upsilon\equiv\half\ {\rm Tr}\big[\sigma^z\,\partial {\bf G}\, {\bf G}^{-1}\big]-\partial\eta=0\ ,\ \ \ \  \ 
\bar{\Upsilon}\equiv\half\ {\rm Tr}\big[\sigma^z\,{\bf G}^{-1}\, {\bar  \partial} {\bf G}\big]+{\bar \partial}\eta=0\ ,
\eea 
where we use $\partial=\frac{1}{2}(\partial_0+\partial_1)$ and $\bar{\partial}=\frac{1}{2}(\partial_0-\partial_1)$.
If  the  infinitesimal gauge transformation of the WZW  field  and
 the massless Gaussian field  is defined as
\bea
\delta {\bf G}=\tfrac{1}{2}\ \big(\sigma^z{\bf G}+{\bf G}\sigma^z\big)\  \delta\omega\ ,\ \ \  \  
\ \ \ \partial_\mu\delta\eta=\epsilon_{\mu\nu}\partial^\nu\, (\delta \omega)\ ,
\eea
then
$\delta{\tilde {\cal L}}$ turns out to be a total derivative
provided the constraints \eqref{aoasoasoas} are imposed.
The  classical  field theory, thus defined,
is equivalent to the gauged WZW model governed by the action \eqref{ahasysysadd}.
In particular for any  field configuration 
satisfying the equations of motion for  \eqref{kakaiaia},\,\eqref{aoasoasoas},
\bea\label{isisaiaa}
{\bf g}= \re^{\frac{1}{2} \omega\sigma^z}\, {\bf G}\, \re^{\frac{1}{2}\omega\sigma^z}\ ,\ \ \ \ \ \ \ \  \ \ 
a_\mu=-\epsilon_{\mu\nu}\partial^\nu \eta+\partial_\mu\omega
\eea
would be a solution of the Euler-Lagrange equations  associated with the action \eqref{ahasysysadd}.
Here $\omega$ is an arbitrary periodic function $\omega(t,x+2\pi)=\omega(t,x)$, which  
appears as a manifestation
of  the gauge
invariance of the model.

\bigskip
To specify  the boundary conditions,
let us first recall some basic facts concerning the  phase space of the  WZW model
(see, e.g., \cite{Witten:1983ar,Faddeev:1987ph,Fateev:1991aw}). 
The latter is conveniently  described in terms of  the left and right WZW currents,\footnote{Here and below we use the
notation ${\bf t}_A$ for the
$2\times 2$ real traceless matrices,
$$
{\bf t}_3=\begin{pmatrix}
1&0\\
0&-1
\end{pmatrix}
\ ,\ \ \ {\bf t}_+=\begin{pmatrix}
0&1\\
0&0
\end{pmatrix}\ ,\ \ \ \ 
{\bf t}_-=\begin{pmatrix}
0&0\\
1&0
\end{pmatrix}\ :\ \ \  \ \ \ \ \ \ \  [{\bf t}_A,{\bf t}_B]={f_{AB}}^C\,{\bf t}_C\ .
$$
Indices are raised and lowered via the Killing form defined as
$$\kappa_{AB}=\half\, {\rm Tr}[\,{\bf t}_A{\bf t}_B\,]\ ,  \ \ \  \ \ \ \ \ \kappa_{AC}\, \kappa^{CB}=\delta_A^B\ .$$}
\bea
\partial{\bf G}{\bf G}^{-1}=L^{A}{\bf  t}_A\ ,\ \ \ \ \ \ \  \ \ {\bf G}^{-1}{\bar \partial}{\bf G}={\bar R}^{A} {\bf t}_A\ ,
\eea
which satisfy the   closed set of  equal-time Poisson Bracket relations:
\bea\label{isisailaoi}
&&\big\{L^A(t,x),L^B(t,y)\big\}=-\half\, \kappa^{AB}\,\delta'(x-y)-\half\, {f^{AB}}_{C}\, L^C(t,x)\,\delta(x-y)\nonumber\\[0.2cm]
&&\big\{{\bar R}^A(t,x),{\bar R}^B(t,y)\big\}=+\half\, \kappa^{AB}\,\delta'(x-y)+\half\, {f^{AB}}_{C}\, {\bar R}^C(t,x)\,\delta(x-y)\ \ \ \ \\[0.2cm]
&&\big\{L^A(t,x),R^B(t,y)\big\}=0\ .\nonumber
\eea
Assuming that the currents are periodic fields,
\bea\label{aiosaisia}
L^A(t,x+2\pi)=L^A(t,x)\ ,\ \ \ \ \  \ \ \ \ \  \ \ {\bar R}^A(t,x+2\pi)={\bar R}^A(t,x)\ ,
\eea
the center of the  Poisson algebra is generated by two elements
\be\label{aoaosao}
{\mathfrak C}={\rm Tr}\bigg[{\overleftarrow{\cal P}}\exp\bigg(+\int_{x_0}^{x_0+2\pi}\rd x\,L^{A}\,{\bf t}_A\bigg)\bigg]\ ,\ \ \ \ 
{\bar {\mathfrak C}}={\rm Tr}\bigg[{\overrightarrow{\cal P}}\exp\bigg(-\int_{-x_0}^{-x_0-2\pi}\rd x\,{\bar R}^{A}\,
{\bf t}_A\bigg)\bigg]\ .
\ee
We will  focus on the field configurations such that  the   values of the central elements
are  restricted by the 
inequalities
\bea\label{jasuasu}
-2<{\mathfrak C},\,{\mathfrak{\bar C}}<2
\eea
and use the parameterization
\bea\label{aisodi10982912}
{\mathfrak C}=2\cos(2\pi P)\ ,\  \ \ \ \ \ \ \ {\mathfrak{\bar C}}=2\cos(2\pi {\bar  P})
\eea
with real $P$ and $\bar{P}$.
In this case   the  path ordered exponentials  inside the traces in \eqref{aoaosao} may be expressed as
\bea\label{ususausa}
{\overleftarrow{\cal P}}\exp\bigg(+\int_{x_0}^{x_0+2\pi}\rd x\,L^{A}\,{\bf t}_A\bigg)&=&{\boldsymbol M}\,\re^{+2\pi\ri P \sigma^y}\,
{\boldsymbol M}^{-1}\\
{\overrightarrow{\cal P}}\exp\bigg(-\int_{-x_0}^{-x_0-2\pi}\rd x\,{\bar R}^{A}\,{\bf t}_A\bigg)&=&
{\bar {\boldsymbol M}}\,\re^{-2\pi\ri {\bar P}\sigma^y }\, {\bar {\boldsymbol M}}^{-1}\,, \nonumber
\eea
where the $2\times 2$ real non-degenerate matrices ${\boldsymbol M }$ and ${\bar {\boldsymbol M}}$
depend on the initial integration point $x_0$.
If we require them to be  ${\rm SL}(2,{\mathbb R})$  matrices,
 then $\re^{+2\pi\ri P \sigma^y}$ and $\re^{-2\pi\ri {\bar P}\sigma^y }$  are uniquely defined.
 At the same time there is an ambiguity in 
${\boldsymbol M }$ and ${\bar {\boldsymbol M}}$ of the form
 ${\boldsymbol M }\mapsto \pm {\boldsymbol M }\, \re^{\ri \phi \sigma^y}$ and ${\bar {\boldsymbol M}}\mapsto
  \pm {\bar {\boldsymbol M}}\,\re^{\ri {\bar \phi} \sigma^y}$ with arbitrary real $\phi$ and ${\bar\phi}$.
  This   can be fixed using the Iwasawa decomposition for
${\rm SL}(2,{\mathbb R})$ matrices, which
  allows one to specify that
  \bea\label{usaususa}
  {\boldsymbol M }=\begin{pmatrix}
  d&0\\
  0&d^{-1}
  \end{pmatrix}\ \begin{pmatrix}
  1& b\\
  0&1
  \end{pmatrix}\ ,\ \ \ \ \ \ \ \ \bar{{\boldsymbol M }}=\begin{pmatrix}
  {\bar d}&0\\
  0&{\bar d}^{-1}
  \end{pmatrix}\ \begin{pmatrix}
  1& {\bar b}\\
  0&1
  \end{pmatrix}\ \ \ \ \quad {\rm with}\ \ \ \quad d,{\bar d}>0\ .
  \eea

\bigskip
The  values  of the currents at $t=0$ are not enough to fully define the  time dependence
of the matrix valued field ${\bf G}(t,x)$.  Indeed the equations  of motion
in the WZW model
are given by
 \bea
 {\bar \partial} L^A=0\ ,\ \ \ \ \ \ \ \ \ \partial {\bar R}^A=0\ .
 \eea
 This implies that 
\bea\label{hsyusyas}
{\bf G}(t,x)={\boldsymbol \Omega}(t+x)\,{\bf G}(0,x_0)\,{\bar{\boldsymbol\Omega}}(t-x)\ ,
\eea
where
\bea\label{ausauasuas}
{\boldsymbol \Omega} (u)&=&{\overleftarrow{\cal P}}\exp\bigg(+\int_{x_0}^{u}\rd x\,L^{A}\,{\bf t}_A\bigg)\\[0.2cm]
{\bar {\boldsymbol \Omega}}({\bar u})&=&{\overrightarrow{\cal P}}\exp\bigg(-\int_{-x_0}^{
{\bar u}}\rd x\,{\bar R}^{A}\,{\bf t}_A\bigg)\ ,
\nonumber
\eea
while  ${\bf G}(0,x_0)$ is an arbitrary ${\rm SL}(2,{\mathbb R})$ matrix. Its   entries, 
together with  the  initial values of the currents, constitute  the full set of the  initial data.
We consider the field configurations at $t=0$ to be such that
\bea\label{saoosao}
{\bf G}(0,x_0)={\boldsymbol M}\ \re^{\ri \alpha\sigma^y }\ {\bar {\boldsymbol M}}^{-1}\ ,
\eea
where ${\boldsymbol M},\, {\bar {\boldsymbol M}}$  are
the same as in \eqref{ususausa},\,\eqref{usaususa}
 and $\alpha$ is some real number.
This is motivated through the following arguments.
Assuming $L^A$ are given,
the path ordered exponent  ${\boldsymbol \Omega} (u)$ \eqref{ausauasuas}
solves  the linear   differential equation
\bea\label{aososao}
\partial {\boldsymbol \Psi}=L^{A}\,{\bf t}_A\, {\boldsymbol \Psi}\ .
\eea
However ${\boldsymbol \Omega}(u)$, apart from the WZW currents,
also depends on
an arbitrarily chosen initial integration  point $x_0$  at which   it becomes   the identity  matrix.
At the same time 
$\bm{\Psi}_P={\boldsymbol \Omega} (u)\,{\boldsymbol M}$  
is  the Floquet solution of  the matrix ODE \eqref{aososao},
which is fixed unambiguously provided 
${\boldsymbol M}$ is taken to be of the form \eqref{usaususa}.
A change in the initial integration point $x_0$ to $x_0'$ would result in 
the transformation 
$\bm{\Psi}_P\mapsto \bm{\Psi}_P\,\re^{\ri \alpha_0\sigma^y }$,
where $\alpha_0=\alpha_0(x_0,x_0')$.
The solutions of the ODE with periodic coefficients
 possess the band structure. Thus the parameter $P$ labeling the 
Floquet solutions $\bm{\Psi}_P$ 
can be defined such that $P\in\mathbb{R}$ and  $2P\notin{\mathbb Z}$,
where the band number  coincides with  the greatest integer less than  $P+\frac{1}{2}$.
 The above carries over to  the Floquet solution 
 $\overline{\bm{\Psi}}_{\bar{P}}=
{\bar {\boldsymbol M}}^{-1}\,{\bar {\boldsymbol \Omega}} (\bar{u})$ of the barred counterpart of the  ODE \eqref{aososao}.
 This way the  construction of the WZW field ${\bf G}(t,x)$ given  by eqs.\,\eqref{hsyusyas} and \eqref{saoosao} 
 involves  the  Floquet solutions  as well as an additional variable 
 $\alpha\sim \alpha+2\pi$. 
  Thus  the algebra of functions on the phase space of the WZW model, generated by the 
  currents $L^A$ and ${\bar R}^A$  subject to the periodic boundary conditions
  \eqref{aiosaisia}, should be extended by the inclusion of  the compact  generalized coordinate $\alpha$. The latter  can be 
 viewed  as  a dynamical variable  
canonically conjugated to the sum $2\pi (P+{\bar P})$.
As for the difference, 
we assume that $\re^{2\pi\ri (P-{\bar  P})}=\re^{2\pi\ri {\tt k}}$,
with  $\half<{\tt k}\leq \half$ being a  fixed parameter.  
Equivalently,
\bea\label{isiasaisi}
P-{\bar P}={\tt k}+{\tt w}\ \ \ \ \  \ \ \ ({\tt w}\in{\mathbb{Z}})
\eea
so that the
integer ${\tt w}$ labels different disjoint components of the phase space.

 \bigskip

The 
 boundary values of the WZW field  at $t=0$, defined by the formulae  \eqref{hsyusyas} and \eqref{saoosao},
satisfy  the relations
\bea\label{isisiisa}
{\bf G}(0,x_0+2\pi)={\boldsymbol M}\, \re^{2\pi\ri{\tt  k} \sigma^y }\, {\boldsymbol M}^{-1}\ {\bf G}(0,x_0)
={\bf G}(0,x_0)\ {\bar {\boldsymbol M}}\ \re^{2\pi\ri {\tt  k}\sigma^y}\ {\bar {\boldsymbol M}}^{-1}\ .
\eea
This  implies 
\bea\label{isisai}
{\rm{Tr}}\Big[\,{\bf G}(t,x+2\pi)\,\big({\bf G}(t,x)\big)^{-1}\,\Big]=2\,\cos(2\pi{\tt k})\ ,
\eea
which  should be imposed 
along  with the periodicity condition  for   the currents \eqref{aiosaisia}.
In fact there is an extra condition which needs to be taken into account.
Substituting the matrix ${\boldsymbol M}$ \eqref{usaususa} into 
eq.\eqref{isisiisa} one finds 
\bea\label{iasisisawwA}
{\rm{Tr}}\Big[\,(-\ri\,\sigma^y)\ {\bf G}(0,x_0+2\pi)\,\big({\bf G}(0,x_0)\big)^{-1}\,\Big]=\sin(2\pi {\tt k})\ \big(d^2+d^{-2}+d^2b^2\big)\ .
\eea
 This results in the inequality
 \bea\label{iasisisaww}
 {\rm{Tr}}\Big[\,(-\ri\,\sigma^y)\ {\bf G}(t,x+2\pi)\,\big({\bf G}(t,x)\big)^{-1}\,\Big]\big/\sin(2\pi {\tt k})>0\ .
 \eea

\bigskip

The constraints  \eqref{aoasoasoas}  will only make sense when  both derivatives $\partial\eta$ and ${\bar \partial}\eta$ are periodic:
\bea\label{aiasiasi}
\partial\eta(t,x+2\pi)=\partial\eta(t,x)\ ,\ \ \ \ \ \ \ \ \ \ \ \ {\bar \partial}\eta(t,x+2\pi)={\bar \partial}\eta(t,x)\ .
\eea
In view of the relation  \eqref{isisaiaa},  the gauge field $a_\mu(x,t)$ in the original
formulation of the gauged WZW model is also periodic,
\bea
a_\mu(t,x+2\pi)=a_\mu(t,x)\,,
\eea
as   was  implicitly  assumed  in our initial  analysis of the model.
The boundary condition \eqref{isisai} as well as  the inequality  \eqref{iasisisaww} 
are invariant under  the gauge transformation and  therefore
  the field 
${\bf g}$ satisfies the similar relations
\begin{subequations}\label{isisaiuy}
\bea\label{isisaiuyA}
&&{\rm{Tr}}\Big[\,{\bf g}(t,x+2\pi)\,\big({\bf g}(t,x)\big)^{-1}\,\Big]=2\,\cos(2\pi{\tt k})\\[0.2cm]\label{isisaiuyB}
&&{\rm{Tr}}\Big[\,(-\ri\,\sigma^y)\,{\bf g}(t,x+2\pi)\,\big({\bf g}(t,x)\big)^{-1}\,\Big]\big/\sin(2\pi {\tt k})>0\ .
\eea
\end{subequations}

\bigskip
Let us make the following important point.
In the case of the gauged ${\rm SL}(2,\mathbb{R})$ WZW model 
with ${\tt k}=0$, the  conditions \eqref{isisaiuy}
yield   ${\bf g}(t,x+2\pi)={\bf g}(t,x)$, i.e., periodicity of
 all the matrix  elements $X,Y,U,V$. In turn one can 
 use  the gauge fixing condition
$X=Y$. 
However for 
${\tt k}\not=0$,
 since  $X$ and $Y$ are no longer periodic fields,  
the same gauge fixing condition is not applicable.
This makes  the model with   ${\tt k}=0$
(which is equivalent to the Lorentzian black hole NLSM) a very special one that  
is not obtainable literally     through
a  naive  ${\tt k}\to 0$ limit.
\bigskip

\bigskip

The Poisson structure of
 the massless Gaussian model, whose Lagrange density is given by
the second term in the r.h.s. of \eqref{kakaiaia},
reads as
\be\label{aisaiasi}
\big\{\partial\eta(t,x),\partial\eta(t,y)\big\}=- \big\{{\bar \partial}\eta(t,x),{\bar \partial}\eta(t,y)\big\}
=\half\ \delta'(x-y)\ ,\ \ \ \ 
\big\{\partial\eta(t,x),{\bar \partial}\eta(t,y)\big\}=0\ .
\ee
With  the boundary conditions \eqref{aiasiasi} imposed,
the center  of this Poisson algebra is generated by
\bea\label{asodioaisd109212}
P_\eta=\int_0^{2\pi}\frac{\rd x}{2\pi}\ { \partial}\eta\ ,\ \ \ \ \ \qquad\qquad
{\bar P}_\eta=\int_0^{2\pi}\frac{\rd x}{2\pi}\  {\bar \partial}\eta\ .
\eea
The general solution of the equation of motion $\partial{\bar\partial}\eta=0$
is
\bea\label{isaisaia}
\eta(t,x)=\tfrac{1}{2}\, \big(f(t+x)-{\bar f}(t-x)\big)
\eea
where, in view of the boundary conditions, the
 arbitrary   functions $f$ and ${\bar f}$ are quasiperiodic:
\bea\label{sasausau}
f(u+2\pi)=f(u)+P_\eta\ ,\ \ \ \ \ \qquad \qquad \bar{f}({\bar u}+2\pi)={\bar f}({\bar u})+{\bar P}_\eta\ .
\eea
\smallskip

The constraints \eqref{aoasoasoas} imposed on the WZW field ${\bf G}$ and
the Gaussian field, combined with \eqref{isaisaia}, yield the relations
\bea\label{isaisaisaisai}
L^3=-\half\, \partial f\ ,\ \ \ \ \ \ \ \ \ \ \ {\bar R}^3=-\half\, {\bar \partial} {\bar f}\ .
\eea
It is  easy to see now that the matrix ${\bf G}$, 
satisfying the equations of motion,
can be brought to the form
\bea\label{iasiisais}
{\bf G}(t,x)=\re^{-\frac{1}{2} f(t+x)\sigma^z}\, \mathfrak{G} (t,x)\, \re^{-\frac{1}{2}{\bar f}(t-x)\sigma^z}\ ,
\eea
where  $\mathfrak{G} $  is such that
\bea\label{iaiassau}
\partial\mathfrak{G} \ \mathfrak{G}^{-1}=
\xi_-{\bf t}_--\xi_+{\bf  t}_+\ ,\ \ \ \ \ \ \ \ \ \ \ \ 
\mathfrak{G}^{-1}\,
{\bar\partial}\mathfrak{G}={\bar \xi}_-{\bf  t}_--{\bar \xi}_+{\bf t}_+
\eea
with
\bea\label{jsusausa}
&&\xi_-=\re^{-f}\, L^-\ ,\ \ \ \qquad\qquad\xi_+=-\re^{+f}\ L^+\nonumber\\[0.2cm]
&&{\bar \xi}_-=\re^{+{\bar f}}\, {\bar R}^-\ ,\ \ \ \qquad\qquad\bar{\xi}_+=-\re^{-{\bar f}}\ {\bar R}^+\ .
\eea
The latter  are real  chiral fields, 
${\bar \partial}\xi_\pm=\partial {\bar\xi}_\pm=0$,
 subject to   the quasiperiodic boundary conditions
\bea\label{aususalk}
 \xi_\pm (u+2\pi)=B^{\pm 1}\ \xi_\pm (u)\ ,\ \ \ \ \ \ 
\qquad
\qquad{\bar \xi}_\pm ({\bar u}+2\pi)={\bar B}^{\pm 1}\ {\bar  \xi}_\pm ({\bar u})\,,
\eea
where $B=\re^{2\pi P_\eta}$ and ${\bar B}=\re^{2\pi {\bar P}_\eta}$.
We set 
$
B={\bar B}
$
 or, equivalently, $P_\eta= {\bar P}_\eta$ 
(assuming that $P_\eta$ and ${\bar P}_\eta$ are real).
In this case, as it follows from 
eqs.\,\eqref{isaisaia} and \eqref{sasausau},
the field $\eta$ is periodic:
\bea\label{hsahsay}
\eta(t,x+2\pi)=\eta(t,x)\,.
\eea
Note that the on-shell gauge potential $a_\mu$, entering into  the initial
formulation of the ${\rm SL}(2,\mathbb{R})$ gauged WZW  model,
satisfies  the condition
\bea\label{isaiasias}
B={\bar B}=\exp\bigg(\oint\rd x^\mu\,a_\mu\bigg)\ .
\eea

\subsection{The phase space}

\bigskip
Consider now  the classical  fields  defined through the relations
\be\label{isisai112assaa}
\arraycolsep=0.7cm
\begin{array}{lll}
W^{(cl)}_2=\xi_+\,\xi_-\ ,& 
W^{(cl)}_3=\tfrac{1}{2}\ \big(\xi_-\,\partial \xi_+-
\xi_+\,\partial \xi_-\big)\ ,
\\[0.4cm]
\overline{W}^{(cl)}_2=\bar{\xi}_+\,\bar{\xi}_-\ ,&
\overline{W}^{(cl)}_3=\tfrac{1}{2}\ \big(
{\bar \xi}_-\,\partial {\bar \xi}_+-
{\bar \xi}_+\,\partial {\bar \xi}_-\big)\ .
\end{array}
\ee
Using \eqref{jsusausa} they can be  rewritten in terms of the WZW currents along with $\partial\eta$ and $\bar{\partial}\eta$:
\be\label{aspod9102123a}
\arraycolsep=0.3cm
\begin{array}{ll}
W^{(cl)}_2=(\partial\eta)^2-\big((L^3)^2+L^+L^-\big) \,, &
W^{(cl)}_3=2\,\partial\eta\, L^+L^-+\half \, (L^+\partial L^--L^-\partial L^+)\ \ \ 
\\[0.4cm]
{\overline W}^{(cl)}_2=({\bar \partial}\eta)^2-\big(({\bar R}^3)^2+{\bar R}^+{\bar R}^-\big)\,,
&
{\overline W}^{(cl)}_3=2\,{\bar \partial\eta}\, {\bar R}^+{\bar R}^-+\half \, ({\bar R}^+\partial{\bar  R}^--
{\bar R}^-{\bar \partial} {\bar R}^+)
\end{array} .
\ee
It is straightforward to check
using the PB relations \eqref{isisailaoi} and \eqref{aisaiasi} that
 the $W$ fields Poisson commute  (in a weak sense) with  the constraints $\bar{\Upsilon}$ and ${ \Upsilon}$ 
\eqref{aoasoasoas},
\be\label{hasyasys}
\begin{array}{l}
\big\{W^{(cl)}_j(t,x),\Upsilon(t,y)\big\}\big|_{\Upsilon=0}=\big\{W^{(cl)}_j(t,x),\bar{\Upsilon}(t,y)\big\}=0\\[0.5cm]
\big\{\overline{W}^{(cl)}_j(t,x),\bar{\Upsilon}(t,y)\big\}\big|_{{\bar \Upsilon}=0}=\big\{\overline{W}^{(cl)}_j(t,x),{\Upsilon}(t,y)\big\}=0
\end{array}\ \ .
\ee
Since $W^{(cl)}_2$ and $\overline{W}^{(cl)}_2$ coincide with  the nonvanishing components of the
stress energy tensor,   the Hamiltonian     commutes  with $\bar{\Upsilon}$ and ${ \Upsilon}$.
Also it is easy to see that
\bea
\big\{\Upsilon(t,x),{ \Upsilon}(t,y)\big\}=
\big\{{\bar \Upsilon}(t,x),{\bar \Upsilon}(t,y)\big\}=\big\{\Upsilon(t,x),{\bar \Upsilon}(t,y)\big\}=0
\eea
and, hence, the system of  constraints \eqref{aoasoasoas}
 are of the first class.
 The  $W$ fields are ``classical observables'' which are
\bigskip

 (i) chiral
 \bea
  W^{(cl)}_j=W^{(cl)}_j(t+x)\ ,\ \ \ \ \ \ \ \  \overline{W}^{(cl)}_j=\overline{W}^{(cl)}_j(t-x)
 \eea

 (ii)  real
 \bea\label{iaosid9891}
\big( W^{(cl)}_j(u)\big)^*=W^{(cl)}_j(u^*)\ ,\ \  \ \ \ \big(\overline{W}^{(cl)}_j({\bar u})\big)^*=\overline{W}^{(cl)}_j({\bar u}^*)
  \eea

(iii) periodic
 \bea\label{oaisd89212}
 W^{(cl)}_j(u+2\pi)=W^{(cl)}_j(u)\ ,\ \ \ \  \ \ \ \ \  \ \ \ \overline{W}^{(cl)}_j({\bar u}+2\pi)=\overline{W}^{(cl)}_j({\bar u})\ .
 \eea
 They generate a closed Poisson algebra in the following sense. Straightforward calculations lead to
\bea\label{jasususa}
&&\big\{W^{(cl)}_2(u_1),W^{(cl)}_2(u_2)\big\}=
-\big(W^{(cl)}_2(u_1)+W^{(cl)}_2(u_2)\big)\ \delta'(u_1-u_2)\nonumber\\[0.4cm]
&&\big\{W^{(cl)}_3(u_1),W^{(cl)}_2(u_2)\big\}= -3\ W^{(cl)}_3(u_1)\ 
\delta'(u_1-u_2)-\partial W^{(cl)}_3(u_1)\ \delta(u_1-u_2)\\[0.4cm]
&&\big\{W^{(cl)}_3(u_1),W^{(cl)}_3(u_2)\big\}=-
\tfrac{1}{4}\,\big(W^{(cl)}_2(u_1)+W^{(cl)}_2(u_2)\big)\ \delta'''(u_1-u_2)
-\delta'(u_1-u_2)\times  \nonumber\\[0.4cm]
&& \Big(W^{(cl)}_4(u_1)+W^{(cl)}_4(u_2)+
2\,W^{(cl)}_2(u_1)\, W^{(cl)}_2(u_2)-\tfrac{3}{20}\,\big(\,\partial^2\, W^{(cl)}_2(u_1)+\partial^2\,
W^{(cl)}_2(u_2)\,\big)\Big)\ .\nonumber
\eea
Here $W^{(cl)}_4$, which appears in the last line,  is expressed in a form similar to \eqref{isisai112assaa}:
\bea
W^{(cl)}_4=\tfrac{2}{5}\ \big(\xi_+\,\partial^2\xi_-+
\xi_-\,\partial^2 \xi_+\big)-\tfrac{6}{5}\ \partial\xi_+\partial\xi_- \ .
\eea
Analogous relations hold true for the ``barred'' fields.
A recursive computation of the Poisson brackets of the $W$ fields  amongst
themselves yields two infinite sets  $\big\{W^{(cl)}_j\big\}_{j=2}^\infty$
and  $\big\{\overline{W}^{(cl)}_j\big\}_{j=2}^\infty$. We'll refer below to the fields from these sets
as the (classical) $W$ currents. All of them
 satisfy the conditions (i)-(iii) as well as \eqref{hasyasys}. 
 The integer $j=2,3,\ldots$   coincides with the Lorentz spin of $W^{(cl)}_j$ while the   Lorentz spin of
$ \overline{W}^{(cl)}_j$ is given by $(-j)$.
The $W$ currents form a quadratic Poisson algebra \cite{Bakas:1991fs}. Since 
 all the Poisson bracket between the $W$  currents of different chirality  vanish, we'll refer 
to it as the classical ${\overline{W}}_\infty\otimes W_\infty$\,-\,algebra.

 \bigskip
 The above considerations 
suggest that the phase space for the gauged ${\rm SL}(2,\mathbb{R})$ WZW model,
subject to the boundary conditions \eqref{aiosaisia},\,\eqref{isisai},\,\eqref{iasisisaww}
 and \eqref{hsahsay}
is made up of the symplectic leaves, $\Gamma_{\bar{P},P,B}$\,, labeled by the real numbers $B$  \eqref{isaiasias}
as well as 
$P$, $\bar{P}$, which  satisfy the relation $P-{\bar P}={\tt k}+{\tt w}$ with ${\tt w}\in\mathbb{Z}$.
On each leaf the symplectic form is non-degenerate.
The algebra of functions on the leaf, $\Gamma_{\bar{P},P,B}^\star$, 
 is generated by the currents $W_j^{(cl)}(u)$ and $\overline{W}_j^{(cl)}(\bar{u})$,
subject to the reality and boundary conditions \eqref{iaosid9891},\,\eqref{oaisd89212}, which form the
 classical ${\overline{W}}_\infty\otimes W_\infty$\,-\,algebra
\bigskip

There are two evident continuous symmetries.
The first one is the ${\rm U}(1)$ invariance w.r.t. a shift of the compact 
variable $\alpha$\ \eqref{saoosao},
 \bea\label{aususuaa}
{\cal U}_\phi\ \ :\ \ \ \ \ \re^{\ri\alpha \sigma^y}\mapsto  \re^{\ri(\alpha + \phi)\sigma^y}\ .
 \eea
The other one corresponds to the transformation
 \bea
 {\cal R}_\theta\ \ :\ \ \ \ \ 
 \eta(t,x)\mapsto\eta(t,x)+\theta\ .
 \eea
Both symmetry transformations are canonical, i.e., they preserve the symplectic structure. They have no effect on the
$W$ currents and  act trivially on the symplectic leaves.
The gauged ${\rm SL}(2,\mathbb{R})$ WZW model also possesses the global symmetries, which leave 
the Lagrange density  \eqref{kakaiaia}, the constraint \eqref{aoasoasoas} as well
 as the boundary conditions imposed on the fields  unchanged.
We'll use the notation ${\cal D}$ for the ${\cal Z}_2$ symmetry, which acts
on the fields and the symplectic leaves as
\be\label{saqqs}
\arraycolsep=0.3cm
{\cal D}\ \ :\ \ \
\begin{array}{ll}
{\bf G}\mapsto  -\sigma^y\, {\bf G}\, \sigma^y\ ,\ \ \ \eta\to-\eta &
\ \ \ 
\\[0.4cm]
 \Gamma_{\bar{P},P,B}\mapsto \Gamma_{\bar{P},P,B^{-1}}
&
\end{array}
\!\!\!\!\!.
\ee
Also, by  ${\cal CP}$ symmetry, we'll mean the invariance  under the transformation
 \bea
 {\cal CP}\ \ :\ \ \ {\bf G}(t,x)\mapsto \big({\bf G}(t,-x)\big)^{-1}\ ,\ \ \ \ \ \ \  \eta(t,x)\mapsto  \eta(t,-x)\ .
 \eea
 Note that  the condition  $P-{\bar P}={\tt k}+{\tt w} \ \ ({\tt w}\in\mathbb{Z})$ is unchanged when
$P\mapsto -{\bar P}$ ,\ ${\bar P}\mapsto -{ P}$.
This suggests that   the action of  ${\cal CP}$ 
 on the  symplectic leaves
  is described as 
 \bea\label{aususuaa1}
 {\cal CP}\ \ :\ \ \ \Gamma_{\bar{P},P,B}\mapsto \Gamma_{-P, -\bar{P},B}\ .
 \eea

\section{Quantum gauged ${\rm SL}(2,\mathbb{R})$ WZW model}

\subsection{BRST quantization}
Once the gauged ${\rm SL}(2,\mathbb{R})$ WZW model
is formulated as a classical dynamical  system possessing constraints 
of the first class
one can consider  the problem of its quantization within 
the BRST approach. Here
we  briefly sketch the
algebraic procedure for the construction of the chiral component of the space of states.

\bigskip
The chiral component of the energy momentum tensor of the quantum theory 
is  split into three terms:
\bea\label{isisaiasaaq}
T_{\scriptscriptstyle{\rm total}}=T_{\scriptscriptstyle{\rm WZW}}+T_{\scriptscriptstyle{\rm Gauss}}+T_{\scriptscriptstyle{\rm ghost}}\ .
\eea
The first one  is \cite{Knizhnik:1984nr}
\bea\label{aspodi0912r43}
T_{\scriptscriptstyle{\rm WZW}}=-\frac{n^2}{n+C_{\tt V}}\  \kappa_{AB}\, L^AL^B\ .
\eea
It is built from the  left currents of the WZW model
\bea
L^{A}(u)=n^{-1}\ \sum_{m=-\infty}^\infty j_m^A\ \re^{-\ri  m u}\ \ \ \ \ \ \ \  \ \ (A=3,\pm)
\eea
whose Fourier coefficients obey  the commutation relations
\bea
\big[\,j^A_m,j^B_r\,\big]=  -n
\ \kappa^{AB}\ \, \tfrac{m}{2}\, \delta_{m+r,0}- \tfrac{\ri }{2}\,  {f^{AB}}_{C}\ j^C_{m+r}\ .
\eea
Here the  level (central element)
  of the  Kac-Moody algebra has been  denoted by $n$.  It is  related to the Plank constant as 
\be
\hbar=\frac{2\pi}{n} \ .
\ee
The constant  $C_{\tt V}$ entering into \eqref{aspodi0912r43}
stands for the so-called dual Coxeter number:
\bea
C_{\tt V}\,\kappa^{AB}=\tfrac{1}{4}\, {f^{AC}}_D\,{f^{BD}}_C
\eea
and in the case under consideration $C_{\tt V}=2$.
The second term in \eqref{isisaiasaaq} represents the contribution of the massless Gaussian field,
\bea
T_{\scriptscriptstyle{\rm Gauss}}=n\ (\partial\eta)^2
\eea
with
\bea\label{aoisido0190921}
\partial\eta(u)=n^{-\frac{1}{2}}\, \sum_{m=-\infty}^\infty a_m\ \re^{-\ri m u}\ :\ \ \ \ \ \big[a_m, a_r\big]=\tfrac{m}{2}\, \delta_{m+r,0}\ .
\eea
Finally $T_{\scriptscriptstyle{\rm ghost}}$ is the chiral component of the energy momentum tensor for the $bc$\,-\,system:
\bea
T_{\scriptscriptstyle{\rm ghost}}=\ri\, b\partial c\ .
\eea
The ghost   fields have 
conformal dimensions  $(\Delta_b,\Delta_c)=(1,0)$ and, as with the chiral fields  $L^A$ and $\partial\eta$,
can also be   expanded in the Fourier series
\bea
b(u)= \sum_{m=-\infty}^\infty b_m\ \re^{-\ri m u}\ \,,\ \ \ \ \ \ \ \ \ c(u)= \sum_{m=-\infty}^\infty c_m\ \re^{-\ri m u}\ \,,
\eea
where
\bea
\big\{b_m, c_r\big\}=\delta_{m+r,0}\ ,\ \ \ \ \ \qquad \big\{b_m, b_r\big\}=\big\{c_m, c_r\big\}=0\ .\nonumber
\eea
The Virasoro  central charge of the $bc$\,-\,system is equal to $(-2)$, so that the total central charge associated with
the energy momentum tensor \eqref{isisaiasaaq} is given by
\bea\label{aoisd989121}
c_{\scriptscriptstyle{\rm total}}=c_{\scriptscriptstyle{\rm WZW}}+c_{\scriptscriptstyle{\rm Gauss}}
+c_{\scriptscriptstyle{\rm ghost}}=\frac{3 n}{n+2}+1-2=
2-\frac{6}{n+2}\ .
\eea

\bigskip
The highest weight  representation for the combined chiral algebra generated  by the
Fourier  coefficients $ j_m^A,\,a_m,b_m,c_m$ is constructed  in the usual manner. 
First of all  one requires  that the  highest state
is  annihilated by   all the positive   frequency modes with $m>0$.
Since  the zero modes of the WZW currents 
satisfy    the  commutation relations
 \bea
  \big[\,j^A_0,j^B_0\,\big]=-\tfrac{\ri }{2}\, {f^{AB}}_{C}\ j^C_0\ ,
 \eea
 the highest states  form a representation of
 the ${\mathfrak{sl}}_2$ algebra. It makes sense to require that 
it is an irreducible one,  characterized by the value of the Casimir operator
\bea
{\hat C}_{\tt G}=-\kappa_{AB}\,j^A_0j^B_0\  ,
\eea
which in the  ${\mathfrak{sl}}_2$ case is usually denoted as $\ell (\ell+1)$. 
Rather than $\ell$ 
 we will employ  the parameter $p=\ell+\half$.
Together with this quantum number 
the highest  states  can be  labeled by the eigenvalues of the zero modes $j^{3}_0$ and $a_0$:
\be
{\hat C}_{\tt G}\, |p,\mu,s\rangle=
\big(p^2-\tfrac{1}{4}\big)\, \, |p,\mu,s\rangle\, ,\ \ \ \  \quad j^{3}_0\,|p,\mu ,s\rangle= \mu\ |p,\mu,s\rangle\, ,\  \ \ 
\quad
a_0\,|p,\mu,s\rangle=\tfrac{s}{\sqrt{n}}\, |p,\mu,s\rangle\ .
\ee
The highest states form a  representation not only of the  ${\mathfrak{sl}}_2$ algebra
but also the simple fermionic one
\bea
\{b_0,c_0\}=1\ ,\ \ \ b_0^2=c_0^2=0\ .
\eea
Thus we  supplement   the set
of conditions defining  them with
\bea\label{aisiaias}
c_0\, |p,\mu,s\rangle_+=0\,,\qquad\qquad
|p,\mu,s\rangle_-\equiv b_0\,|p,\mu,s\rangle_+\ .
\eea
The   highest weight representation  is built by the action of 
the negative frequency modes
  $ j_m^A,\,a_m,b_m,c_m$ with $m<0$ 
on the  highest state  multiplet.
The corresponding linear space  will be denoted by
 ${\cal A}_{p,s}$. The latter possesses a  grading induced by 
the Virasoro algebra generator $L_0^{\scriptscriptstyle{\rm (total)}}$.
 For given  ${\tt L}=0,1,2,\ldots\,$, the level subspace 
${\cal A}_{p,s}^{({\tt L})}$  is  finite dimensional and all its states
 have  the same conformal dimension $\Delta_{p,s}+{\tt L}$ with 
\bea
\Delta_{p,s}=\frac{p^2-\tfrac{1}{4}}{n+2}+\frac{s^2}{n}\ .
\eea
Note that the  conformal dimensions of the primary states
 do not depend on the quantum number $\mu$.

\bigskip
The parameter
  $p$ and its barred counterpart $\bar{ p}$ are related to  the central
elements \eqref{aoaosao}-\eqref{aisodi10982912}  of the  Poisson algebra of the WZW currents. In particular
the sum $p+\bar{ p}$ can be identified with the eigenvalues of the
operator $-\ri\, \frac{\partial}{\partial\alpha}$
with $\alpha$ being the dynamical variable from \eqref{saoosao}. 
Then the compactness condition $\alpha\sim \alpha+2\pi$ 
yields the quantization rule 
 $p+\bar{ p}\in {\mathbb Z}$. This, in  view of  the  classical relation \eqref{isiasaisi}, leads to\footnote{%
Here we identify the difference $p-\bar{ p}$  with $(n+2)\,(P-\bar{P})$.
Within the semi-classical analysis where  $n\gg 1$,  $n$ and $n+2$ are indistinguishable.
The finite renormalization $n\mapsto n+2$ 
may be advocated for using  similar arguments as those in the work \cite{Dijkgraaf:1991ba}.
}
\bea\label{rcaji8912A}
p=\half\, \big({\tt u}+(n+2)\, ({\tt k}+{\tt w})\, \big)\,,
\qquad 
\bar{p}=\half\, \big({\tt u}-(n+2)\, ({\tt k}+{\tt w})\, \big)
\ \ \ \  \quad \big({\tt u}, {\tt w}\in{\mathbb Z}\big)\, .
\eea
At the same time $s$ may take any real value,
\bea\label{rcaji8912B}
s\in{\mathbb R}\ .
\eea

\bigskip
The central r$\hat{\rm o}$le in the BRST approach belongs to
the BRST charge and the ghost number operator. These obey the relations
\bea
\hat{Q}_{\scriptscriptstyle{\rm BRST}}^2=0\ ,\ \ \ \ \qquad
\big[\hat{Q}_{\scriptscriptstyle{\rm BRST}},\,\hat{q}_{\scriptscriptstyle{\rm ghost}}\big]=\hat{Q}_{\scriptscriptstyle{\rm BRST}}\ .
\eea
In the
case at hand they read explicitly as
\bea\label{asussauas}
&&\!\!\!\!\!\!\!\!\!\!\!\!\!\!\!\!\!\!\!\!\!\!
\hat{Q}_{\scriptscriptstyle{\rm BRST}}=\frac{1}{\hbar}\int_0^{2\pi}\rd u\ \big(L^3-\partial\eta\big)\,c(u)=\big(j^3_0-\sqrt{n}\, a_0\big)\, c_0+\sum_{m\not=0} \big(j^3_m-\sqrt{n}\, a_m\big)\, c_{-m}\\[0.2cm]
&&\!\!\!\!\!\!\!\!\!\!\!\!\!\!\!\!\!\!\!
\hat{q}_{\scriptscriptstyle{\rm ghost}}=\int_0^{2\pi}\frac{\rd u}{2\pi}\ b c(u)=\ b_0c_0+\sum_{m=1}^\infty
\big(\,b_{-m}c_m-c_{-m}b_m\, \big) \ .\nonumber
\eea
It is easy to see that both  operators commute with the zero mode of the current $L^3(u)$:
\bea
\big[ j^3_0,\hat{Q}_{\scriptscriptstyle{\rm BRST}}\big]=\big[ j^3_0,{\hat q}_{\scriptscriptstyle{\rm ghost}}\big]=0\ .
\eea

Let $\mu$ be an eigenvalue of $ j^3_0$ corresponding to one of the states from the highest
state multiplet of ${\cal A}_{p,s}$. Consider the 
eigenspace ${\cal A}_{p,\mu,s}\subset {\cal A}_{p,s}$  such that
\bea\label{ysysaya}
 j^3_0\, {\cal A}_{p,\mu,s}=\mu\ {\cal A}_{p,\mu, s}\ .
\eea
Since both the  BRST charge and ghost number operator act invariantly in this subspace, one can introduce 
the component of the factor space 
$ {\rm Ker}[Q_{\scriptscriptstyle{\rm BRST}}]/{\rm Im}[Q_{\scriptscriptstyle{\rm BRST}}]$
with zero ghost number,
 \bea\label{iaosd9102812}
 \widetilde{\cal A}_{p,\mu,s}\,\subset\,
 {\rm Ker}[Q_{\scriptscriptstyle{\rm BRST}}]/{\rm Im}[Q_{\scriptscriptstyle{\rm BRST}}] |_{{\cal A}_{p,\mu,s}}\  :
\ \ \ \ \ \ 
{\hat q}_{\scriptscriptstyle{\rm ghost}} \,\widetilde{\cal A}_{p,\mu,s}= 0\ .
 \eea
The operators
$\hat{Q}_{\scriptscriptstyle{\rm BRST}},\,\hat{q}_{\scriptscriptstyle{\rm ghost}}$ and  $j^3_0$,
all
commute with the energy momentum tensor
\bea\label{aosoaoas}
\big[\hat{Q}_{\scriptscriptstyle{\rm BRST}},T_{\scriptscriptstyle{\rm total}}(u)\big]=
\big[\hat{q}_{\scriptscriptstyle{\rm ghost}},T_{\scriptscriptstyle{\rm total}}(u)\big]=
\big[j^3_0,T_{\scriptscriptstyle{\rm total}}(u)\big]=0\ .
\eea
This implies that $\widetilde{\cal A}_{p,\mu,s}$ is a  
 naturally graded 
 space  and similar to ${\cal A}_{p,s}$ admits the decomposition
 \bea
 \widetilde{\cal A}_{p,\mu,s}=\bigoplus_{{\tt L}=0}^{\infty}\, \widetilde{\cal A}_{p,\mu,s}^{({\tt L})}\ .
 \eea
 The 
 dimensions of the  level subspaces $\widetilde{\cal A}_{p,\mu,s}^{({\tt L})}$
  depend
 essentially   on  whether or not   $\mu-s$ vanishes.
This difference 
is
the coefficient in front of the ghost zero mode  $c_0$ in \eqref{asussauas}
 when the  action of  the BRST charge
is restricted 
to  the eigenspace ${\cal A}_{p,\mu,s}$.
Consider the highest states $|p,\mu,s\rangle_\pm$. If $\mu\ne s$,
then the state $|p,\mu,s\rangle_+$ is annihilated by the BRST charge.
On the other hand  $\hat{Q}_{\scriptscriptstyle{\rm BRST}}|p,\mu,s\rangle_-\ne 0$
and is proportional to $|p,\mu,s\rangle_+$. This implies that
the level subspace $\widetilde{\cal A}_{p,\mu,s}^{(0)}$ with $\mu\ne s$  is trivial.
In the case when $\mu=s$ both highest states are annihilated by the BRST charge.
However only $|p,s,s\rangle_+$ has zero ghost number so that
$\dim\big(\widetilde{{\cal A}}_{p,s,s}^{(0)}\big)=1$. 
Recall that $|p,s,s\rangle_+$ is a state from 
a ${\mathfrak{sl}}_2$ irrep characterized by  $p$.
The eigenvalues of $j_0^3$ for the other highest states from the multiplet
are given by  $\mu=s+\ri\,m$, where $m$ is a nonzero integer, and hence
the difference $\mu-s$ for these states would be nonvanishing.
Proceeding further, it is straightforward to check at least for small values of ${\tt L}=0,1,2,\ldots\,$, that
all the spaces $\widetilde{\cal A}_{p,\mu,s}^{({\tt L})}$ are trivial for $\mu\ne s$, while
the dimensions of $\widetilde{\cal A}_{p,s,s}^{({\tt L})}$ with generic $p$ is equal to the number of bipartitions of ${\tt L}$.

  \subsection{$W_\infty$\,-\,algebra}

  Perhaps the  easiest way to explore the linear structure of the factor space $ \widetilde{\cal A}_{p,\mu,s}$
  is to bosonize  the $\widehat{{\mathfrak {sl}}}(2,{\mathbb R})$ current algebra \cite{Zam1986,Wakimoto,Gerasimov:1989mz}.
  This allows one to isolate the physical states  in  ${\cal A}_{p,s}$
  and to show that 
  $\dim\big(\widetilde{\cal A}_{p,s,s}^{({\tt L})}\big)$ 
  coincides with the corresponding dimensions of the
  level subspaces of  the highest weight representation of the quantum $W_\infty$\,-\,algebra.
The bosonization of $\widehat{{\mathfrak {sl}}}(2,{\mathbb R})$ requires three chiral Bose
fields. The spin 2 current which commutes with $L^3(u)$ is identified with 
  $W_2(u)$. It can be arranged so that
\be\label{w2iosdi}
W_2=(\partial\vartheta)^2+(\partial \varphi)^2+\frac{\ri}{\sqrt{n+2}}\ \partial^2\varphi\ ,
\ee 
which involves only two of the chiral fields.
By construction the field $W_2$ commutes with the BRST charge and $\hat{q}_{{\rm ghost}}$.
Using the Operator Product Expansion (OPE),
\be
\varphi(u)\,\varphi(0)=-\tfrac{1}{2}\log(u)+O(1) \,,\qquad\qquad
\vartheta(u)\,\vartheta(0)=-\tfrac{1}{2}\log(u)+O(1) \ ,
\ee
a simple calculation shows that
\be\label{oias90190321}
 W_2(u)\,W_2(0)=\frac{c}{2 u^4}-\frac{2}{u^2}\ W_2(0)-\frac{1}{u}\ \partial W_2(0)+O(1)\,,
\ee
where  the central charge
\be\label{aoisdo90121}
c=2-\frac{6}{n+2}
\ee
coincides with $c_{\scriptscriptstyle{\rm total}}$  from eq.\,\eqref{aoisd989121}.
The spin 3 $W$ current is also obtained from the commutativity condition
with $L^3(u)$. However, it contains an ambiguity related to the addition of 
 $\partial W_2(u)$. 
The requirement that  $W_3(u)$ be a primary conformal field, i.e.,
\be
W_2(u)\,W_3(0)=-\frac{3}{u^2}\ W_3(0)-\frac{1}{u}\ \partial W_3(0)+O(1)
\ee
fixes it up to an overall multiplicative factor.
We take
\be\label{W3def1a}
W_3=
\frac{6n+8}{3n+6}\, (\partial \vartheta)^3+2\,
 (\partial \varphi)^2\partial \vartheta+\ri\sqrt{n+2}\ \partial^2 \varphi\,\partial\vartheta
-\frac{\ri n}{\sqrt{n+2}}\ \partial\varphi\,\partial^2\vartheta+\frac{n}{6(n+2)}\  \partial^3\vartheta \ .
\ee
As in  the classical case, the currents $W_2$ and $W_3$ generate a closed algebra,
whose commutation relations may be conveniently encoded via OPEs. 
In particular, the spin 4 current is obtained via the 
OPE of $W_3$ with itself:
 \bea\label{aiisaisa1a2a}
&&W_3(u)\,W_3(0)=-\frac{c(c+7)(2c-1) }{9(c-2)u^6}+\frac{ (c+7)(2c-1)}{3(c-2)u^4}\ 
\big(W_2(u)+W_2(0)\big) -\frac{1}{u^2}\times \\[0.2cm]
&& \Big( W_4(u)+W_4(0)\
+W^2_2(u)+W^2_2(0)
+\frac{2c^2+22c-25}{30 (c-2)}\,\big (\partial^2W_2(u)+\partial^2W_2(0)\big)\Big)+O(1)\ .\nonumber
\eea
The definition of $W_4$, in principle, is not unique 
and contains the freedom  in the addition of any of the spin 4 fields 
$\partial^2 W_2$, $\partial W_3$ as well as the composite field $W_2^2$.
Fixing $W_4$ as in eq.\,\eqref{aiisaisa1a2a}, it turns out that
 the singular terms $\propto u^{-6}$ and $u^{-3}$ are absent in the OPE
\be\label{8989aioiaso}
 W_2(u)\,W_4(0)=\frac{(c+10)(17c+2)}{15 (c-2)\, u^4}
 \ W_2(0)-\frac{4}{u^2}\ W_4(0)-\frac{1}{u}\ \partial W_4(0)+O(1)\,.
\ee
 This way, by recursively computing OPEs one can generate
the full set of quantum $W$ currents $\{W_j(u)\}_{j=2}^\infty$.
In the  classical limit with $n\to\infty$,
\bea\label{90s9889dfsa}
W_j\to n^{j/2}\ W_j^{(cl)}
\eea
 and the quantum $W_\infty$\,-\,algebra  
becomes the classical Poisson algebra,
whose first few PB relations are given in \eqref{jasususa} \cite{Bakas:1991fs}.
\bigskip

The quantum $W$ currents, similar to the classical ones, are periodic fields and 
can be expanded in a Fourier series,
 \be\label{aoisd182918}
 W_{j}(u)=-\frac{c}{24}\ \delta_{j,2}+\sum_{m=-\infty}^\infty {\widetilde W}_j(m)\ \re^{-\ri m u} \ .
 \ee
Note that the expansion coefficients
for the $W_2$ current, defined in such a way,
satisfy the Virasoro algebra commutation relations:
\be
[L_m,L_r]=(m-r)\, L_{m+r}+\tfrac{c}{12}\, m(m^2-1)\, \delta_{m+r,0}\ \qquad \qquad
\big( L_m\equiv{\widetilde W}_2(m)\big)\ .
\ee
The Fourier modes  ${\widetilde W}_j(m)$
are used in the construction of the highest weight irreps of the $W_\infty$\,-\,algebra.
The highest state is defined by 
the conditions
\be\label{aios3298}
\widetilde{W}_j(m)\,|\bm{\omega}\rangle=0 \qquad\qquad (\forall m>0)\,,\qquad\qquad
\widetilde{W}_j(0)\,|\bm{\omega}\rangle=\omega_j\,|\bm{\omega}\rangle\,,
\ee
where $\bm{\omega}=(\omega_2,\omega_3)$ is the highest weight.
Notice that the component $\omega_2$ is equal to the conformal dimension of the 
state.
It turns out that $|\bm{\omega}\rangle$ is fully specified by the relations \eqref{aios3298}
 with $j=2,3$.
The linear span of  states
\be
\widetilde{W}_{2}(-l_1)\ldots\widetilde{W}_{2}(-l_m)\,
\widetilde{W}_{3}(-l_{1}')\ldots\widetilde{W}_{3}(-l_{m'}')|\bm{\omega}\rangle
\ee
with $1\le l_1\le l_2\le \ldots\le l_m$ and  $1\le l_{1}'\le l_{2}'\le \ldots\le l_{m'}'$
form a representation of the $W_\infty$\,-\,algebra which is usually referred to as the Verma module.
The latter is a graded linear space, and the dimensions of its 
level subspace with  ${\tt L}=\sum_i l_i+\sum_{i'} l'_{i'}$ is given by ${\rm par}_2({\tt L})$ --
the number of bipartitions of ${\tt L}$.
For generic values of the highest weight $\bm{\omega}$ and the central charge $c$, the Verma module
is an irrep of the $W_\infty$\,-\, algebra. However, under certain conditions imposed on the parameters,
 it may contain null vectors 
-- highest states occurring at non-zero levels. 
In this case the  irrep can be obtained
from the Verma module by factoring out 
all of the  invariant subspace(s) generated by the null vector(s).
We'll parameterize the central charge by $n$ according to \eqref{aoisdo90121}, and the highest weight
$\bm{\omega}$ by the pair $(p,s)$ as
\bea\label{Deltvarpi1a}
\omega_2&=&\frac{p^2-\frac{1}{4}}{n+2}+\frac{s^2}{n}\equiv \Delta_{p,s}  \\[0.2cm]
\omega_3&=&
\frac{2 s}{\sqrt{n}}\,\Big(\,\frac{p^2}{n+2}+\frac{(3n+4)\,s^2}{3n\,(n+2)}-\frac{2n+3}{12\,(n+2)}\,\Big)\nonumber
\ .
\eea
In turn  the highest weight irrep of the $W_\infty$\,-\,algebra
will be denoted by  ${\cal W}_{p,s}\equiv{\cal W}_{-p,s}$.
\bigskip

The motivation for swapping from $\bm{\omega}$ to the pair of parameters $(p,s)$
 is based on the following.
The bosonization formulae \eqref{w2iosdi} and \eqref{W3def1a} introduce the structure of the $W_\infty$
Verma module in the Fock space ${\cal F}_{\bf P}$. The latter is
the space of the representation of two independent copies of the Heisenberg algebra, 
generated by the Fourier modes of $\partial\vartheta$ and $\partial\varphi$,
whose commutation relations are the same as in eq.\,\eqref{aoisido0190921}.
The space ${\cal F}_{\bf P}$ with ${\bf P}=(\frac{p}{\sqrt{n+2}},\frac{s}{\sqrt{n}})$ 
 is characterized by the highest weight -- the values of the zero modes of
the chiral fields $\partial\vartheta$, $\partial\varphi$ in ${\cal F}_{\bf P}$:
\be
\int_0^{2\pi}\frac{\rd u}{2\pi}\ \partial\varphi\,\big|_{{\cal F}_{{\bf P}}}=\frac{p}{\sqrt{n+2}} \,,\qquad \qquad
\int_0^{2\pi}\frac{\rd u}{2\pi}\ \partial\vartheta\,\big|_{{\cal F}_{{\bf P}}}=\frac{s}{\sqrt{n}}\ .
\ee
Then 
eqs.\,\eqref{w2iosdi} and \eqref{W3def1a} yield that the highest weight
$\bm{\omega}=(\omega_2,\omega_3)$  is related to  $p$ and $s$ as
in \eqref{Deltvarpi1a}.

\subsection{Space of states}
So far we have been mainly focused on just one chirality.
Of course, all the above follows through for the other chirality as well.
This way one concludes that the algebra of extended conformal symmetry
for the gauged ${\rm SL}(2,\mathbb{R})$
WZW model
coincides with $\overline{W}_\infty\otimes W_\infty $.
In turn the full space
of states, ${\cal H}^{({\rm cont})}$, can be decomposed into the irreps 
$\overline{\cal W}_{\bar{ p},{s}}\otimes {\cal W}_{p,s}$
of this algebra. At this point we put forward the conjecture that 
for generic values of the twist parameter ${\tt k}$, 
 the linear structure of ${\cal H}^{({\rm cont})}$ is described by 
\be\label{oasoid1092021}
{\cal H}^{({\rm cont})}=\bigoplus_{{\tt u},{\tt w}\in\mathbb{Z}}
\int^{\oplus}_{\mathbb{R}}\!\rd s\
\overline{{\cal W}}_{\bar{p},s}\otimes {\cal W}_{p,s}\qquad  
{\rm with}
\qquad 
\begin{array}{l}{  p}=\frac{1}{2}\,{\tt u}+\frac{1}{2}\,(n+2)\,({\tt k}+{\tt w})\\[0.2cm]
                                       \bar{p}=\frac{1}{2}\,{\tt u}-\frac{1}{2}\,(n+2)\,({\tt k}+{\tt w})
\end{array}\,.
\ee
The subscript ``${\rm cont}$'' is used to emphasize the presence of the direct integral in the above linear decomposition.
\bigskip

Let's clarify the condition that $-\frac{1}{2}<{\tt k}\le \frac{1}{2}$ be  generic. 
What we mean by this is that $(n+2)\,{\tt k}\notin \mathbb{Z}$.
Then, if $s$ is  real,  each of the irreps 
${\cal W}_{p,s}$ and $\overline{{\cal W}}_{\bar{p},s}$ coincide with the Verma module.
The corresponding character,
\be\label{chdef1a}
{\rm ch}_{p,s}({\tt q})\equiv{\rm Tr}_{{\cal W}_{p, s}}\Big[\,{\tt q}^{\widetilde{W}_2(0)-\frac{c}{24}}\,\Big]\,,
\ee
is given by 
\be\label{iasodi1092091}
{\rm ch}_{p,s}({\tt q})=\frac{{\tt q}^{-\frac{1}{12}+\frac{s^2}{n}+\frac{p^2}{n+2}}}{({\tt q},{\tt q})_\infty^2}
\  \ \qquad\qquad \big(\,p,s\ \ {\rm generic}\big)\ ,
\ee
where $({\tt q},{\tt q})_\infty^{-2}$  is the generating function for ${\rm par}_2({\tt L})$:
\be\label{aisuasau}
\frac{1}{({\tt q},{\tt q})_\infty^{2}}\equiv
\prod_{m=1}^\infty\frac{1}{(1-{\tt q}^{m})^{2}}=\sum_{{\tt L}=0}^\infty {\tt par}_2({\tt L})\,{\tt q}^{\tt L}
 \ .
\ee
\bigskip

Each term in the linear decomposition \eqref{oasoid1092021} can be interpreted as being the result
of  the quantization of the algebra of functions on the classical symplectic leaf $\Gamma_{\bar{P},P,B}$.
One can make the following identification 
\be
B=\re^{\frac{4\pi s}{n}}\ ,
\ee 
while in  the classical limit $\frac{\bar{p}}{n+2}\to {\bar P}$, $\frac{{p}}{n+2}\to { P}$.
The classical theory possesses a set of global symmetries, which were described in 
eqs.\,\eqref{aususuaa}-\eqref{aususuaa1} and are inherited by the quantum model.
Regarding the continuous symmetries, 
the operators $\hat{{\cal U}}_{\phi}$  and $\hat{{\cal R}}_{\theta}$,
being restricted to a $\overline{W}_\infty\otimes W_\infty$ irrep,
are proportional to the identity operator:
\begin{subequations}
\bea
\hat{{\cal U}}_{\phi}\big(\overline{{\cal W}}_{\bar{p},s}\otimes {\cal W}_{p,s}\big)&=&\label{OIASDIO120191}
\re^{\ri(\bar{ p}+p)\phi}\ \overline{{\cal W}}_{\bar{p},s}\otimes {\cal W}_{p,s}\\[0.2cm]
\hat{{\cal R}}_{\theta}
\big(\overline{{\cal W}}_{\bar{p},s}\otimes {\cal W}_{p,s}\big)&=&
\re^{2\ri s\theta}\ \label{OIASDIO120192}
\overline{{\cal W}}_{\bar{p},s}\otimes {\cal W}_{p,s}\ \,.
\eea
\end{subequations}
The generator of the ${\cal Z}_2$ symmetry, $\hat{{\cal D}}$, acts as the intertwiner,
\be
\hat{{\cal D}}\,: \  \overline{{\cal W}}_{\bar{p},s}\otimes {\cal W}_{p,s}\mapsto
\overline{{\cal W}}_{\bar{p},-s}\otimes {\cal W}_{p,-s}
\ee
and satisfies the following relations with the $W$ currents
\be
\hat{{\cal D}}\,{W}_j({u})\,\hat{{\cal D}}=(-1)^j\,{W}_j({u}) \,,\qquad\ \ \ \ 
\hat{{\cal D}}\,\overline{W}_j(\bar{u})\,\hat{{\cal D}}=(-1)^j\,\overline{W}_j(\bar{u})\ .
\ee
Similarly,  the action of the ${\cal CP}$ transformation is described by the formula
\be\label{isoaido1212}
{\cal CP}\,: \ \ \ \ \ \ \ \ 
\overline{{\cal W}}_{\bar{p},s}\otimes{\cal W}_{p,s}\mapsto
\overline{{\cal W}}_{-p,s}\otimes{\cal W}_{-\bar{p},s}\ ,
\qquad\qquad {\cal \hat{C}\hat{P}}\,{W}_j(u)={\overline{W}}_j(u)\,{\cal \hat{C}\hat{P}}\ .
\ee
\bigskip

In the classical theory the $W$ currents are real fields (see eq.\,\eqref{iaosid9891}).
Upon quantization, this translates to the conjugation conditions
\be\label{oasoidoi1982981}
\big({W}_j({u})\big)^\star={W}_j(u^*)\ ,\qquad\qquad
\big(\overline{W}_j(\bar{u})\big)^\star=\overline{W}_j(\bar{u}^*)\ .
\ee
The latter induce an inner product for the states belonging to the irreps
$\overline{{\cal W}}_{\bar{p},s}\otimes {\cal W}_{p,s}$.
In the case when the central charge $c<2$ it turns out that such an inner product is not a positive definite one 
\cite{Dixon:1989cg}.
This way the space of
states ${\cal H}^{({\rm cont})}$ is equipped with the structure of a pseudo-Hilbert space.
The conjugation conditions \eqref{oasoidoi1982981} are not sufficient to fix the
inner product of the $\overline{W}_\infty\otimes W_\infty$  primary states, $\Psi_{{\tt u},{\tt w},s}$\,,
which,
in general, has the form
\be\label{iaosido891221aa}
\big\langle{\Psi}_{{\tt u}',{\tt w}',s'}\,,
{\Psi}_{{\tt u},{\tt w},s}\big\rangle= N_{{\tt u},{\tt w},s}\ \delta_{{\tt u}',{\tt u}}\,\delta_{{\tt w}',{\tt w}}\, 
\delta(s'-s)\ .
\ee
Notice that, though the coefficients 
 $N_{{\tt u},{\tt w},s}$  contain the ambiguity
\be
N_{{\tt u},{\tt w},s}\mapsto N_{{\tt u},{\tt w},s}\ |C_{{\tt u},{\tt w},s}|^2 \ ,
\ee
a change in the normalization of the primary states 
has no effect on the sign of $N_{{\tt u},{\tt w},s}$
for given ${\tt u}$, ${\tt w}$ and $s$.
Thus, despite the ambiguity, the ``norms'' $N_{{\tt u},{\tt w},s}$
are an important characteristic of the quantum theory.

\bigskip

 \section{Lorentzian black hole NLSM}

\subsection{Space of states}
The field theory governed by the action \eqref{iasdioi120143}
with periodic boundary conditions imposed on
 $U$ and $V$ corresponds to 
the gauged WZW model with ${\tt k}=0$.
However,
as was already pointed out, even  at the classical level
this case requires some special attention.
A brief examination of formula \eqref{oasoid1092021}
specialized to ${\tt k}=0$  shows that  
  each term
$\overline{{\cal W}}_{\bar{p},s}\otimes {\cal W}_{p,s}$
in the  decomposition 
of ${\cal H}^{({\rm cont})}$  appears together with
$\overline{{\cal W}}_{-\bar{p},s}\otimes {\cal W}_{-p,s}$.
Both of them
are equivalent representations of the
 $\overline{W}_\infty\otimes W_\infty$\,-\,algebra. 
This signalizes  the presence of an additional global symmetry that
would commute with all of the $W$ currents
\be
\hat{{\cal C}}\,{W}_j({u})={W}_j({u})\,\hat{{\cal C}} \,,\qquad\ \ \ \ 
\hat{{\cal C}}\,\overline{W}_j(\bar{u})=\overline{W}_j(\bar{u})\,\hat{{\cal C}}\ .
\ee
We will refer to it as
 ${\cal C}$ conjugation. 
The space of states  ${\cal H}^{({\rm cont})}$ is splitted into the two components
\be\label{aisodioas9012}
{\cal H}^{({\rm cont})}={\cal H}^{({\rm cont})}_{{\rm even}}\bigoplus{\cal H}^{({\rm cont})}_{{\rm odd}}
\ee
having definite ${\cal C}$ parity $+1$ (even) and $-1$ (odd). 
As linear spaces, each of the components are identical.\footnote{%
The  space 
$\overline{{\cal W}}_{\bar{p},s}\otimes {\cal W}_{p,s}$ with $p=-\bar{p}=\frac{1}{2}(n+2)\,{\tt w}$
turns out to be ${\cal C}$ invariant and the splitting into its ${\cal C}$ even and ${\cal C}$ odd components
requires some clarification. This will be addressed in sec.\,\ref{sec53}.
}
Together with the $W$ currents, $\hat{{\cal C}}$ also commutes
 with
the symmetry operators $\hat{{\cal D}}$, $\hat{{\cal C}}\hat{{\cal P}}$ and $\hat{\cal R}_\theta$:
\be
\big[\hat{\cal C},\,\hat{{\cal D}}\big]=
\big[\hat{\cal C},\,\hat{{\cal C}}\hat{{\cal P}}\big]=
\big[\hat{\cal C},\,\hat{\cal R}_\theta\big]= 0\ .
\ee
However with  $\hat{{\cal U}}_\phi$
it satisfies the relation
\be
\hat{\cal C}\,\hat{{\cal U}}_\phi\,\hat{\cal C}=\hat{{\cal U}}_{-\phi}\ .
\ee
Hence, since $\phi\sim\phi+2\pi$, the operator $\hat{{\cal C}}$  commutes only with $\hat{{\cal U}}_\pi$ -- the generator
of a $180^\circ$ rotation.

\bigskip

At the classical level, the action of the $\,\hat{{\cal U}}_\pi$ transformation,
as follows from the general formula \eqref{aususuaa} and \eqref{saoosao}, reduces to flipping the sign
of the matrix $\bf{G}$:
\be\label{saqqs444}
\arraycolsep=0.3cm
{\cal U}_{\pi}\ \ :\ \ \
\begin{array}{ll}
{\bf G}\mapsto  -  {\bf G}\ ,\qquad\qquad\eta\mapsto\eta &
\ \ \ 
\\[0.4cm]
 \Gamma_{\bar{P},P,B}\mapsto \Gamma_{\bar{P},P,B}
&
\end{array}\!\!\!\!\!\! .
\ee
At the same time, the action of the ${\cal C}$ conjugation can be identified 
as
\be\label{saqqddsds}
\arraycolsep=0.3cm
{\cal C}\ \ :\ \ \
\begin{array}{ll}
{\bf G}\mapsto  -\sigma^z\, {\bf G}\,\sigma^z\ ,\qquad\qquad\eta\mapsto\eta &
\ \ \ 
\\[0.4cm]
 \Gamma_{\bar{P},P,B}\mapsto \Gamma_{-\bar{P},-P,B}
&
\end{array}
\ee
which is similar to the relation \eqref{saqqs} for the ${\cal Z}_2$ symmetry transformation ${\cal D}$.\footnote{%
 For ${\tt k}\ne 0$ the ${\cal C}$ invariance of the model
is broken by the condition \eqref{iasisisaww}.
}
Recall that applying the gauge fixing condition $X=Y$ (see \eqref{iaisaiassas})
in the original formulation of the gauged ${\rm SL}(2,\mathbb{R})$ 
WZW model, results in two identical copies of the  Lorentzian black hole NLSM  corresponding to 
the cases $X=Y>0$ and $X=Y<0$. The ${\cal C}$ conjugation intertwines these two copies.
By ``symmetrizing'' them one can arrange so that
${\cal C}$, by definition, would act as the identity operator in the Lorentzian black hole NLSM.
\bigskip

To get more  insight into the symmetries and the global structure of the phase space of the
classical model, it is useful to consider
 elementary solutions of the classical equations of motion.
These may be constructed  by setting
$f={\bar f}=0$ as well as $\xi_\pm=P$ and ${\bar \xi}_\pm={\bar P}$
in the general relations \eqref{isaisaisaisai}-\eqref{jsusausa}, yielding
\bea\label{jsausay}
U(t,x)=V(t,x)=\sin\big(\,(P+{\bar P})\, t+(P-{\bar P})\, x\, \big)\,.
\eea
One can easily see that the equations of motion corresponding to the action  \eqref{iasdioi120143} are indeed satisfied.
The periodic boundary condition  for $U$ and $V$  requires that the difference $P-{\bar P}$
be an integer,
\be\label{iasodiao3109}
P-\bar{P}=0,\pm 1,\pm 2,\ldots\ ,
\ee
which coincides with  \eqref{isiasaisi} for the case ${\tt k}=0$.
\bigskip

 The explicit solutions provide some hints concerning
the  action of the global symmetries
on the phase space.
There are two evident space-time symmetry transformations of the model \eqref{iasdioi120143}.
Namely, time-reversal and parity conjugation,
which are defined as
\bea\label{iaosid892}
&&{\cal T}\ : \ \ \  U(t,x)\mapsto \phantom{-}U(-t,x)\,,\qquad  V(t,x)\mapsto \phantom{-} V(-t,x)
\\[0.2cm]
&&{\cal P}\ : \ \  \ U(t,x)\mapsto -U(t,-x)\,,\qquad V(t,x)\mapsto -V(t,-x)\,.\nonumber
\eea
The extra sign in the definition of 
${\cal P}$  is a matter of convention
since the transformation ${\cal U}_\pi:$ $U\mapsto -U$,
 $V\mapsto -V$
is also a symmetry.
The solutions \eqref{jsausay}
are unchanged under the ${\cal P T}$ transformation. More generally, 
we will assume that two solutions related via ${\cal P T}$ belong
to the same symplectic leaf, i.e.,
\be\label{oasiod192aa}
{\cal PT}\ : \ \ \ \Gamma_{\bar{P},P,B}\mapsto \Gamma_{\bar{P},P,B}\ .
\ee
The action of ${\cal P}$ and ${\cal T}$  on the fundamental  fields, as described by formula \eqref{iaosid892},  induces
the action of these transformations 
on $\Gamma_{\bar{P},P,B}$. 
We make the assumption that two solutions 
related through separate ${\cal P}$ or ${\cal T}$
transformations
 belong to different symplectic leaves.
A brief examination of \eqref{jsausay}
motivates that 
\bea\label{aisodi1902}
&&{\cal T}\ : \ \ \Gamma_{\bar{P},P,B}\mapsto \Gamma_{-P,-\bar{P},B}\nonumber \\[0.2cm]
&&{\cal P}\ : \ \ \Gamma_{\bar{P},P,B}\mapsto \Gamma_{P,\bar{P},B}\ .
\eea
An immediate consequence is that ${\cal PT}$ maps
 $\Gamma_{\bar{P},P,B}$ to  $\Gamma_{-\bar{P},-P,B}$.
In view of the condition \eqref{oasiod192aa}  
the following identification must be made
\be\label{isaisisaias}
\Gamma_{\bar{P},P,B}\equiv \Gamma_{-\bar{P},-P,B}\ .
\ee
Notice that the latter is required for setting the ${\cal C}$ conjugation \eqref{saqqddsds} 
to be the identity transformation for the Lorentzian black hole NLSM.
Then, without loss of generality, one can assume that
\be\label{iasoid18392}
P+\bar{P}\ge 0\ .
\ee
It should be emphasized that the identification \eqref{isaisisaias} is expected to hold true only in the case ${\tt k}=0$,
as otherwise neither ${\cal P}$ nor ${\cal C}$ separately are symmetries of the model.
\bigskip

Taking into account \eqref{iasoid18392} one arrives at the conjecture
that the space of states of the Lorentzian black hole
NLSM is given by
\be\label{iaosidoi192009AA}
{\cal H}^{({\scriptscriptstyle{\rm LBH}})}=
\bigoplus_{{\tt u}= 0}^{\infty}\Bigg[
\bigoplus_{{\tt w}=-\infty}^{\infty}\int^{\oplus}_{\mathbb{R}}\!\rd s\
\overline{{\cal W}}_{\bar{p},s}\otimes {\cal W}_{p,s}\Bigg]\qquad  
{\rm with}
\qquad 
\begin{array}{l}{  p}=\frac{1}{2}\,{\tt u}+\frac{1}{2}\,(n+2)\,{\tt w}\\[0.2cm]
                                       \bar{p}=\frac{1}{2}\,{\tt u}-\frac{1}{2}\,(n+2)\,{\tt w}
\end{array}\,.
\ee
At first glance, in view of the above discussion of the ${\cal C}$ conjugation,
the formula seems to be a straightforward consequence of \eqref{oasoid1092021}.
However, one should  keep in mind that the notation ${\cal W}_{p,s}$
 stands for the \emph{irreducible} highest weight 
representation of the $W_\infty$\,-\,algebra.
For $(n+2)\,{\tt k}\notin \mathbb{Z}$ and  $s\in\mathbb{R}$ each of the irreps 
${\cal W}_{\bar{p},s}$ and ${\cal W}_{p,s}$ coincides with the Verma module.
However for ${\tt k}=0$ some of the Verma modules become reducible. It turns out
that  (see, e.g., \cite{Jayaraman:1989tu,Bazhanov:2019xvyA})
\be\label{oasid89189212}
{\cal V}er_{ \rho, s}={\cal W}_{ \rho,s}\bigoplus {\cal W}_{ \rho+m(n+2),s}\ \ \ \ \ \ \ 
 {\rm for}\ \ \  \ \rho=\tfrac{1}{2}\,\big(r-m\,(n+2)\,\big)\, ,\ \   m,r=1,2,\ldots
\ee
Note that the character of ${\cal W}_{ \rho,s}$  in the direct sum  reads as
\be\label{charadegen1aaa}
{\rm ch}_{\rho,s}({\tt q})={\tt q}^{-\frac{1}{12}+\frac{s^2}{n}+\frac{\rho^2}{n+2}}
\  \frac{1-{\tt q}^{mr}}{({\tt q},{\tt q})_\infty^{2}}\ \ ,\  \ \  \ \qquad
\begin{array}{l}
\rho= \frac{1}{2}\,\big(r-m\,(n+2)\,\big)\,,\ \ \ m,r=1,2,\ldots\  \\[0.2cm]
s\in\mathbb{R}\,,\ \ n\,{\rm-\, generic}\,,
\end{array}
\ee
while the character of  $ {\cal W}_{ \rho+m(n+2),s}$ is described by eq.\,\eqref{iasodi1092091}.
This way the ${\cal C}$ even component of the space  \eqref{oasoid1092021} as ${\tt k}\to 0$
admits the decomposition
\bea\label{ioasid1892981}
{\cal H}^{({\rm cont})}_{\rm even}={\cal H}^{({\scriptscriptstyle{\rm LBH}})}
\bigoplus {\cal H}^{(\rm null)}\ .
\eea
Here ${\cal H}^{({\scriptscriptstyle{\rm LBH}})}$ is given by \eqref{iaosidoi192009AA}, while 
the space  ${\cal H}^{(\rm null)}$ is a direct sum of two components,
\bea\label{aosid9182233AA}
{\cal H}^{(\rm null)}={\cal H}^{(\rm null)}_{+}\bigoplus{\cal H}^{(\rm null)}_{-} \ .
\eea
The latter  are decomposed identically into the irreps of the  algebra of extended conformal symmetry
\bea\label{aopsid9120asas}
{\cal H}^{(\rm null)}_{\pm}
=\bigoplus_{{\tt v},{\tt w}=1}^{+\infty}\int^{\oplus}_{\mathbb{R}}\!\rd s\
\overline{{\cal W}}_{{\rho},s}\otimes {\cal W}_{\rho,s}\ \ \ \ \ \ \ \ \ \ \ \  \big(\,\rho=\tfrac{1}{2}\,{\tt v}+\tfrac{1}{2}\,(n+2)\,{\tt w}\, \big)\ .
\eea
The highest state 
in either one $\overline{{\cal W}}_{{\rho},s}$ or ${\cal W}_{\rho,s}$,
appearing in the integrand, coincides with the null
vector in the original Verma module (see \eqref{oasid89189212}).
\bigskip

There is a simple minded argument in support of 
the dropping of   the ``null'' component 
in \eqref{ioasid1892981}  for the construction  of
the space of states of the Lorentzian black hole NLSM.
The space ${\cal H}^{({\rm null})}$  admits an evident symmetry, which
interchanges the subspaces ${\cal H}^{({\rm null})}_+$
and ${\cal H}^{({\rm null})}_-$ in \eqref{aosid9182233AA}.
On the other hand, there is no clear manifestation of such an additional ${\cal Z}_2$
symmetry for the model described by the classical action \eqref{iasdioi120143}.

\subsection{Minisuperspace approximation\label{sec213}}
For a better qualitative understanding of the quantum NLSM it is useful to consider the model within the
so-called minisuperspace approximation. This entails 
taking into account only those
field configurations that do not depend on the space co-ordinate $x$,
such as the classical solutions \eqref{jsausay}  with $P=\bar{P}$.
We still keep $U$ and $V$ to satisfy the constraint $0\le UV<1$ corresponding  to the
union of regions  III and IV in fig.\,\ref{fig20}.
For a  preliminary analysis 
it is convenient to parameterize $U,\,V$ from this domain as
\bea\label{iaoisod1}
U=\re^{\Theta}\ \sin(\Phi)\ ,\ \ \ \ \ V=\re^{-\Theta}\ \sin(\Phi)\ ;\ \ \ \  \Phi\in(-\tfrac{\pi}{2},\tfrac{\pi}{2})\ ,
\ \ \ \Theta\in(-\infty,\infty)\ .
\eea
Then the minisuperspace version of the  classical action   \eqref{iasdioi120143}  reads as
\bea\label{aksisa}
{ S}^{({\rm ms})}_{\scriptscriptstyle{\rm LBH}}
=\frac{\pi}{\hbar}\int \rd t \ \big(\,\dot{\Phi}^2-\tan^2(\Phi)\ \dot{\Theta}^2\,\big)\ .
\eea
Since the generalized coordinate  $\Theta$ is cyclic, its conjugate  momentum 
$\Pi_\Theta=-\tan^2(\Phi)\, \dot{\Theta}$  is an integral of motion. 
The  effective  Lagrangian (the Routhian)  for  
the non-cyclic degree of freedom  is given by
\bea
L_{\rm eff}=\frac{1}{2}\ \big(\dot{\Phi}^2-V_{\rm eff}(\Phi)\,\big)\ , \ \ \ \  \       
V_{\rm eff}(\Phi)=-\Pi_\Theta^2\ \cot^2(\Phi) \ .
\eea
 The latter  describes a 1D particle falling to the origin $\Phi=0$.
An elementary calculation shows  that for any value $\Pi_\Theta\not= 0$ the particle, starting  its motion at
 $t=0$,  reaches  the origin 
in a finite amount of time $t_{\rm fall}<+\infty$.   For $t>t_{\rm fall}$ the motion remains undetermined.
Thus  the action \eqref{aksisa} specifies the time evolution of the mechanical system   only within 
a finite time interval
(except for the trajectories with  $\Pi_\Theta=0$). To continue the classical trajectories  for $t>t_{\rm fall}$
the unbounded effective potential should be somehow regularized. There are of course numerous ways
of doing this.
A simple minded one is to  replace  $ V_{\rm eff}(\Phi)=-\Pi_\Theta^2\ \cot^2(\Phi)$
by a  smooth  potential $V^{(\rm reg)}_{\rm eff}(\Phi)$, which together with 
its 
derivative is bounded from below 
within the infinitesimal interval $\Phi\in (-\epsilon,\epsilon)$.
Outside this interval 
$V^{(\rm reg)}_{\rm eff}(\Phi)=V_{\rm eff}(\Phi)$.
To keep the original symmetry of the potential we  assume  that the regularized one is an
even function:
\bea\label{aisaisai}
V^{(\rm reg)}_{\rm eff}(\Phi)=V^{(\rm reg)}_{\rm eff}(-\Phi)\ .
\eea
Then the motion of $\Phi$
becomes   globally defined and periodic  for any values of $\Pi_\Theta\ne 0$.

\bigskip
With basic intuition from quantum mechanics,
we can  predict  the symmetry properties of the minisuperspace stationary wave functions.
First of all, that the regularized potential is an even function of $\Phi$   implies that 
the stationary states may be assigned a parity $\sigma=\pm 1$,
\bea\label{usassayt}
\hat{\cal U}_\pi\,\Psi^{(\sigma)}(U,V)\equiv\Psi^{(\sigma)}(-U,-V) =\sigma\, \Psi^{(\sigma)}(U,V)\ ,
\eea
where we now switch to the original target space coordinates $(U,V)$.
This relates the values of the wave function in the domains III 
and IV from fig.\,\ref{fig20}.
Next, $\Psi^{(\sigma)}$ can be chosen to be  an eigenfunction  of the operator
$\hat{\Pi}_\Theta=\frac{\hbar}{\ri}\,\partial_\Theta=\frac{\hbar}{\ri}\,(U\partial_U-V\partial_V)$:
\bea\label{aisaiasi1a}
\hat{\Pi}_\Theta\,  \Psi^{(\sigma)}_s=2\hbar s\ \Psi^{(\sigma)}_s\ .
\eea
It follows that
\bea\label{hassysa}
\Psi^{(\sigma)}_s(U,V)= \bigg(\frac{U}{V}\bigg)^{\ri s}\ F^{(\sigma)}_s(UV)\ .
\eea
The minisuperspace approximation ignores  the presence of the oscillatory  modes
so that  the  wave functions $\Psi^{(\sigma)}_s$ are  expected to correspond
 to   $\overline{W}_\infty\otimes W_\infty$ primary  states,
characterized by $p=\bar{ p}$ and $s$ (it is instructive to compare eqs.\,\eqref{usassayt} and \eqref{aisaiasi1a} with
\eqref{OIASDIO120191} and \eqref{OIASDIO120192}, respectively).
In turn the minisuperspace energy  becomes  
$\Delta_{p,s}+\Delta_{{\bar p},s}=2\Delta_{p,s}$ in the leading non-vanishing order of $\hbar=\frac{2\pi}{n}$
(the approximation is  reliable only in the limit  $n\to\infty$). Namely,
\bea\label{iassausa}
E^{\rm (ms)}=\tfrac{\hbar}{\pi}\ \big(p^2+s^2-
\tfrac{1}{4}\big)\ .
\eea
At this point $p$ can be thought  of as a real number  
parameterizing the minisuperspace
energy $E^{\rm (ms)}$  and the corresponding wavefunction $\Psi^{(\sigma)}_{p,s}$.
Since the highest weight  is an even function of $p$,
\bea
\Psi^{(\sigma)}_{-p,s}(U,V)=\Psi^{(\sigma)}_{p,s}(U,V)\,.
\eea
As was emphasized previously, one can assume that
$p={\bar p}\geq 0$.

\bigskip
Though the 
highest weight of the $W_\infty$ irrep $\bm{\omega}=(\omega_2,\omega_3)$ 
is not sensitive to the sign of $p$, 
as follows from \eqref{Deltvarpi1a}
it does depend on the sign of  $s$: $\omega_3(p,-s)=-\omega_3(p,s)$.
 Thus  
 the primary states characterized by $(p,s)$ and $(p,-s)$
are distinguishable. They are related
through the ${\cal Z}_2$ transformation, so that
\bea\
{\hat {\cal D}}\, \Psi^{(\sigma)}_{p,s}(U,V)=\Psi^{(\sigma)}_{p,-s}(U,V)\ .
\eea
On the other hand, by definition, this symmetry  interchanges $U$ and $V$:
\bea
{\hat {\cal D}}\, \Psi^{(\sigma)}_{p,s}(U,V)\equiv\Psi^{(\sigma)}_{p,s}(V,U)\ .
\eea
Combining the above two relations with \eqref{hassysa} one  concludes that
\bea\label{isaisisaa}
\Psi^{(\sigma)}_{p,s}(U,V)= \bigg(\frac{U}{V}\bigg)^{\ri s}\ F^{(\sigma)}_{p,s}(UV)\ ,\ \ \ \ \ {\rm where}\ \ \ \ 
 F^{(\sigma)}_{p,s}(z)= F^{(\sigma)}_{-p,s}(z)= F^{(\sigma)}_{p,-s}(z)\ .
\eea

\bigskip

Having described the symmetry properties
of the stationary wave functions, we turn to deriving them explicitly.
In the work \cite{Dijkgraaf:1991ba}, a minisuperspace analysis was performed for the NLSM 
\eqref{iasdioi120143} with the fields $U$, $V$ belonging to region I  from fig.\,\ref{fig20} (or equivalently II).
Though this is not the domain of interest,  
we can still follow the same line of arguments of that paper.
In particular, up to a trivial factor,  the minisuperspace  Hamiltonian
coincides with the ``dilatonic'' Laplacian:
\bea\label{isaiasias2a}
\hat{H}^{\rm (ms)}=-\frac{\hbar }{4\pi}\ \triangle_D\ ,\ \ \ \ \ \triangle_D=\frac{1}{\re^{D}\sqrt{-G}}\ \partial_i
\big(\re^D\sqrt{-G}\,G^{ij}\partial_j\big)\ ,
\eea
where  the metric is the one in  \eqref{hsasaysaty} and the dilaton field is given by
\bea
 D= \log(1-UV)\ .
\eea
The stationary Schr${\rm{\ddot o}}$dinger equation $\hat{H}^{\rm (ms)}\,\Psi=E^{\rm ( ms)}\, \Psi$
reads explicitly as
\bea\label{ajsaasu}
-\big(\, (1-UV)\ \partial_U\partial_V-\tfrac{1}{2}
\ ( U\partial_U+V\partial_V)\, \big)\, \Psi=\tfrac{\pi}{\hbar} \,E^{\rm (ms)}\ \Psi\ .
\eea
Using the general form   \eqref{hassysa}  for the stationary  wave functions  and
parameterizing the energy as in \eqref{iassausa},  it is straightforward to
show that $F(z)=z^{-\ri s}\, F^{(\sigma)}_{p,s}(z)$ obeys the 
 Gauss  hypergeometric equation
\bea\label{isaisai}
z\,(1-z)\,F''+\big(1+2\ri s-2\,(1+\ri s)\,z\,\big)\,
F'-\big( \tfrac{1}{2}+\ri s+{   p}\big)\big( \tfrac{1}{2}+\ri s-{   p}\big)\,F=0\, .
\eea
Keeping in mind our preliminary analysis,
the ODE  \eqref{isaisai}  is applicable only in the
domain $\epsilon^2<z<1$ with  a 
small  regularization parameter $\epsilon\ll 1$ (recall that $z=UV=\sin^2(\Phi)$).

\bigskip
The function
$F^{(\sigma)}_{p,s}(z)$ \eqref{isaisisaa}  is a certain linear combination of 
$z^{\pm \ri s}\ {}_2F_{1}\big(\tfrac{1}{2}\pm \ri s+{   p},
\tfrac{1}{2}\pm\ri s-{   p}, 1\pm 2\ri s, z)$, which
can be specified as follows.  Applying the
elementary identity
\be\label{hasysa}
 \re^{D}\sqrt{-G}\, \big(\Psi^*_1\,  \hat{H}^{\rm( ms)}\,\Psi_2-\Psi_2 \,\hat{H}^{\rm( ms)} \, \Psi^*_1\big)=
\frac{\hbar}{4\pi }\ \partial_i \Big[   \re^{D} \sqrt{-G}\, G^{ij} \big(\Psi_2 \partial_j\Psi_1^*-\Psi_1^* \partial_j\Psi_2\big)\Big]
\ee
to the pair of stationary wave functions $\Psi_1$, $\Psi_2$
corresponding to the energies $E^{\rm ( ms)}_1$, $E^{\rm ( ms)}_2$ and then integrating
the result  over the domain ${\rm B}_\epsilon: \ \epsilon^2<UV<1$, one obtains
\be\label{hsasaysay}
\big(E^{\rm ( ms)}_2-E^{\rm ( ms)}_1\,\big)\ 
\int_{{\rm B}_\epsilon}\rd U\rd V\  \re^{D}\sqrt{-G}\  \Psi^*_1  \Psi_2=\frac{\hbar}{4\pi }\
\int_{\partial {\rm B}_\epsilon}\rd\ell \,  
\re^{D}\, \big(\Psi_2 \partial_n\Psi_1^*-\Psi_1^* \partial_n\Psi_2\big)\ .
\ee
Here the integral in the r.h.s. is taken over the boundary of ${\rm B}_\epsilon$, 
which is the union of $UV=\epsilon^2$ and
$UV=1$. Also, 
$\partial_n$ stands for the normal derivative to $\partial {\rm B}_\epsilon$. 
 As was discussed before, the wave functions possess a definite parity. Due to this
either the wave function or its normal derivative vanishes at $UV=0$. Hence as $\epsilon\to 0$
the horizon  $UV=0$
 does not contribute to the r.h.s. of  eq.\,\eqref{hsasaysay}. Further, since the dilaton factor $\re^D$ 
vanishes at
 the black hole singularity  $UV=1$
one could make 
 the whole boundary integral vanish
 by imposing that both the eigenfunctions and their normal derivatives remain finite at 
$UV=1$. In this case the wave functions corresponding to different energies would be orthogonal w.r.t. the
inner product
\bea\label{ususu}
\big\langle\Psi_1,\Psi_2\big\rangle= \int_{0< UV<1}\rd U\rd V\  \re^{D}\sqrt{-G}\  \Psi^*_1  \Psi_2\ .
\eea
This suggests to take $F^{(\sigma)}_{p,s}(z)$ in \eqref{isaisisaa}
as
\bea\label{isaias}
F^{(\sigma)}_{p,s}(z)=z^{\ri s}\ 
{}_{2}F_{1}\big(\tfrac{1}{2}+\ri s+{   p},
\tfrac{1}{2}+\ri s-p, 1; 1-z)\ \ \ \ \ \ \ (\epsilon^2<z<1)
\eea
or, equivalently,
\bea
F^{(\sigma)}_{p,s}(z)&=&A_{p,+s}\  z^{+\ri s}\ {}_2F_{1}\big(\tfrac{1}{2}+ \ri  s+{   p},
\tfrac{1}{2}+\ri s-{   p}, 1+ 2 \ri s; z)\\[0.2cm]
&+
&A_{p,-s}\  z^{ -\ri s}\ {}_2F_{1}\big(\tfrac{1}{2}- \ri s+{   p},
\tfrac{1}{2}-\ri s-p, 1- 2 \ri s; z)\ ,\nonumber
\eea
where
\bea
A_{p,s}=\frac{\Gamma(-2\ri s)}
{\Gamma(\frac{1}{2}-\ri s-{   p}) \Gamma(\frac{1}{2}-\ri s+{   p})}\ .
\eea
\bigskip

For $z\ll 1$ it is convenient to use the variable $y$ such that $z=\re^{y}$. Then
$F^{(\sigma)}_{p,s}$ asymptotically approaches to a superposition of two plane waves
 \bea
 F^{(\sigma)}_{p,s}\asymp A_{p,+s}\ 
\re^{+\ri s y}+A_{p,-s}\ \re^{-\ri s y}\ \ \ \ \ \ \ \ \  \big( 1\ll (-y)<2\,\log(1/\epsilon)\big)\ .
 \eea
The regularized   interaction discussed before in the domain $(-y)>2\,\log(1/\epsilon)$  results in
 a quantization condition for $s$ 
 \bea\label{ioasio1290}
 \epsilon^{-4\ri s}\ \re^{\frac{\ri }{2}\delta^{\rm (ms)}(p,s)}\asymp \sigma\ .
 \eea
 The phase shift $\delta^{\rm (ms)}$ here depends on the precise form of the regularized potential.
As $\epsilon\to 0$, the spectrum of $s$ becomes continuous and is
characterized by the  density of states
\bea\label{oas091errr}
\rho^{\rm(ms)}(s)=\tfrac{2}{\pi}\ \log(1/\epsilon)+\tfrac{1}{4\pi}\ \partial_s\,\delta^{\rm(ms)}(p,s)\ .
\eea
The corresponding minisuperspace wave functions would be orthogonal w.r.t. the inner product \eqref{ususu}:
\bea\label{asusausa}
\big\langle\Psi_{p',s'}^{(\sigma')},\Psi_{p,s}^{(\sigma)}\,\big\rangle\propto
\delta_{p',p}\ \delta_{\sigma',\sigma}\ \delta(s'-s)\ .
\eea
Here we use the Dirac $\delta$-function for $s$ since the latter
can be any real number. 
At the same time the Kronecker symbol indicates that $p$ belongs to  some  discrete set.
The quantization of $p$ seems rather natural once we note that  the term 
$\frac{\hbar}{\pi}\,(p^2-\frac{1}{4}) $ in 
the formula for the minisuperspace energy  \eqref{iassausa} can be interpreted as the contribution
of the non-cyclic degree of freedom $\Phi$, which executes  periodic motion in the regularized 
effective potential. This is consistent with our discussion of the quantization of the 
Lorentzian black hole NLSM. Setting  ${\tt w}=0$ in formula \eqref{iaosidoi192009AA}
giving the admissible values of $p$ and $\bar{ p}$, one has
$
2{ p}=2\bar{ p}={\tt v}=0,1,2,\ldots\ $.
Also $\delta_{\sigma',\sigma}$ in \eqref{asusausa} can be ignored --
 the sign factor $\sigma$ is not an independent quantum number
and is defined by the parity of the integer ${\tt v}$ (see eq.\,\eqref{OIASDIO120191} with $\phi=\pi$).

\section{Low energy states of the ${\cal Z}_2$ invariant spin chain
in the scaling limit. Continuous spectrum}
In the  seminal work \cite{Baxter:1971}, Baxter  introduced a
multiparametric, integrable,
statistical system  that
covers a variety of  classes of critical behaviour.
In particular, it was observed in \cite{Jacobsen:2005xz} that the ${\cal Z}_2$ invariant spin chain,
corresponding to a certain specialization of the parameters
of the general Baxter model, is critical and   possesses a continuous  spectrum of scaling dimensions. 
The spin chain was  subsequently
 studied in the works  
\cite{Ikhlef:2008zz,IJS2,Ikhlef:2011ay,Frahm:2012eb,Candu:2013fva,Frahm:2013cma,Bazhanov:2019xvy}. 
In the recent paper \cite{Bazhanov:2019xvyA}
a systematic analysis, including a  study of  the
finite size corrections, was performed.
Arguments were presented 
that the low energy states, in a suitably defined scaling limit,
organize into the space ${\cal H}^{({\rm cont})}$, whose linear structure
 is described by eq.\,\eqref{oasoid1092021}.
This leads to the idea that the critical behaviour of the spin chain is governed by
the gauged  WZW model.
Here, accepting the conjecture, we use the 
results obtained for the lattice system to
move forward in the study of the field theory.

\subsection{Global symmetries and Hermitian structure}
The subject of our interest is a spin $\frac{1}{2}$ chain of length $N$,
which is always an even number, governed by the Hamiltonian\footnote{%
This form for the Hamiltonian, up to 
an overall multiplicative factor and an additive constant, 
appeared in ref.\cite{IJS2}.
The one defined by eq.\,(2) in the work \cite{Bazhanov:2019xvy} coincides with
$
\hat{\mathsf{V}}\,\mathbb{H}\,\hat{\mathsf{V}}^{-1}
$,
where $\mathbb{H}$ is as in \eqref{aioiisa}, while 
$\hat{\mathsf{V}}=\prod_{m=1}^{N/2}\,\exp\big(\frac{\ri\pi}{4}\,\sigma^z_{2m-1}\big)$.
}
\bea\label{aioiisa}
{\mathbb H}&=&\frac{1}{\sin(2\gamma)}\
\sum_{m=1}^{N}\,\Big(2\sin^2(\gamma)\ \sigma^z_m\,\sigma^z_{m+1}-
\big(\sigma^x_m\,\sigma^x_{m+2}+\sigma^y_m\,\sigma^y_{m+2}+
\sigma^z_m\,\sigma^z_{m+2}\big)\nonumber
\\[0.2cm]
&-&\ri\sin(\gamma) \big(\sigma_m^x\sigma_{m+1}^x+
\sigma_m^y\sigma_{m+1}^y\big)
\big(\sigma^z_{m-1}-\sigma^z_{m+2}\big)
\,\Big)
+N\cot(2\gamma)\, \hat{{\bf 1}}\,.
\eea
The operators $\sigma^A_m$
stand for the Pauli matrices that act non-trivially in the $m$-th factor
of the tensor product
 \be\label{vec1}
{\mathscr V}_N=\mathbb{C}^2_N\otimes
 \mathbb{C}^2_{N-1}\otimes\cdots\otimes\mathbb{C}^2_1\qquad \qquad\qquad (N\,-\,{\rm even})\ .
\ee
They are taken to 
satisfy the quasiperiodic boundary conditions:
\bea\label{sisisaisu}
\sigma^{\pm}_{N+\ell}=\re^{\pm 2\ri\pi{\tt k}}\ \sigma^{\pm}_{\ell}\ ,\ \ \ \ \ \ \ 
\sigma^{z}_{N+\ell}=\sigma^{z}_{\ell}\ \ \ \ \ \ \ \ \ 
(\ell=1,2)\ ,
\eea
where
\be
\sigma^\pm_m=\tfrac{1}{2}\,(\sigma^x_m\pm\ri\sigma^y_m)\ .
\ee
The Hamiltonian commutes with the $z$ projection of the total spin operator
\be
\mathbb{S}^z=\frac{1}{2}\sum_{m=1}^N\sigma^z_m\,:\ \ \ \ \ \ \ \ 
\big[\mathbb{S}^z,\,{\mathbb H}\big]=0\ ,
\ee
which  is the infinitesimal generator of the ${\rm U}(1)$ symmetry.
The action of the finite rotation $\hat{{\cal U}}_\phi$  on the local
spin operators  is given by
\be\label{aisod8212}
\hat{{\cal U}}_\phi\,\sigma^\pm_m\ \hat{{\cal U}}_\phi^{-1}=\re^{\pm\ri\phi}\ \sigma^\pm_m\,,
\qquad\qquad
\hat{{\cal U}}_\phi\,\sigma^z_m\,\hat{{\cal U}}_\phi^{-1}=\sigma^z_m
\ .
\ee
 Another evident symmetry of \eqref{aioiisa} and \eqref{sisisaisu} is
${\cal C}{\cal P}$-invariance.
The corresponding transformation
 is described through the formula
\bea
\hat{\cal C}\hat{\cal P}\,\sigma^{{{\pm}}}_m\,
\hat{\cal C}\hat{\cal P}=\sigma^{{{\mp}}}_{N+1-m}\, , \ \ \ \ \ \ \ 
\hat{\cal C}\hat{\cal P}\,\sigma^{{{z}}}_m\,
\hat{\cal C}\hat{\cal P}=-\sigma^{{{z}}}_{N+1-m}\ \ \ \ \ \ (\,m=1,\ldots, N\,)\ .
\eea
A characteristic property of the  model
is the presence of an additional ${\cal Z}_2$ symmetry. The adjoint action of
its generator $\hat{{\cal D}}$ on $\sigma^A_m$ is more involved and
for odd $m$  reads as
\begin{subequations}\label{Dadjoint1}
\bea\label{Dadjoint1a}
\hat{\cal D}\,\sigma^\pm_m\,\hat{\cal D}&=&\frac{1}{\cos(\gamma)}\ \Big( \sigma^{\pm}_{m+1}-\ri\sin(\gamma)\,
\sigma^z_{m+1}\sigma^{\pm}_m\,\Big)\\
\hat{\cal D}\,\sigma^z_m\,\hat{\cal D}&=&
\frac{1}{\cos^2(\gamma)}\,\Big(
\sigma^z_{m+1}-\sin^2(\gamma)\ \sigma^z_m+2\ri\sin(\gamma)\
\big(\sigma^+_{m+1}\,\sigma^-_m+\sigma^-_{m+1}\,\sigma^+_m\big)\,\Big)\,,\nonumber
\eea
while for even $m$:
\bea\label{Dadjoint2a}
\hat{\cal D}\,\sigma^\pm_m\,\hat{\cal D}&=&\frac{1}{\cos(\gamma)}\ \Big(\sigma^{\pm}_{m-1}+\ri\,\sin(\gamma)\,
\sigma^{\pm}_m\sigma^{z}_{m-1}\,\Big)\\
\hat{\cal D}\,\sigma^z_m\,\hat{\cal D}&=&
\frac{1}{\cos^2(\gamma)}\,\Big(
\sigma^z_{m-1}-\sin^2(\gamma)\ \sigma^z_m-2\ri\,\sin(\gamma)\
\big(\sigma^-_{m}\,\sigma^+_{m-1}+\sigma^+_{m}\,\sigma^-_{m-1}\big)\,\Big)\ . \nonumber
\eea
\end{subequations}
The lattice system also possesses the time-reversal symmetry
generated by  the anti-unitary transformation $\hat{{\cal T}}$, such that
\be
\hat{{\cal T}}\bm{\Psi}=\bigg(\prod_{m=1}^N\sigma^x_m\bigg)\,\bm{\Psi}^*\,,\qquad\qquad\qquad
\bm{\Psi}\in\mathscr{V}_N\ .
\ee
\bigskip

With the anisotropy parameter in the domain $0<\gamma<\pi$,
the spin chain \eqref{aioiisa},\,\eqref{sisisaisu} 
is critical.
However, different types of  critical behaviour occur depending on whether
$\gamma\in(0,\frac{\pi}{2})$ or $\gamma\in(\frac{\pi}{2},\pi)$.
The latter case was considered in ref.\cite{IJS2}.
The relation between the spin chain and the gauged WZW model, which was proposed in the work
\cite{Bazhanov:2019xvyA}, occurs when $0<\gamma<\frac{\pi}{2}$. 
Then $\gamma$ is related to the parameter $n\equiv2\pi/\hbar$
from the field theory side as
\be
\gamma=\frac{\pi}{n+2}\qquad\qquad \qquad (\,0<n<+\infty\,)\ .
\ee
The twist parameter ${\tt k}\in\big(-\frac{1}{2},\frac{1}{2}\,\big]$ in \eqref{sisisaisu} is identified with 
that entering into the boundary condition \eqref{isisai}.
\bigskip

The Hamiltonian \eqref{aioiisa} is not Hermitian 
w.r.t. the usual matrix Hermitian conjugation ${\cal O}^\dag=\big({\cal O}^T\big)^*$.
 Nevertheless one can introduce
a conjugation, 
\be\label{asido91029012}
\hat{{\mathsf O}}^\star=
\hat{{\mathsf X}}^{-1}_\star\ \hat{{\mathsf O}}^\dag\  \hat{{\mathsf X}}_\star\ ,
\ee
for which the
Hamiltonian satisfies
\be\label{APSOIDioi12989}
\mathbb{H}^\star=\mathbb{H}\ .
\ee 
The expression for  the matrix
$\hat{{\mathsf X}}_\star=\hat{{\mathsf X}}_\star^\dag$ is given by formula (19.62) in 
\cite{Bazhanov:2019xvyA}. It should be kept in mind that the $\star$\,-\,conjugation does not
correspond to any positive definite inner product.
A manifestation of this is that some  of the eigenvalues  of the $2^N$ dimensional matrix $\mathbb{H}$ are complex.
The analysis of \cite{Bazhanov:2019xvyA} shows that the conjugation \eqref{asido91029012}
in the scaling limit induces the field theory conjugation for the $W$ currents \eqref{oasoidoi1982981} in
the space ${\cal H}^{({\rm cont})}$. Moreover it yields that the normalization
 of the $W_\infty\otimes W_\infty$
primary states can be chosen such that their ``norms'', i.e., the coefficients $N_{{\tt u},{\tt w},s}$ entering into eq.\,\eqref{iaosido891221aa}, 
are given by
\be\label{iaosido891221}
N_{{\tt u},{\tt w},s}=
\frac{\Gamma(1+\frac{2\bar{p}}{n+2})\,\Gamma(1+\frac{2 p}{n+2})}
{\Gamma(1+2\bar{p})\,\Gamma(1+2 p)}\qquad  \quad
{\rm with}
\qquad \quad
\begin{array}{l}{ p}=\frac{1}{2}\,{\tt u}+\frac{1}{2}\,(n+2)\,({\tt k}+{\tt w})\\[0.2cm]
                                       \bar{p}=\frac{1}{2}\,{\tt u}-\frac{1}{2}\,(n+2)\,({\tt k}+{\tt w})
\end{array}\,.
\ee
Note that the norms are independent of $s$, while
the quantum number ${\tt u}$ is identified with the eigenvalue of the operator
$\mathbb{S}^z$:
\be\label{iaosdi90129012}
{\tt u}=S^z=0,\pm1,\pm2,\ldots\ .
\ee

\bigskip

Similar to the gauged WZW model with ${\tt k}=0$,
the spin chain subject to periodic boundary conditions
possesses an extra symmetry
-- that of ${\cal C}$ conjugation. For the finite lattice system
the matrix $\hat{{\cal C}}$ is given by
\be\label{oaspodp9012}
\hat{{\cal C}}=c_N
\prod_{m=1}^N\,(\eta_m)^{\frac{1}{2}\sigma^z_m}\,\sigma^x_m\,,
\qquad \quad {\rm where} \qquad\quad \eta_m=(-1)^{m+1}\,\ri
\ee
and the choice of the overall sign factor, $c_N^2=1$, is a matter of convention.\footnote{%
We found it convenient to set
\be\label{aosid9818921}
c_N=\begin{cases}(-1)^{N/4} & \quad N/2\,-\,{\rm even}\\[0.2cm]
1 & \quad N/2\,-\,{\rm odd}
\end{cases}\  .
\ee
For $N/2$ even  the ground state, i.e., the state with the lowest possible 
energy (ordered w.r.t the real part),
of the lattice Hamiltonian with periodic boundary conditions
is non-degenerate. With this  choice of $c_N$  its ${\cal C}$ parity is equal to $+1$.
When $N/2$ is odd there are two ground states, forming a  ${\cal Z}_2$ doublet, which are distinguished
by their ${\cal C}$ parity.
}
Since $\hat{{\cal C}}$ anticommutes with the $z$ projection of the total spin,
\be
\hat{{\cal C}}\,\mathbb{S}^z=-\mathbb{S}^z\,\hat{{\cal C}}\ ,
\ee
the ${\cal C}$ even and ${\cal C}$ odd components of the space of states of 
the spin chain would not be invariant w.r.t. the action of the ${\rm U}(1)$ 
transformation \eqref{aisod8212}, except the case $\phi=\pi$. Nevertheless 
$|S^z|=0,1,2,\ldots$ is a well defined quantum number
for the states from each component.
Taking the scaling limit  results in the spaces 
${\cal H}^{({\rm cont})}_{\rm even}$ and ${\cal H}^{({\rm cont})}_{\rm odd}$, which
appear in eq.\,\eqref{aisodioas9012}.
Recall that of special interest is the subspace 
${\cal H}^{({\scriptscriptstyle{\rm LBH}})}$ \eqref{iaosidoi192009AA} of 
${\cal H}^{({\rm cont})}_{\rm even}$, which is expected to serve as the space
of states for the Lorentzian black hole NLSM. The Hermitian structure  of ${\cal H}^{({\scriptscriptstyle{\rm LBH}})}$
is specified  through the conjugation conditions \eqref{oasoidoi1982981} of the $W$ currents 
as well as the norms of the primary $W_\infty\otimes W_\infty$ states occurring in its decomposition.
The latter may be obtained from \eqref{iaosido891221} via a taking of the limit ${\tt k}\to 0$.
However, special care is needed for the states with ${\tt u}=0$ and ${\tt w}\ne 0$,
as the $\Gamma$\,-\,functions in that formula become singular. 
The issue is treated in the work \cite{Bazhanov:2019xvyA} with the result that
\be\label{iaosido891221aa1A}
\big\langle{\Psi}_{{\tt u}',{\tt w}',s'}\,,
{\Psi}_{{\tt u},{\tt w},s}\big\rangle_{\scriptscriptstyle{\rm  LBH}}=N_{{\tt u},{\tt w}}^{({\scriptscriptstyle{\rm  LBH}})}
\ \delta_{{\tt u}',{\tt u}}\,\delta_{{\tt w}',{\tt w}}\, 
\delta(s'-s)\, \,,
\ee
where
\be\label{aosidi39034sa}
N_{0,0}^{({\scriptscriptstyle{\rm  LBH}})}=1\,,\qquad N_{{\tt u},{\tt w}}^{({\scriptscriptstyle{\rm  LBH}})}=
\begin{cases}
(-1)^{\tt w}\ \,\dfrac{\sin(\pi(n+2){\tt w})}{\pi (n+2)}\ \ \ &
\quad {\tt u}=0,\,{\tt w}\ne 0\\[0.6cm]
\dfrac{\Gamma(1-{\tt w}+\frac{{\tt u}}{n+2})\,
\Gamma(1+{\tt w}+\frac{{\tt u}}{n+2})}{\Gamma(1+{\tt u}-(n+2)\,{\tt w})\,
\Gamma(1+{\tt u}+(n+2)\,{\tt w}) } 
\  \ \  &\quad {\tt u}\geq 1,\,{\tt w}\in\mathbb{Z}
\end{cases}
\ee
Then ${\cal H}^{({\scriptscriptstyle{\rm LBH}})}$  is a pseudo-Hilbert space equipped with a non-positive
definite inner product.
This would reflect the fact that 
the target space for the  NLSM \eqref{iasdioi120143} has Lorentzian signature.

\subsection{Density of states}
The distinguishing feature of the spin chain \eqref{aioiisa},\,\eqref{sisisaisu}
in the parametric domain $0<\gamma<\frac{\pi}{2}$  is 
that the spectrum of the rescaled energy $({\cal E}-{\cal E}_{{\rm vac}})\,N$,
though it remains discrete for finite $N$,
becomes densely distributed as $N\to\infty$.
The analysis
of the scaling limit leads to a certain 
density matrix for the gauged ${\rm SL}(2,\mathbb{R})$ WZW model as well as the Lorentzian black hole
NLSM.
Here we give a brief summary of the
relevant results obtained in refs.\cite{Jacobsen:2005xz,Ikhlef:2008zz, Ikhlef:2011ay,
Frahm:2012eb,Frahm:2013cma,Candu:2013fva,Bazhanov:2019xvy,Bazhanov:2019xvyA}.

\bigskip

A key r$\rm{\hat{o}}$le in the description of the scaling limit
of the ${\cal Z}_2$ invariant spin chain belongs to the so-called
quasi-shift operator $\mathbb{B}$. The latter was first introduced in
ref.\cite{Ikhlef:2011ay}. The expression for this operator,
in the conventions adopted in this paper,
is given by eq.\,(8.4) of the work \cite{Bazhanov:2019xvyA}.
The quasi-shift operator belongs to the commuting family of matrices,
which includes the Hamiltonian $\mathbb{H}$, the $z$ projection of the total spin 
$\mathbb{S}^z$ and
the lattice translation operator $\mathbb{K}$, whose matrix elements read as
\be\label{Kformula1}
\big({{\mathbb K}}\big)_{a_{N} a_{N-1}\ldots
  a_1}^{b_{N}b_{N-1}\ldots b_1}=\re^{\ri\pi{\tt k}\,(a_1+a_2)}
\,\delta_{a_N}^{b_{N-2}}\,\delta_{a_{N-1}}^{b_{N-3}}\,\ldots\,
\delta_{a_1}^{b_{N-1}}\ .
\ee
Here the indices $a_m$ and $b_m$ take the values $\pm$ and
label the states in the space \eqref{vec1}.
\bigskip

Each eigenstate belonging to the low energy part of the spectrum
can be assigned, together with $S^z$,
the ``winding number'' ${\tt w}=0,\pm1,\pm2,\ldots$ and a pair of non-negative integers
$(\bar{\tt L},{\tt L})$, which are referred to as the levels.
In view of the conjectured relation with the gauged WZW model, we swap the notation $S^z$ in favour of
 ${\tt u}$ (see \eqref{iaosdi90129012}) and also use $p$ and $\bar{p}$
defined through eq.\,\eqref{rcaji8912A}.
The extensive numerical work performed in refs.\cite{Jacobsen:2005xz,Ikhlef:2008zz, Ikhlef:2011ay,
Frahm:2012eb,Frahm:2013cma,Candu:2013fva,Bazhanov:2019xvy}
suggests that
 the large $N$ behaviour of the eigenvalues of the Hamiltonian  $\mathbb H$ and the lattice translation operator
$\mathbb K$
is described by the formula
\begin{subequations}\label{tower1}
  \bea\label{tower1a}
  {\cal E}&=&e_\infty\,N  +\frac{4\pi v_{\tt F}}{N }\ \bigg(\frac{{p}^2+{\bar p}^2}{n+2}+\frac{2 b^2}{n}-\frac{1}{6}+
 {\tt L}+\bar{\tt L}\bigg)+o\big(N^{-1-\varepsilon}\big)\\[0.2cm]
\label{tower1b}
K&=& \exp\bigg(\frac{4\pi\ri}{N}\,\bigg( \frac{{p}^2-{\bar p}^2}{n+2}+ {\tt L}-\bar{\tt L}\bigg)\bigg)\ .
  \eea
\end{subequations}
Here
 \bea\label{uassaysa}
e_{\infty}\,=\, -\frac{2 v_{{\rm F}}}{\pi}\ \int_0^\infty{\rm d}t\ \frac{\sinh\big(\frac{2 t}{n}\big)}
{\sinh\big(\frac{(n+2)t}{n}\big)\,\cosh(t)}\ , \qquad \qquad \qquad
v_{{\rm F}}\,=\,\frac{2(n+2)}{n}\ ,
\eea
while the correction
 term $o\big(N^{-1-\epsilon}\big)$ contains an infinitesimally small positive $\varepsilon>0$
 (for a  more detailed  description of the correction term see ref.\cite{Bazhanov:2019xvy}).
The structure \eqref{tower1} looks typical for the low energy spectrum of a critical 1D system,
where the states organize into the conformal towers \cite{Cardy:1986ie}. However, an unusual feature
is the presence of the $N$\,-\,dependent term $\propto b^2$ with $b=b(N)$. The latter turns out
to be related to the eigenvalue of the quasi-shift operator computed on the state
\be\label{poapso1a}
b(N)=\frac{n}{4\pi}\,\log(B)\ ,\qquad\qquad
\mathbb{B}\,\bm{\Psi}=B\,\bm{\Psi}\  .
\ee
\bigskip

It is important to keep in mind the following point.
In writing the asymptotic formulae \eqref{tower1} as well as \eqref{poapso1a} 
we have implicitly assigned an $N$ dependence to a stationary state
$\bm{\Psi}=\bm{\Psi}_N$.
For a general  lattice system there are obvious difficulties in  doing this,
i.e., forming Renormalization Group (RG) trajectories
for individual  states. Of course, 
 since the space of states  of a finite lattice model has different dimensions for different lattice sizes,
the problem  only makes sense for the low energy part of the spectrum. 
It is clear how to assign an $N$  dependence to the ground state  or, for that matter, the lowest energy
state in  a disjoint sector of the space of states 
(say in a sector with given value of $S^z$ for the case under consideration). 
However 
 forming  individual   RG flow trajectories 
 for low energy stationary states that are 
densely distributed does not seem to be a trivial task.
For the ${\cal Z}_2$ invariant spin chain  the problem was
discussed in ref.\cite{Bazhanov:2019xvy} and essentially exploits the integrable 
structure.
\bigskip

It turns out that
the large $N$ behaviour of $b=b(N)$ for a state $\bm{\Psi}=\bm{\Psi}_N$
can be described through the asymptotic relation
\bea\label{quantC1}
 \epsilon^{-4\ri b}\ \re^{\frac{\ri}{2}\delta_{\bm{\Psi}}(b)}\,=\sigma+
O\big(\log (1/\epsilon)^{-\infty}\big)\,,\qquad\qquad \sigma=(-1)^{\frac{N}{2}-S^z}\ .
\eea
Here $\delta_{\bm{\Psi}}(b)$ depends on the  stationary state under consideration
and,  for future convenience, we swap $N$ for the parameter $\epsilon$, defined as
\be\label{ioasd89121}
\epsilon^{-1}=
\frac{2^{\frac{n+2}{n}}\,\Gamma\big(\frac{3}{2}+\frac{1}{n}\big)}
{\sqrt{\pi}\,\Gamma\big(1+\frac{1}{n}\big)}\  N\,.
\ee
Formula \eqref{quantC1} resembles
the quantization condition   of a  
particle in a potential well of length $\propto\log(1/\epsilon)$
with $\delta_{\bm{\Psi}}$ being the phase shift picked up by the particle at the turning points.
It has the same form as the quantization condition 
\eqref{ioasio1290} appearing in our discussion of the Lorentzian black
hole NLSM. For the ``primary'' states, corresponding to vanishing levels ${\tt L}=\bar{\tt L}=0$ in 
eq.\,\eqref{tower1}, the explicit formula for the phase shift was proposed in ref.\cite{Ikhlef:2011ay}:
\be\label{oisaoid132}
\re^{\frac{\ri}{2}\delta_{\bm{\Psi}}(s)}=
\ \frac{\Gamma(\frac{1}{2}+p-{\ri s})\,\Gamma(\frac{1}{2}+{\bar p}-{\ri s})}
{\Gamma(\frac{1}{2}+p+{\ri s})\,\Gamma(\frac{1}{2}+{\bar p}+{\ri s})}\qquad
\qquad
\qquad (\,{\tt L}=\bar{\tt L}=0\,)\ .
\ee
In the later work \cite{Kotousov:2019nvt} a closed form expression for
 $\delta_{\bm{\Psi}}$ was obtained for an arbitrary low energy state.

\bigskip

There is a class of low energy states such that $\Im m\big(b(N)\big)\to 0$ as $N\to\infty$.
In the scaling limit they form the space ${\cal H}^{({\rm cont})}$, whose linear structure
is described through the  decomposition
\eqref{oasoid1092021} into the highest weight irreps of the $\overline{W}_\infty\otimes W_\infty$\,-\,algebra 
with $c=2-\frac{6}{n+2}$. As usual, the exact knowledge of the phase shift
is sufficient to derive the density of states that occurs in the continuous limit. In particular,
let $\rho_{\bar{p},p}^{(\bar{\tt L},{\tt L})}(s)\,\Delta s$ 
be the number of states with given $p$, $\bar{p}$, ${\tt L}$ and $\bar{\tt L}$ such that
$\Re e\big(b(N)\big)\in (s,s+\Delta s)$. Then as $\epsilon\propto N^{-1}\to 0$, 
the density of states is given by
\bea\label{aisodio12311}
&&\rho_{\bar{p},p}^{(\bar{\tt L},{\tt L})}(s)=\frac{2}{\pi}\ 
{\rm par}_2({\tt L})\,{\rm par}_2(\bar{\tt L})\ 
\log\big(1/\epsilon\big)+\tilde{\rho}_{\bar{p},p}^{(\bar{\tt L},{\tt L})}(s)+o(1)\ ,
\eea
where the finite part reads as
\bea\label{aasas312321A}
\tilde{\rho}_{\bar{p},p}^{(\bar{\tt L},{\tt L})}(s)
&=&\frac{1}{2\pi\ri}\ \partial_s
\log\bigg[\,
\big(\mathfrak{D}^{(\bar{\tt L})}_{\bar{p}}(s)\big)^{{\rm par}_2({{\tt L}})}\
\big(\mathfrak{D}^{({{\tt L}})}_{{p}}(s)\big)^{{\rm par}_2(\bar{{\tt L}})}\,\bigg]
\eea
with
\be\label{aasas312321B}
{\mathfrak D}^{(\ell)}_\rho(s)=
\bigg(\frac{\Gamma(\frac{1}{2}+\rho-\ri s)}{\Gamma(\frac{1}{2}+\rho+\ri s)}
\bigg)^{{\tt par}_2(\ell)}\  
\prod_{a=0}^{\ell-1}\,
\Bigg[\frac{\big(\tfrac{1}{2}+a+\rho-\ri s\big)\,\big(\tfrac{1}{2}+a-\rho-\ri s\big)}
{\big(\tfrac{1}{2}+a+\rho+\ri s\big)\,\big(\tfrac{1}{2}+a-\rho+\ri s\big)}\Bigg]^{{\rm par}_2(\ell)-d_{a}(\ell)} \ .
\ee
Recall that ${\rm par}_2(\ell)$ denotes the number of bipartitions of $\ell$.
The integers $d_{a}(\ell)$ appearing in the exponent in the last formula
are defined through their  generating function,
\be\label{Zdef1b}
\chi_{a}({\tt q})\equiv \frac{1}{({\tt q},{\tt q})^{2}_\infty}\  
\sum_{m=0}^\infty (-1)^{m}\ {\tt q}^{m a+\frac{m(m+1)}{2}}=
\sum_{\ell=0}^\infty {d}_{a}(\ell)\,{\tt q}^\ell\ .
\ee
\bigskip

Introduce the density matrix $\hat{\rho}$, which is an operator
acting in ${\cal H}^{({\rm cont})}$ that commutes with the CFT Hamiltonian
and total momentum operator
\be
\hat{H}_{\rm CFT}=L_0+\bar{L}_0-\frac{c}{12}\ , \qquad \qquad
\hat{P}_{\rm CFT}=L_0-\bar{L}_0\ .
\ee
Being restricted to the level subspace of the highest weight representation
$\overline{{\cal W}}_{{\bar{p}},s}\otimes {\cal W}_{p,s}$,
the operator $\hat{\rho}$
is given by
\be\label{aoisd98923}
\hat{\rho}\,\big|_{\overline{{\cal W}}_{{\bar{p}},s}^{(\bar{\tt L})}\otimes{\cal W}_{p,s}^{({\tt L})}}=
\bigg[\,\frac{2}{\pi}\,\log(1/\epsilon)+\frac{\tilde{\rho}_{\bar{p},p}^{(\bar{\tt L},{\tt L})}(s)}
{{\rm par}_2(\bar{\tt L})\,{\rm par}_2({\tt L})}\,
\bigg]\ \bar{\tt q}^{\Delta_{\bar{p},s}-\frac{c}{24}+\bar{{\tt L}}}\
{\tt q}^{\Delta_{p,s}-\frac{c}{24}+{\tt L}}\ \hat{{\bf 1}}\ ,
\ee
where $\Delta_{p,s}$ is as in \eqref{Deltvarpi1a},
while
${\tt q}$ and $\bar{\tt q}={\tt q}^*$ are two complex conjugated numbers such
that $|{\tt q}|<1$.
The contribution of the low energy states forming ${\cal H}^{({\rm cont})}$ 
in the scaling limit
to the  spin chain partition function reads as
\bea\label{iaosido12032}
Z^{({\rm cont})}&=&
\sqrt{\frac{n}{\Im m(\tau)}}\ \ 
\frac{\log\big(1/\epsilon\big)}{\pi\, (\bar{{\tt q}},{\bar{\tt q}})_\infty^{2}({\tt q},{\tt q})_\infty^{2}}\ 
\sum_{{\tt u},{\tt w}=-\infty}^\infty
\bar{{\tt q}}^{-\frac{1}{12}+\frac{\bar{p}^2}{n+2}}\ 
{\tt q}^{-\frac{1}{12}+\frac{p^2}{n+2}}\\[0.2cm]
&+&
\sum_{{\tt u},{\tt w}=-\infty}^\infty\int_{-\infty}^{+\infty}\rd s\
\sum_{{\tt L},\bar{\tt L}\ge 0}\tilde{\rho}_{\bar{p},p}^{(\bar{\tt L},{\tt L})}(s)\ 
\bar{{\tt q}}^{-\frac{1}{12}+\frac{s^2}{n}+\frac{\bar{p}^2}{n+2}+\bar{{\tt L}}}\ 
{\tt q}^{-\frac{1}{12}+\frac{s^2}{n}+\frac{p^2}{n+2}+{\tt L}}\ ,\nonumber
\eea
where $\Im m(\tau)=-\frac{1}{4\pi}\log({\tt q}\bar{\tt q})$.
This can be equivalently 
expressed as the trace  of the density matrix:
\be\label{aoisd812912}
{ Z}^{({\rm cont})}={\rm Tr}_{{\cal H}^{({\rm cont})}}(\hat{\rho})\ .
\ee
\bigskip

It is possible to perform the sum over  ${\tt L}$ and $\bar{\tt L}$  
in the second line of the  formula  \eqref{iaosido12032} and show that
\bea\label{apsido10291}
\sum_{{\tt L},\bar{\tt L}\ge 0}
\tilde{\rho}_{\bar{p},p}^{(\bar{\tt L},{\tt L})}(s)\ 
\bar{{\tt q}}^{\bar{{\tt L}}} {\tt q}^{{\tt L}}=-
\frac{r_{\bar p}(s,\bar{\tt q})+r_{ p}(s,{\tt q})}{\pi\, (\bar{{\tt q}},{\bar{\tt q}})_\infty^{2}({\tt q},{\tt q})_\infty^{2}}
\eea
with
\bea\label{iaosid1298}
r_{ p}(s,{\tt q})&=&\frac{1}{2}\ \sum_{\sigma=\pm}\psi\big(\tfrac{1}{2}+p+\ri\sigma s\big)\\
&+& \oint_{|z|<1}\frac{\rd z}{2\pi\ri}\ \frac{({\tt q},{\tt q})^2_\infty}{
(z,{\tt q})_\infty (z^{-1}\,{\tt q},{\tt q})_\infty}\ 
\ \frac{1}{2}\sum_{\sigma,\sigma'=\pm }\Phi(z,1,\tfrac{1}{2}+\sigma' p+\ri\sigma s)\  .\nonumber
\eea
Here $\psi(\alpha)=\partial_\alpha\log \Gamma(\alpha)$,  while  $\Phi(z,1,\alpha)$ stands for the  Lerch transcendent,
\bea
\Phi(z,s,\alpha)=\sum_{m=0}^\infty\frac{z^m}{(m+\alpha)^s}\ .
\eea
The relation \eqref{apsido10291} is useful for the numerical computation of $Z^{({\rm cont})}$.
\bigskip

\section{Density matrix for the Lorentzian black hole NLSM\label{sec53}}
Since ${\cal H}^{({\scriptscriptstyle{\rm LBH}})}$ \eqref{iaosidoi192009AA}
 is expected to be the space of states of the Lorentzian black hole NLSM,
the equilibrium density matrix 
\be
\hat{\rho}_{\scriptscriptstyle{\rm  LBH}}:\  \ \ \ \ \ \ {\cal H}^{(\scriptscriptstyle{\rm LBH})}\mapsto 
{\cal H}^{({\scriptscriptstyle{\rm LBH}})}\ ,\qquad
\qquad \big[\hat{\rho}_{\scriptscriptstyle{\rm  LBH}},\,\hat{H}_{\rm CFT}\big]=
\big[\hat{\rho}_{\scriptscriptstyle{\rm  LBH}},\,\hat{P}_{\rm CFT}\big]=0
\ee
is of special interest.
The  space ${\cal H}^{({\scriptscriptstyle{\rm LBH}})}$ admits the decomposition
\be
{\cal H}^{({\scriptscriptstyle{\rm LBH}})}=\bigoplus_{{\tt u}=0}^\infty\,
\bigoplus_{{\tt w}=-\infty}^\infty\,{\cal H}_{{\tt u},{\tt w}}^{({\scriptscriptstyle{\rm LBH}})}\ .
\ee
The operator $\hat{\rho}_{\scriptscriptstyle{\rm  LBH}}$  acts invariantly in the 
sectors ${\cal H}_{{\tt u},{\tt w}}^{({\scriptscriptstyle{\rm LBH}})}$,
which are the
 linear span of states 
with given quantum numbers ${\tt u}$ and ${\tt w}$.
In a similar manner, 
\be\label{aiosd90121}
{\cal H}^{({\rm cont})}=\bigoplus_{{\tt u}=-\infty}^\infty\,
\bigoplus_{{\tt w}=-\infty}^\infty
{\cal H}_{{\tt u},{\tt w}}^{({\rm cont})}
\ee
and
the sectors ${\cal H}_{{\tt u},{\tt w}}^{({\rm cont})}$
are invariant subspaces for 
$\hat{\rho}$ \eqref{aoisd98923}
for any  value of ${\tt k}$
including ${\tt k}=0$.

\bigskip
When ${\tt u}=0$ and for any ${\tt w}\in\mathbb{Z}$, the highest weight representations
 $\overline{{\cal W}}_{\bar{p},s}\otimes {\cal W}_{p,s}$
with ${\tt k}\ne0$ remain irreducible at ${\tt k}= 0$.
In the last case, the irreps are also ${\cal C}$ invariant.
As a result, the subspace ${\cal H}_{0,{\tt w}}^{({\scriptscriptstyle{\rm LBH}})}$
coincides with the 
${\cal C}$ even component
of ${\cal H}_{0,{\tt w}}^{({\rm cont})}$.
 The latter occurs in the
scaling limit of the low energy states with $S^z=0$ and given ${\tt w}$.
It turns out that for fixed $N\gg 1$ the difference between the
number of ${\cal C}$ even and odd such low energy  stationary states 
  is of order one. Furthermore, in the limit $N\to\infty$
the density of ${\cal C}$ even and odd states with fixed ${\tt L}$, $\bar{\tt L}$,
${p}=-\bar{p}=(n+2)\,{\tt w}$ is the same and coincides with
$\frac{1}{2}\,\rho_{\bar{p},p}^{(\bar{\tt L},{\tt L})}$ described by 
eqs.\,\eqref{aisodio12311}-\eqref{aasas312321B} (for further details 
see sec.\,17.4 from ref.\cite{Bazhanov:2019xvyA}). 
Thus  the restriction of the density matrix  
$\hat{\rho}_{\scriptscriptstyle{\rm  LBH}}$
to the level subspaces of the  irreps occurring in the decomposition  
of ${\cal H}_{0,{\tt w}}^{({\scriptscriptstyle{\rm LBH}})}$
is given by
\bea\label{aoisd9892322222}
\hat{\rho}_{\scriptscriptstyle{\rm  LBH}}\,\big|_{
\overline{{\cal W}}_{{\bar{p}},s}^{(\bar{\tt L})}\otimes{\cal W}_{p,s}^{({\tt L})}}&=&
\frac{1}{2}\
\bigg[\,\frac{2}{\pi}\,\log(1/\epsilon)+\frac{\tilde{\rho}_{\bar{p},p}^{(\bar{\tt L},{\tt L})}(s)}
{{\rm par}_2(\bar{\tt L})\,{\rm par}_2({\tt L})}\,
\bigg] \\[0.2cm] 
&\times & \bar{\tt q}^{\Delta_{\bar{p},s}-\frac{c}{24}+\bar{{\tt L}}}\
{\tt q}^{\Delta_{p,s}-\frac{c}{24}+{\tt L}}\ \hat{{\bf 1}}\qquad
\qquad \qquad \big(\,p=-\bar{p}=\tfrac{1}{2}\,(n+2)\,{\tt w}\, , \ {\tt w}\in{\mathbb Z}\,\big)\ . \nonumber
\eea
\bigskip

For ${\tt w}=0$
the  irreps  $\overline{{\cal W}}_{\bar{p},s}\otimes {\cal W}_{p,s}$,
like in the previous case, remain irreducible at ${\tt k}=0$.
Also the subspaces ${\cal H}^{({\scriptscriptstyle{\rm LBH}})}_{{\tt u},0}$
and ${\cal H}^{({\rm cont})}_{{\tt u},0}$ for ${\tt u}>0$ are equivalent.
Hence the operator
$\hat{\rho}_{\scriptscriptstyle{\rm  LBH}}$ restricted to the level subspaces 
of the corresponding irreps  
is given by the same formula as \eqref{aoisd98923}:
\bea
\hat{\rho}_{\scriptscriptstyle{\rm  LBH}}\,
\big|_{\overline{{\cal W}}_{{\bar{p}},s}^{(\bar{\tt L})}\otimes{\cal W}_{p,s}^{({\tt L})}}&=&
\bigg[\,\frac{2}{\pi}\,\log(1/\epsilon)+\frac{\tilde{\rho}_{\bar{p},p}^{(\bar{\tt L},{\tt L})}(s)}
{{\rm par}_2(\bar{\tt L})\,{\rm par}_2({\tt L})}\,
\bigg]\   \\[0.2cm]
&\times &
\bar{\tt q}^{\Delta_{\bar{p},s}-\frac{c}{24}+\bar{{\tt L}}}\
{\tt q}^{\Delta_{p,s}-\frac{c}{24}+{\tt L}}\ \hat{{\bf 1}}\  
\qquad\qquad \quad\qquad\big(\,p=\bar{p}=\tfrac{1}{2},1,\tfrac{3}{2},2,\ldots\,\big)\ .
\nonumber
\eea
However some care is needed in specializing
the density of states 
to $p=\bar{p}=\frac{1}{2}\,{\tt u}$ for odd ${\tt u}$. 
In this case, as follows from 
eqs.\,\eqref{aasas312321A} and \eqref{aasas312321B} the  function
$\tilde{\rho}_{\bar{p},p}^{(\bar{\tt L},{\tt L})}(s)$
contains a simple pole at  $s=0$,
making its
 integration 
over $s$ ambiguous.
The ambiguity
can be resolved by starting with non-zero ${\tt k}$ and performing
the limit ${\tt k}\to 0$.  Using the Sokhotski-Plemelj formula one finds
\bea\label{aposdoio12099102}
\tilde{\rho}_{\bar{p},p}^{(\bar{\tt L},{\tt L})}(s)&=&
{\rm P.V.}\,\Big(\tilde{\rho}_{\bar{p},p}^{(\bar{\tt L},{\tt L})}(s)\Big) \\[0.2cm]
&+&\big(\,{\rm par}_2({\tt L})\,
d_{p-\frac{1}{2}}(\bar{\tt L}) \,-\,
{\rm par}_{2}(\bar{\tt L})\,d_{p-\frac{1}{2}}({\tt L}) \,\big)\,\delta(s)\qquad
  \big(\,p=\bar{p}=\tfrac{1}{2},\tfrac{3}{2},\tfrac{5}{2},\ldots\,\big) \nonumber 
\eea
where the symbol ${\rm P.V.}$ stands for the principal value, while the integers $d_a({\tt L})$ are 
defined through their generating function in eq.\,\eqref{Zdef1b}.

\bigskip

When ${\tt u}> 0$ and ${\tt w}\ne 0$
the subspaces ${\cal H}_{{\tt u},{\tt w}}^{({\scriptscriptstyle{\rm LBH}})}$ and 
${\cal H}_{{\tt u},{\tt w}}^{({\rm cont})}$ do not coincide. In turn,
the operator $\hat{\rho}_{\scriptscriptstyle{\rm  LBH}}$ can not be obtained through
a ${\tt k}\to 0$ limit of $\hat{\rho}$. Instead, one 
should return to the lattice system and compute the density of 
low energy stationary states that become part of 
 ${\cal H}_{{\tt u},{\tt w}}^{({\scriptscriptstyle{\rm LBH}})}$ in the scaling limit.
To express the result, together with the function $\mathfrak{D}^{({\ell})}_{{\rho}}(s)$
\eqref{aasas312321B}, we use
\bea
\widetilde{\mathfrak D}^{(\ell)}_{\rho}(s)&=&
\prod_{a=0}^{\ell-1}
\bigg(\frac{\frac{1}{2}+a+\rho-\ri s}
{\frac{1}{2}+a+\rho+\ri s}\bigg)^{\tilde{d}_{a}(\ell\,|+\rho)}  \ 
\bigg(\frac{\frac{1}{2}+a-\rho-\ri s}
{\frac{1}{2}+a-\rho+\ri s}\bigg)^{\tilde{d}_{a}(\ell\,|-\rho)} \\[0.2cm]
&\times &
\bigg(\frac{\Gamma(\frac{1}{2}+\rho-\ri s)}{\Gamma(\frac{1}{2}+\rho+\ri s)}
\bigg)^{{\tt par}_2({\ell})-{\tt par}_2({\ell}-mr)}\,,\  \qquad {\rm where}\qquad
\rho=\tfrac{1}{2}\,r-\tfrac{1}{2}\,(n+2)\,m\ .\nonumber
\eea
The generating function for the exponents $\tilde{d}_{a}(\ell\,|\pm\rho)$ entering into the above product 
is given by
\be
\sum_{\ell=0}^\infty \tilde{d}_{a}(\ell\,|\pm\rho)\,{\tt q}^\ell\,=\,
\frac{1}{({\tt q},{\tt q})^{2}_\infty}\ \sum_{j=1}^\infty (-1)^{j+1}\ {\tt q}^{ja+\frac{j(j+1)}{2}}\,\big(\,1-
{\tt q}^{(m\pm j)\,r}\,\big)\ .
\ee
Note that we take by definition ${\rm par}_2(\ell-mr)=0$ when $\ell<mr$.
With this notation, the density matrix restricted to the subspaces 
${\overline{{\cal W}}_{{\bar{p}},s}^{(\bar{\tt L})}\otimes{\cal W}_{p,s}^{({\tt L})}}$ with
\be
p=\tfrac{1}{2}\,{\tt u}+\tfrac{1}{2}\,(n+2)\,{\tt w} \ ,\qquad\qquad
\bar{p}=\tfrac{1}{2}\,{\tt u}-\tfrac{1}{2}\,(n+2)\,{\tt w}\qquad {\rm and}\qquad 
{\tt u}>0\,,\ {\tt w}\ne 0
\ee
is given by
\be
\hat{\rho}_{\scriptscriptstyle{\rm  LBH}}\,
\big|_{\overline{{\cal W}}_{{\bar{p}},s}^{(\bar{\tt L})}\otimes{\cal W}_{p,s}^{({\tt L})}}\,=\,
\bigg(\,\frac{2}{\pi}\,\log(1/\epsilon)+
\frac{1}{2\pi\ri}\,\partial_s\big(\,f^{(\bar{{\tt L}})}_{\bar{p}}(s)+f^{({\tt L})}_p(s)\,\big)
\bigg)\times
\bar{\tt q}^{\Delta_{\bar{p},s}-\frac{c}{24}+\bar{{\tt L}}}\
{\tt q}^{\Delta_{p,s}-\frac{c}{24}+{\tt L}}\ \hat{{\bf 1}}
\ee
where
\be\label{aoisd8912kn211}
f^{(\ell)}_{\rho}(s)\,=\, \begin{cases}
\dfrac{\log\mathfrak{D}^{({\ell})}_{\rho}(s)}{{\rm par}_2(\ell)}
 & {\rm for} \qquad m<0 \\[0.4cm]
\dfrac{\log\widetilde{\mathfrak{D}}^{({\ell})}_{\rho}(s)}{
{\rm par}_2(\ell)-{\rm par}_2(\ell-mr)}\ \ \ \ \ \ 
 & {\rm for} \qquad m>0
\end{cases}\quad
\qquad \big(\,\rho = \tfrac{1}{2}\,r-\tfrac{1}{2}\,(n+2)\,m\,\big)\ .
\ee

\section{Low energy states of the ${\cal Z}_2$ invariant spin chain
in the scaling limit. Discrete spectrum}
The low energy  spectrum of the ${\cal Z}_2$ invariant
spin chain consists of two classes of states which 
are distinguished by the large $N$ behaviour of 
the eigenvalue of the quasi-shift operator. 
Up till now we have been focused on the states, where
the imaginary part of $b(N)$ \eqref{poapso1a}
vanishes as $N\to\infty$. In the scaling limit these organize into the
space ${\cal H}^{({\rm cont})}$. 
For $(n+2)\,{\tt k}\notin\mathbb{Z}$
the linear 
structure of this space is given by \eqref{oasoid1092021}, while its Hermitian structure
is specified through the conjugation conditions \eqref{oasoidoi1982981} along with
the inner product of the $\overline{W}_\infty\otimes W_\infty$ primary states 
\eqref{iaosido891221aa},\,\eqref{iaosido891221}.
For the states from the second class $\lim_{N\to\infty}b(N)$
is a pure imaginary number whose admissible values 
form a discrete set. The energy-momentum spectrum 
is still described by the large $N$ asymptotic formula \eqref{tower1}.
We'll refer
to the space into which the states organize in the scaling limit as ${\cal H}^{({\rm disc})}$.
Here we quote the results of ref.\cite{Bazhanov:2019xvyA} regarding its linear and Hermitian structure for
$(n+2)\,{\tt k}\notin\mathbb{Z}$.

\subsection{Decomposition into the irreps of the algebra of extended conformal symmetry}
The space ${\cal H}^{({\rm disc})}$ is splitted into two sectors
\be\label{eqstart1A}
{\cal H}^{({\rm disc})}={\cal H}^{({\rm disc},+)}\bigoplus{\cal H}^{({\rm disc},-)} \ .
\ee
In turn for each of the spaces,
\be
{\cal H}^{({\rm disc},\pm)}=\bigoplus_{{\tt u},{\tt w}\in\mathbb{Z}}\,{\cal H}^{({\rm disc},\pm)}_{{\tt u},{\tt w}}
\ee
where, similar to \eqref{aiosd90121}, the components 
${\cal H}^{({\rm disc},\pm)}_{{\tt u},{\tt w}}$ are formed by
the linear span of states with fixed value of  $S^z={\tt u}$ and  winding number ${\tt w}$.
For ${\cal H}^{({\rm disc},+)}_{{\tt u},{\tt w}}$ the linear decomposition into
irreps of the $\overline{W}_\infty\otimes W_\infty$\,-\,algebra is given by
\be\label{8se94398ir}
 {\cal H}_{{\tt u},{\tt w}}^{({\rm disc},+)}=
\bigoplus_{\sigma=\pm1} \Big({\cal H}^{(1,+)}_{{\tt u},{\tt w},\sigma}\,\oplus\,
{\cal H}^{(2,+)}_{{\tt u},{\tt w},\sigma}\Big)
\ee
with
\be\label{iao989382931}
{\cal H}^{(1,+)}_{{\tt u},{\tt w},\sigma}=\bigoplus_{a\in\Sigma(p)}
\overline{{\cal W}}_{\bar{p},\sigma\ri\mathfrak{q}_a}\otimes{\cal W}_{p,\sigma\ri\mathfrak{q}_a}\,,
\qquad \qquad
{\cal H}^{(2,+)}_{{\tt u},{\tt w},\sigma}=
\bigoplus_{a\in{\Sigma(\bar{p})}}
\overline{{\cal W}}_{\bar{p},\sigma\ri\bar{\mathfrak{q}}_a}\otimes{\cal W}_{p,\sigma\ri\bar{\mathfrak{q}}_a}\ .
\ee
Here $\mathfrak{q}_a$ and $\bar{\mathfrak{q}}_a$
 are defined as
\be
\mathfrak{q}_a=-p-\tfrac{1}{2}-a,\qquad\qquad
\bar{\mathfrak{q}}_a=-\bar{p}-\tfrac{1}{2}-a\ .
\ee
The summation in \eqref{iao989382931} 
is taken over the non-negative integer
$a$ restricted to the set
\be\label{sigmaP1a}
\Sigma(p)=\Big\{a:\ a\in\mathbb{Z}_+,\ 
-p-\tfrac{n+2}{4}\le a<-\tfrac{1}{2}-p\Big\}
\ee
as well as
 $\Sigma(\bar{p})$, which is given by the same formula with $p$ substituted by $\bar{p}$.
\bigskip

The 
linear structure of ${\cal H}^{({\rm disc},-)}_{{\tt u},{\tt w}}$ is more involved.
To describe it, in addition to $p$, $\bar{p}$, $\mathfrak{q}_a$ and 
$\bar{\mathfrak{q}}_a$, we use the notation
\be\label{sdkkjaqw}
\arraycolsep=0.8cm
\begin{array}{ll}
p_+=\tfrac{1}{2}\,{\tt u}+\tfrac{1}{2}\,(n+2)({\tt k}+{\tt w}+1)\,, & 
\bar{p}_+=\tfrac{1}{2}\,{\tt u}-\tfrac{1}{2}\,(n+2)({\tt k}+{\tt w}+1) \\[0.4cm]
p_-=\tfrac{1}{2}\,{\tt u}+\tfrac{1}{2}\,(n+2)({\tt k}+{\tt w}-1)\,, & 
\bar{p}_-=\tfrac{1}{2}\,{\tt u}-\tfrac{1}{2}\,(n+2)({\tt k}+{\tt w}-1) \\[0.4cm]
\mathfrak{q}_a'=-p-\tfrac{n+1}{2}-a\,, &  
\bar{\mathfrak{q}}_a'=-\bar{p}-\tfrac{n+1}{2}-a\ .
\end{array}
\ee
Then
\be\label{ioasiod1298aaab}
 {\cal H}_{{\tt u},{\tt w}}^{({\rm disc},-)}=\bigoplus_{\sigma=\pm1}\Big(\,
{\cal H}_{{\tt u},{\tt w},\sigma}^{(1,-)}\oplus{\cal H}_{{\tt u},{\tt w},\sigma}^{(2,-)}
\oplus{\cal H}_{{\tt u},{\tt w},\sigma}^{(3,-)}\oplus{\cal H}_{{\tt u},{\tt w},\sigma}^{(4,-)}\,\Big)
\ee
and the decomposition 
of each of the four spaces ${\cal H}^{(i,-)}_{{\tt u},{\tt w},\sigma}$  into 
the irreps of the $\overline{W}_\infty\otimes{W}_\infty$\,-\,%
algebra reads explicitly as
\bea\label{iaosido89812aaa}
&&{\cal H}_{{\tt u},{\tt w},\sigma}^{(1,-)}=\bigoplus\limits_{a\in\Sigma_1(p)}
\overline{{\cal W}}_{\bar{p}_+,\sigma\ri\mathfrak{q}_a'}\otimes
{\cal W}_{p_+,\sigma\ri\mathfrak{q}_a'}\,,\qquad 
{\cal H}_{{\tt u},{\tt w},\sigma}^{(2,-)}=
\bigoplus\limits_{a\in\Sigma_2(p)}
\overline{{\cal W}}_{\bar{p},\sigma\ri\mathfrak{q}_a}\otimes
{\cal W}_{p_+,\sigma\ri\mathfrak{q}_a'}
\nonumber
\\[-0.4cm]
&& \hspace{12.5cm}\\
&&{\cal H}_{{\tt u},{\tt w},\sigma}^{(3,-)}=\bigoplus\limits_{a\in\Sigma_2(\bar{p})}
\overline{{\cal W}}_{\bar{p}_-,\sigma\ri\bar{\mathfrak{q}}_a'}\otimes
{\cal W}_{p,\sigma\ri\bar{\mathfrak{q}}_a}\,,\qquad \ \,
{\cal H}_{{\tt u},{\tt w},\sigma}^{(4,-)}=\bigoplus\limits_{a\in\Sigma_1(\bar{p})}
\overline{{\cal W}}_{\bar{p}_-,\sigma\ri\bar{\mathfrak{q}}_a'}\otimes
{\cal W}_{p_-,\sigma\ri\bar{\mathfrak{q}}_a'}\nonumber
\eea
The summation index $a$  takes negative integer values and runs over the sets
\bea\label{oaspdoaposd}
\Sigma_1(p)&=&\Big\{a:\ a\in\mathbb{Z}_-,\ 
-p-\tfrac{n+2}{4}\le a<-\tfrac{1}{2}-p\ { \&}\  a<-{\tt u}\Big\}
\nonumber\\[-0.2cm]
\\[-0.1cm]
\Sigma_2(p)&=&\Big\{a:\ a\in\mathbb{Z}_-,\ 
-p-\tfrac{n+2}{4}\le a<-\tfrac{1}{2}-p\ \&\  a\ge -{\tt u}\Big\}
\nonumber
\eea
and $\Sigma_1(\bar{p})$, $\Sigma_2(\bar{p})$, which are defined by the analogous formulae.
\bigskip

All the  chiral irreps 
appearing in the decomposition of ${\cal H}^{({\rm disc})}$
are of the form 
 \be
{\cal W}_{\rho,\nu}\,:\ \ \ \ \  \quad \Im m(\rho)=  \Re e(\nu)=0\,, \ \ \qquad
\rho+\tfrac{1}{2}\pm\ri\nu\in\mathbb{Z}
\ee
 with some choice of the sign $\pm$.
In this case the irrep of the $W_\infty$\,-\,algebra
  does not coincide with
the Verma module 
as the latter contains null vector(s). It turns out that if either $ \rho+\tfrac{1}{2}+\ri\nu=-a\in\mathbb{Z}$ or
 $ \rho+\tfrac{1}{2}-\ri\nu=-a\in\mathbb{Z}$
and $2\rho\notin\mathbb{Z}$, the character of ${\cal W}_{\rho,\nu}$
is given by \cite{Jayaraman:1989tu}
\be\label{charadegen1a}
{\rm ch}_{\rho,\nu}({\tt q})=\frac{{\tt q}^{-\frac{1}{12}+\frac{\nu^2}{n}+\frac{\rho^2}{n+2}}}{ ({\tt q},{\tt q})_\infty^{2}}
\ \
\sum_{m=0}^\infty (-1)^{m}\ {\tt q}^{m|a+\frac{1}{2}|+\frac{m^2}{2}}\qquad\qquad
\begin{array}{l}
\rho+\tfrac{1}{2}\pm\ri\nu\in\mathbb{Z}\  \\[0.2cm]
\rho\ \ {\rm generic}
\end{array}\ \,.
\ee
Note that when $2\rho,2\ri \nu\in\mathbb{Z}$, while $2(\rho+\ri \nu)$ is an odd integer then,
assuming also that $n$ is irrational,%
\footnote{For integer $n=2,3,\ldots$ the corresponding formula for the character
was first obtained in ref.\cite{Gepner} (see also \cite{Jayaraman:1989tu}).
In addition note that the formulae 
\eqref{iasodi1092091},\,%
\eqref{charadegen1aaa},\,\eqref{charadegen1a},\,\eqref{charadegen1v}
for the characters,
which 
assume that $c=2-\frac{6}{n+2}<2$,  can be applied 
to the case $c>2$ if one makes the formal substitutions 
$n\to-n-2$, $\rho\to \ri s$, $\nu\to\ri p$.
The central charge and highest weight of the  irrep would be parameterized as
in \eqref{Deltvarpi1aas}
and \eqref{Deltvarpi1aasa} below  (see refs.\cite{Griffin:1990fg,Bakas:1991fs}).}
\be\label{charadegen1v}
{\rm ch}_{\rho,\nu}({\tt q})=\frac{{\tt q}^{-\frac{1}{12}+\frac{\nu^2}{n}+\frac{\rho^2}{n+2}}}
{({\tt q},{\tt q})_\infty^{2}} \ 
\sum_{m=0}^\infty (-1)^{m}\, {\tt q}^{\frac{m^2}{2}}\big(\, {\tt q}^{m |\, |\rho|-|\nu|\, |}-{\tt q}^{(m+1)
(|\rho|+|\nu| +1 )-\frac{1}{2}}\,\big)\ ,
\ee
where $\Im m(\rho)=\Re e(\nu)=0$ such that
$
|\rho|\pm|{\mathfrak \nu}|\in\tfrac{1}{2}+{\mathbb Z}
$.

\bigskip
The linear decomposition of ${\cal H}^{({\rm disc})}$ 
described above together with the formula \eqref{charadegen1a} for the
character are sufficient to compute the partition function
\be
{ Z}^{({\rm disc})}={\rm Tr}_{{\cal H}^{({\rm disc})}}\Big[\,
\bar{\tt q}^{\bar{L}_0-\frac{c}{24}}\
{\tt q}^{L_0-\frac{c}{24}}\Big]\ .
\ee
To write the result in a compact way
we borrow the notation $\chi^d_{(\mathfrak{j}, a-\mathfrak{j})}({\tt q})$
from ref.\cite{Ribault:2003ss}. Up to a simple factor, this function coincides with  $\chi_{a}({\tt q})$ 
defined in eq.\,\eqref{Zdef1b},
\be
\chi^d_{(\mathfrak{j}, a-\mathfrak{j})}({\tt q})\equiv
{\tt q}^{-\frac{1}{12}-\frac{(\mathfrak{j}+\frac{1}{2})^2}{n}+\frac{({\mathfrak j}-a)^2}{n+2}}\
\chi_{a}({\tt q}) \qquad\qquad\qquad (a\in\mathbb{Z})\ .
\ee
It is related to the character \eqref{charadegen1a} as
\be
\chi^d_{(\mathfrak{j}, a-\mathfrak{j})}({\tt q})=
{\rm ch}_{a-\mathfrak{j},\ri(\mathfrak{j}+\frac{1}{2})}({\tt q})  \times \begin{cases}
1  & \quad{\rm for}\qquad  a\ge 0 \\[0.2cm]
{\tt q}^{-a}& \quad{\rm for}\qquad a<0
\end{cases}\ \  .
\ee
Also  introduce the notation ${\mathfrak  J}({\tt v},{\tt u})$
for the finite set of all  real numbers  belonging to the half-open segment
$[-\frac{n+1}{2},-\frac{1}{2})$ such that
\be\label{iaosid982981}
{\mathfrak  J}({\tt v},{\tt u})\equiv\Big\{{\mathfrak{j}}:\  {\mathfrak{j}}\in\big[-\tfrac{n+1}{2},-\tfrac{1}{2}\big)\ \&\ {\mathfrak{j}}-\tfrac{1}{2}\,{\tt v}-\tfrac{1}{2}\,(n+2)({\tt k}+{\tt u})
\in{\mathbb Z}\,\Big\}\ .
\ee
Then the calculation of the
 trace over the space ${\cal H}^{({\rm disc})}={\cal H}^{({\rm disc},+)}\bigoplus{\cal H}^{({\rm disc},-)}$ 
yields\footnote{%
In the  formula \eqref{oaspdo121}
for $Z^{({\rm disc})}$ the integers ${\tt v}$ and ${\tt u}$ are formal summation variables,
which can not be identified with the eigenvalue of $\mathbb{S}^z$ and the winding number ${\tt w}$.
In turn the notation ${\tt  p}$ and $\bar{{\tt  p}}$ in \eqref{aoisdo1902asas}
 should not be confused with $p$ and $\bar{p}$ from \eqref{rcaji8912A}. 
}
\be\label{oaspdo121}
{ Z}^{({\rm disc})}=
2\sum_{{\tt v},{\tt u}\in{\mathbb Z}}\ 
\sum_{\mathfrak{j}\in {\mathfrak  J}({\tt v},{\tt u}) }\ 
\chi^d_{(\mathfrak{j},\bar{{\tt p}})}(\bar{{\tt q}})\,\chi^d_{(\mathfrak{j},-{\tt p})}({\tt q})\,,
\ee
where
\be\label{aoisdo1902asas}
\bar{{\tt  p}}=\tfrac{1}{2}\,{\tt v}-\tfrac{1}{2}\,(n+2)\,({\tt k}+{\tt u})\,,\qquad\qquad
{\tt p}=\tfrac{1}{2}\,{\tt v}+\tfrac{1}{2}\,(n+2)\,({\tt k}+{\tt u})\ .
\ee
The overall factor of $2$ in the formula for ${ Z}^{({\rm disc})}$
occurs due to the global ${\cal Z}_2$ invariance of the model.
\bigskip

The following comment is in order here.
For arbitrary values of ${\tt k}$, the inclusion of 
the endpoints into  the interval for $\mathfrak{j}$  in \eqref{iaosid982981} 
 has no effect on the set $\mathfrak{J}({\tt v},{\tt u})$.
However for ${\tt k}=0$ and with  $n$ generic,  $\mathfrak{j}$ may
coincide with $-\frac{n+1}{2}$ or $-\frac{1}{2}$. 
Taking the limit ${\tt k}\to 0$ of $Z^{({\rm disc})}$ one finds that in order for 
\eqref{oaspdo121} to correctly describe the contribution of the discrete spectrum
to the partition function $Z^{({\rm disc})}$ with ${\tt k}=0$
one  of the endpoints  in \eqref{iaosid982981} must be included.  
The choice of whether to include $\mathfrak{j}=-\frac{n+1}{2}$ or
$\mathfrak{j}=-\frac{1}{2}$ does not matter, since they  correspond to the contribution of
the same states to $Z^{({\rm disc})}$.

\bigskip

\subsection{Hermitian structure}
The space  ${\cal H}^{({\rm cont})}$ is built out of the
$\overline{W}_\infty\otimes W_\infty$  irreps whose highest weights,
$(\bar{\omega}_2,\bar{\omega}_3;\omega_2,\omega_3)$, are real. In this case the  inner product
was introduced via the conjugation conditions \eqref{oasoidoi1982981} for the $W$ currents.
On the other hand, for the irreps appearing in the decomposition of $  {\cal H}^{({\rm disc})}$,
though 
$\omega_2$, $\bar{\omega}_2$ are real,  $\omega_3$ and $\bar{\omega}_3$ are pure imaginary.
As was discussed in the work \cite{Bazhanov:2019xvyA},
the natural inner product for such irreps is the one  that is induced by the conjugation conditions
\bea\label{aiosido1898291}
&&\big[W_j(u)\big]^{\text{\tiny\ding{105}}}=(-1)^j\,W_j(u^*)\,,\qquad\qquad
\big[\overline{W}_j(\bar{u})\big]^{\text{\tiny\ding{105}}}=
(-1)^j\,\overline{W}_j(\bar{u}^*)\ .
\eea
It turns out that the   latter occur in the scaling limit
of the finite length spin chain with the matrix conjugation 
being defined as
\be\label{oaisod91289aaa}
\hat{{\mathsf O}}^{\text{\tiny\ding{105}}}
=\hat{{\mathsf X}}_{\text{\tiny\ding{105}}}^{-1}\ \hat{{\mathsf O}}^\dag\  
\hat{{\mathsf X}}_{\text{\tiny\ding{105}}}^{}\ ,\qquad\qquad
\hat{{\mathsf O}}\in{\rm End}\big(\mathscr{V}_N\big)\ .
\ee
Here $\hat{{\mathsf X}}_{\text{\tiny\ding{105}}}=\hat{{\mathsf X}}_{\text{\tiny\ding{105}}}^\dag$ 
is related to the matrix $\hat{{\mathsf X}}_\star$
appearing in eq.\,\eqref{asido91029012} via the generator of the ${\cal Z}_2$ symmetry:
\be
\hat{{\mathsf X}}_{\text{\tiny\ding{105}}}=\hat{{\mathsf X}}_\star\,\hat{\cal D}\ .
\ee
Since the  Hamiltonian commutes with $\hat{\cal D}$,
it follows from \eqref{APSOIDioi12989} that
it is Hermitian w.r.t. the  ${\text{\tiny\ding{105}}}$\,-\,conjugation as well:
\be
\mathbb{H}^{\text{\tiny\ding{105}}}=\mathbb{H}^\star=\mathbb{H}\,.
\ee
Note that for the lattice translation operator \eqref{Kformula1},
\be
\mathbb{K}^{\text{\tiny\ding{105}}}=\mathbb{K}^\star=\mathbb{K}^{-1}\ .
\ee
\bigskip

Let 
\be
\Psi_{\bar{\rho},\rho,\bar{\nu},\nu}\in{\cal W}_{\bar{\rho},\bar{\nu}}\otimes {\cal W}_{\rho,\nu}
\subset {\cal H}^{({\rm disc})}
\ee
be the $\overline{W}_\infty\otimes W_\infty$ primary state, 
where the admissible set of values for $(\bar{\rho},\rho,\bar{\nu},\nu)$ is described by the formulae
 \eqref{eqstart1A}-\eqref{oaspdoaposd}.
The inner product of these states, together with the conjugation conditions
 \eqref{aiosido1898291} for the $W$ currents, fully specify 
the structure  of the pseudo-Hilbert space for ${\cal H}^{({\rm disc})}$.
The results of \cite{Bazhanov:2019xvyA} imply that
the normalization of the primary $\overline{W}_\infty\otimes W_\infty$ states
 can be chosen in such a way that
\be\label{otho99raaa77}
\big\langle{\Psi}_{\bar{\rho},'\rho',\bar{\nu}',\nu'} ,\,
{\Psi}_{\bar{\rho},\rho,\bar{\nu},\nu} \big\rangle_{\rm disc}=
\sigma\, f_{\bar{\rho},\bar{\nu}} \ f_{\vphantom{\bar{\rho}}\rho,\nu} \ 
\delta_{\bar{\rho}',\bar{\rho}}\,
\delta_{\rho',\rho}\, 
\delta_{\bar{\nu}',\bar{\nu}}\,
\delta_{\nu',\nu}\ ,
\ee
where
\be\label{aisod09123}
f_{\vphantom{\bar{\rho}}\rho,\nu}=
\frac{\Gamma(1+2 \rho)\,\Gamma\big(\tfrac{1}{2}+\rho-|{\nu}|\big)}{2\pi\,\Gamma(1+\frac{2 \rho}{n+2})}
\ \times \ \begin{cases}
(-1)^{a}\,a!\,
 &\ \   {\rm if} \quad \frac{1}{2}+{\rho}+|{\nu}|=-a=0,-1,\ldots\\[0.3cm]
\dfrac{2\pi}{\Gamma(\frac{1}{2}+{\rho}+|{\nu}|)} &\ \   {\rm otherwise}
\end{cases}
\ee
 \,

\noindent
Here the sign factor  $\sigma=(-1)^{N/2-S^z}$ is the same one that enters into the
quantization condition \eqref{quantC1} and depends on whether, in constructing
the RG trajectories, $N/2-S^z$ is kept to be an even or an odd number.
Let's reiterate that the above formulae are not literally applicable  when $(n+2)\,{\tt k}\in\mathbb{Z}$,
including the case ${\tt k}=0$.

\bigskip

The fact that the Hermitian structures  for the spaces 
${\cal H}^{({\rm cont})}$ and ${\cal H}^{({\rm disc})}$ 
correspond to different conjugation conditions in the
algebra of extended conformal symmetry
suggests that the states from these two spaces can not be interpreted
simultaneously as normalizable states within a single conformal field theory.
In this paper we argue that for generic values of ${\tt k}$, ${\cal H}^{({\rm cont})}$
serves as the pseudo-Hilbert space for the gauged ${\rm SL}(2,\mathbb{R})$ WZW
model.
Perhaps the simplest idea for the  field theory whose quantization 
results in the pseudo-Hilbert space ${\cal H}^{({\rm disc})}$, is the model described 
by the same Lagrangian density and constraints \eqref{kakaiaia},\,\eqref{aoasoasoas}
as well as the boundary conditions \eqref{aiosaisia} for the WZW currents and \eqref{aiasiasi}
for   $\partial_\mu\eta$.
However  the  fields now are subject to different reality conditions. The
classical counterpart to \eqref{aiosido1898291} reads as
\bigskip
 \bea\label{xiasissssai}
\big(W^{(cl)}_j\big)^*=(-1)^j\ W^{(cl)}_j\ ,\ \ \ \ \ \ \ \big(\overline{W}^{(cl)}_j\big)^*=(-1)^j\ \overline{W}^{(cl)}_j\ .
\eea
In view of eq.\,\eqref{aspod9102123a} this would follow from the reality conditions
\be\label{pasodpo11212}
\big(L^3\big)^*=-L^3\,,\qquad
\big(L^{\pm}\big)^*=L^{\mp}\,,\qquad
 \big(R^3\big)^*=-R^3\,,\qquad
\big(R^{\pm}\big)^*=R^{\mp}
\ee
imposed on the classical  WZW currents and
\be
\big(\partial\eta\big)^*=-\partial\eta\,,\qquad \qquad
\big(\bar{\partial}\eta\big)^*=-\bar{\partial}\eta
\ee
for the Gaussian field.
Furthermore  $\ri\eta$ is expected to be a real and compactified field,
\be
\ri\eta\sim\ri\eta+2\pi\ .
\ee
 The latter  implies that  the zero mode momenta 
$P_\eta$ and $\bar{P}_\eta$ \eqref{asodioaisd109212} are no longer equal, but instead
\be
\ri\,(P_\eta-\bar{P}_\eta)\in\mathbb{Z} \ .
\ee
Notice that $B=\re^{2\pi P_\eta}$ and $\bar{B}=\re^{2\pi\bar{P}_\eta}$
appearing in the boundary conditions \eqref{aususalk} still coincide. 
Such reality and boundary conditions for the 
currents correspond to the ${\rm SU}(2)$ WZW model  
gauged over the compact subgroup. However, they are not
enough to fully specify the field theory. In the ${\rm SL}(2,\mathbb{R})$  case
there were the additional constraints \eqref{isisai} and \eqref{iasisisaww}
whose motivation relied on the fact that the WZW field ${\bf G}\in{\rm SL}(2,\mathbb{R})$.
At the moment, it is not clear to us what extra conditions need to imposed for
the ${\rm SU}(2)$ case.

\section{Density matrix for the Euclidean black hole NLSM}

In the work \cite{Ikhlef:2011ay} the 
authors put forward the pioneering conjecture that the
Euclidean black hole NLSM is
the  CFT governing the scaling limit of the ${\cal Z}_2$ invariant
spin chain in the domain 
of the anisotropy parameter $\gamma\in(0,\frac{\pi}{2})$. 
This is not quite in line with what is proposed here.
Nevertheless the study of the spin chain yields a certain density matrix
whose trace coincides with the modular invariant partition function
$Z_{\rm \scriptscriptstyle EBH}$. This makes it  a good candidate for
the equilibrium density matrix $\hat{\rho}_{\rm \scriptscriptstyle EBH}$.

\bigskip

The classical field theory is described by the action \eqref{iasdioi1201435},
where $U$ and $U^*$ are a pair of complex conjugated  fields.
Usually they are assumed to be periodic, however,
it is useful to consider the more general quasiperiodic boundary conditions
\bea\label{iasod129102}
U(t,x+2\pi)=\re^{2\pi\ri {\tt k}}\ U(t,x)\ .
\eea
Thus defined, the model possesses a ${\rm U}(1)$ symmetry with the corresponding  Noether current  given by
\bea\label{hsysya}
I_\mu=\frac{1}{2\ri}\ \frac{U^*\partial_\mu U-U^*\partial_\mu U}{1+UU^*}\ .
\eea
There is also an infinite set of chiral currents, which form the 
classical $\overline{W}_\infty\otimes W_\infty$ Poisson algebra. 
The  quantization of the latter leads to 
the   algebra of extended conformal symmetry with central charge $c>2$.
The Hilbert space  of the quantum theory can be classified according to the highest weight irreps of the
$\overline{W}_\infty\otimes W_\infty$\,-\,algebra.
It is convenient to parameterize the central charge and the highest weight 
of the irreps, $\bm{\omega}=(\omega_2,\omega_3)$,  using $n$, $s$ and $p$ as
 \bea\label{Deltvarpi1aas}
 c=2+\frac{6}{n}\, >\, 2
 \eea
and
 \bea\label{Deltvarpi1aasa}
\omega_2&=&\frac{s^2+\frac{1}{4}}{n}+\frac{p^2}{n+2}\equiv \Delta_{p,s}^{(c>2)}  \\[0.2cm]
\omega_3&=&
\frac{2p}{\sqrt{n+2}}\,\Big(\,\frac{s^2}{n}+\frac{(2+3 n)\,p^2}{3n\,(n+2)}-\frac{2n+1}{12\, n}\,\Big)\nonumber\ .
\eea
To avoid confusion let us emphasize that in these relations $n>0$, $s$ and $p$ are formal parameters,
without the meaning that was assigned to them in the previous sections.
The Hilbert space of the NLSM
 contains both a continuous ${\cal H}_{{\rm\scriptscriptstyle EBH}}^{({\rm cont})}$ 
and a discrete component ${\cal H}_{{\rm\scriptscriptstyle EBH}}^{({\rm disc})}$.
The linear
decomposition of the continuous one into the irreps of the $\overline{W}_\infty\otimes W_\infty$\,-\,algebra is given by
\cite{Dijkgraaf:1991ba,ZAM,Maldacena:2000hw,Maldacena:2000kv,Hanany:2002ev,Ribault:2003ss,Schomerus:2005aq}
 \be\label{iaosasdidoi192009AA}
{\cal H}_{{\rm\scriptscriptstyle EBH}}^{({\rm cont})}=
\bigoplus_{{\tt v},{\tt w}=-\infty}^{+\infty}\ \int^{\oplus}_{s>0}\!\rd s\
\overline{{\cal W}}_{\bar{p},s}^{(c>2)}\otimes {\cal W}_{{ p},s}^{(c>2)}\,,\qquad
 {\rm where}
\qquad 
\begin{array}{l} p=\frac{1}{2}\, {\tt v}+\frac{1}{2}\,(n+2)\,({\tt k}+{\tt w})\\[0.2cm]
                                       \bar{p}=\frac{1}{2}\,{\tt v} -\frac{1}{2}\,(n+2)\,({\tt k}+{\tt w})
\end{array}.
 \ee
Here ${\tt v}$ 
is the eigenvalue of the ${\rm U}(1)$ conserved charge 
$\hbar^{-1}\oint\rd x I_0$ associated with the Noether current \eqref{hsysya}.
It takes integer values 
 provided that the Planck constant is 
identified with $n$ as 
\be
\hbar=\frac{2\pi}{n+2}\ .
\ee
The integer ${\tt w}$
may be interpreted as 
a winding number related to the fact that the boundary condition \eqref{iasod129102}
is invariant w.r.t. the substitution ${\tt k}\mapsto {\tt k}+{\tt w}$
 with ${\tt w}\in\mathbb{Z}$. Let us note that
the highest weight \eqref{Deltvarpi1aasa} is not sensitive to the sign of $s$.
Due to this the direct integral in  \eqref{iaosasdidoi192009AA}
is restricted to positive values of $s$.
 For the states at the level $\bar{\tt  L}$ and ${\tt L}$ 
in the irrep
$\overline{{\cal W}}_{\bar{p},s}^{(c>2)}\otimes {\cal W}_{{ p},s}^{(c>2)}$, the corresponding energy 
$E=\Delta+{\bar\Delta}-\frac{c}{12}$
in terms of the parameters $n$, $s$ and $p$ reads  as
 \bea\label{iaosid1212}
 E=-\frac{1}{6}+\frac{2s^2}{n}+\frac{p^2+{\bar p}^2}{n+2}+{\tt L}+{\bar {\tt L}}\ .
 \eea
It is worth mentioning that
the space of states ${\cal H}_{{\rm\scriptscriptstyle EBH}}={\cal H}_{{\rm\scriptscriptstyle EBH}}^{({\rm cont})}\bigoplus
{\cal H}_{{\rm\scriptscriptstyle EBH}}^{({\rm disc})}$
is equipped with a positive definite inner product \cite{Dixon:1989cg} such that the Fourier modes
of the $W$ and $\overline{W}$ currents, generating the  $\overline{W}_\infty\otimes W_\infty$\,-\,algebra 
with $c>2$, satisfy the conjugation conditions 
 \bea
\big[\widetilde{W}_j(m)\big]^\dagger=\widetilde{W}_j(-m)\,,\qquad\qquad
 \big[\,  \widetilde{\overline{W}}_j(m)\, \big]^{\dagger}=
 \widetilde{\overline{W}}_j(-m)\ .
\eea
\bigskip

In the works \cite{Maldacena:2000kv,Hanany:2002ev} 
a modular invariant partition function was found for the Euclidean black hole NLSM
with periodic boundary conditions imposed on the fundamental field (${\tt k}=0$).
It is straightforward to generalize the result to the case of twisted boundary conditions.
This yields\footnote{%
The formulae for the partition function presented in the works 
\cite{Maldacena:2000kv,Hanany:2002ev} is twice $Z_{{\rm \scriptscriptstyle EBH}}$ 
given by \eqref{parfun2} with ${\tt k}=0$.
This is related to the fact that the corresponding NLSM was obtained
by gauging the  U(1) symmetry, ${\bf g}\mapsto {\bf h}\,{\bf g}\,{\bf h}$ $({\bf h}=\re^{\frac{\ri\phi}{2}\sigma^y})$,
of the ${\rm SL}(2,{\mathbb R})$ WZW model.
This  results in two  copies of the Euclidean black hole NLSM.
A similar occurrence happens for 
the Lorentzian black hole NLSM, as mentioned in sec.\,\ref{sec22}.}
\be\label{parfun2}
Z_{{\rm \scriptscriptstyle EBH}}=
\frac{\sqrt{n(n+2)}}{\Im m(\tau)}\,\sum_{{\tt a},{\tt b}\in\mathbb{Z}}\
\int_{D_{\epsilon}}\rd^2 z\  
\re ^{-\frac{\pi(n+2)}{\Im m(\tau)}|z+{\tt a}+({\tt b}+{\tt k})\,\tau|^2
+\frac{2\pi}{\Im m(\tau)}(\Im m(z))^2}\,
\left|\frac{\eta(\tau)}{\vartheta_1(z |\tau)}\right|^2\,,
\ee
where   $\vartheta_1$ and $\eta$ are the standard 
elliptic theta and Dedekind eta functions:
\bea
\vartheta_1(u|\tau)&=&2 {\tt q}^{\frac{1}{8}}\,\sin(\pi u)
\ (\re^{2\pi\ri u}\,{\tt q},{\tt q})_\infty \ (\re^{-2\pi\ri u}\,{\tt q},{\tt q})_\infty \ ({\tt q},{\tt q})_\infty
  \\[0.2cm]
\eta(\tau)&=&{\tt q}^{\frac{1}{24}}\,({\tt q},{\tt q})_{\infty} \ \qquad \qquad \qquad\qquad\qquad\qquad
\qquad \qquad (\,{\tt q}=\re^{2\pi\ri \tau}\,)\ .\nonumber
\eea
Note that the dependence on the twist parameter ${\tt k}$ manifests itself only as
 a shift of the summation variable ${{\tt b}}\mapsto{\tt b}+{\tt k}$, 
which appears in the exponent in the integrand  in \eqref{parfun2}.
The integral is taken over the parallelogram $D$
in the complex $z$ plane 
with vertices at $z=\pm \half \pm \half\,\tau$. However since the
integrand is singular at $z=0$, a small neighbourhood
around the origin, whose size is controlled by the
parameter $\epsilon$, should be excluded from the integration domain.
We found it convenient to take
\be\label{iaosid182981}
D_\epsilon=D\big/\big\{z:\ |z|<\tfrac{1}{2\pi}\, \re^{-\gamma_{{\rm E}}}\,\epsilon\,\big\}\ ,
\ee
where $\gamma_{\rm E}$ denotes the Euler constant.
Then  as $|{\tt q}|\to 0$
\be
Z_{{\rm \scriptscriptstyle EBH}}=\frac{1}{2\pi}\, \sqrt{\frac{n}{\Im m(\tau)}}\  |{\tt q}|^{-\frac{1}{6}} \ 
\Big(\log(4\re^{\gamma_{\rm E}}/\epsilon)+o\big(|{\tt q}|^0\big)\Big)\ .
\ee
\bigskip

Through  a numerical study, we found the following relation
between the partition function of the Euclidean black hole NLSM
and that 
which occurs in the scaling limit of the 
${\cal Z}_2$ invariant spin chain:
\be\label{asiodio1902}
2\,Z_{{\rm \scriptscriptstyle EBH}}=Z^{({\rm cont})}+Z^{({\rm disc})}\,.
\ee
Here $Z^{({\rm cont})}$ and $Z^{({\rm disc})}$ are 
given by eqs.\,\eqref{iaosido12032} and \eqref{oaspdo121}, respectively.
Both sides of the above formula contain a divergent part $\propto\log(1/\epsilon)$.
For the Euclidean black hole NLSM this parameter regularizes the integral in \eqref{parfun2}, while
for the lattice model $\epsilon\propto N^{-1}$  as in eq.\,\eqref{ioasd89121}. 
To perform a numerical check of \eqref{asiodio1902}  the divergent part needs
to be subtracted.
For this purpose,
introduce
the regularized partition function of  the Euclidean black hole NLSM as
\be\label{asodi1288912}
Z_{{\rm \scriptscriptstyle EBH}}^{({\rm reg})}= \lim_{\epsilon\to 0}\ 
\big(\,Z_{{\rm \scriptscriptstyle EBH}} - Z^{({\rm sing})}_{\epsilon}
\,\big)
\ee
with
\be\label{ia89989888912kja}
Z^{({\rm sing})}_{\epsilon}=\sqrt{\frac{n}{\Im m(\tau)}}\ \ 
\frac{\log(4\re^{\gamma_{\rm E}}/\epsilon)+\frac{1}{2}\log\big(\Im m(\tau)\big)}
{2\pi\, (\bar{{\tt q}},{\bar{\tt q}})_\infty^{2}({\tt q},{\tt q})_\infty^{2}}\ 
\sum_{{\tt u},{\tt w}\in\mathbb{Z}}
\bar{{\tt q}}^{-\frac{1}{12}+\frac{\bar{p}^2}{n+2}}\ 
{\tt q}^{-\frac{1}{12}+\frac{p^2}{n+2}}
\ee
and recall that ${\tt q}=\re^{2\pi\ri\tau}$, $\bar{{\tt q}}=\re^{-2\pi\ri\tau^*}$.
Here an extra term $\propto \log\big(\Im m(\tau)\big)$ was included into the definition of 
$ Z^{({\rm sing})}_{\epsilon}$ in order to ensure that the regularized partition function
is invariant under modular transformations in the case when ${\tt k}=0$ (for ${\tt k}\ne 0$
the partition function is not a modular invariant quantity).
In turn,
we define the regularized part of $Z^{({\rm cont})}$ to be
\bea\label{aoisdo091212}
Z_{{\rm reg}}^{({\rm cont})}&=&
\sum_{{\tt u},{\tt w}\in\mathbb{Z}}\ \int_{-\infty}^{+\infty}\rd s\
\sum_{{\tt L},\bar{\tt L}\ge 0}\tilde{\rho}_{\bar{p},p}^{(\bar{\tt L},{\tt L})}(s)\ 
\bar{{\tt q}}^{-\frac{1}{12}+\frac{s^2}{n}+\frac{\bar{p}^2}{n+2}+\bar{{\tt L}}}\ 
{\tt q}^{-\frac{1}{12}+\frac{s^2}{n}+\frac{p^2}{n+2}+{\tt L}} \nonumber \\[0.2cm]
&-&
\sqrt{\frac{n}{\Im m(\tau)}}\ \ 
\frac{\log(4\re^{\gamma_{\rm E}})+\frac{1}{2}\log\big(\Im m(\tau)\big)}
{\pi\, (\bar{{\tt q}},{\bar{\tt q}})_\infty^{2}({\tt q},{\tt q})_\infty^{2}}\ 
\sum_{{\tt u},{\tt w}\in\mathbb{Z}}
\bar{{\tt q}}^{-\frac{1}{12}+\frac{\bar{p}^2}{n+2}}\ 
{\tt q}^{-\frac{1}{12}+\frac{p^2}{n+2}}\ .
\eea
Here  the density of states 
$\tilde{\rho}_{\bar{p},p}^{(\bar{\tt L},{\tt L})}(s)$
is given by eqs.\,\eqref{aasas312321A}-\eqref{Zdef1b}.
Then \eqref{asiodio1902} is equivalent to
\be\label{aoisdo091212A}
2\,Z_{{\rm \scriptscriptstyle EBH}}^{({\rm reg})}= Z_{{\rm reg}}^{({\rm cont})}+
Z^{({\rm disc})}\ .
\ee
The numerical data in support of this relation is presented in tabs.\,\ref{tab01} and \ref{tab02}. 
\bigskip

\begin{table}
\centering
\scalebox{1}{
\begin{tabular}{|c|c|c|c|c|}
\hline
 & & & & \\[-0.3cm]
$\tau$ &$Z^{\rm (cont)}_{\rm reg}$ &$Z^{\rm (disc)}$&
$Z^{\rm (cont)}_{\rm reg}+Z^{\rm (disc)}$&
$2\,Z_{{\rm \scriptscriptstyle EBH}}^{({\rm reg})}$\\[.2cm]
\hline
\hline
 & & & & \\[-0.3cm]
$\tau=.9 \ri$ &$ -3.9509313$ &$ 0.0210525$ &$-3.9298787$&$-3.9298786$
\\[.2cm]
$-1/\tau$&  $-3.9358543 $& $0.0059766$ &$-3.9298776$ & $-3.9298787$ \\[.2cm]
$\tau+1$& $-3.9509313$ &$ 0.0210525$ & $-3.9298787$ &$-3.9298786$\\[.2cm]
\hline
\hline
 & & & & \\[-0.3cm]
$\tau=.2+.9\ri$&$-3.8983544 $&$0.0065418 $&$-3.8918125$&$-3.8918125$\\[.2cm]
$-1/\tau$&$ -3.8925978 $&$ 0.0007853 $&$
   -3.8918125 $&$-3.8918124$\\[.2cm]
$\tau+1$&$ -3.8983544 $&$ 0.0065418 $&$
   -3.8918125 $&$-3.8918124$\\[.2cm]
\hline
\hline
 & & & & \\[-0.3cm]
$\tau=.66\ri$&$-4.4682528$&$0.0943594$&$-4.3738934$&$-4.3738934$\\[.2cm]
$-1/\tau$&$-4.3744476$&$0.0005542$&$-4.3738934$&$-4.3738933 $\\[.2cm]
$\tau+1$&$-4.4682528$&$0.0943594  $&$-4.3738934  $&$-4.3738933  $\\[.2cm]
\hline
\hline
 & & & & \\[-0.3cm]
$\tau=.5\ri$&$-5.7668560$&$0.2960118$&$-5.4708441$&$-5.4708421  $\\[.2cm]
$-1/\tau  $&$-5.4708761$&$0.0000322$&$-5.4708439$&$-5.4708437  $\\[.2cm]
$\tau+1$&$-5.7668560$&$0.2960118$&$-5.4708441$&$ -5.4708421 $\\[.2cm]
\hline
\hline
 & & & & \\[-0.3cm]
$\tau=.33\ri$&$-12.070612$&$1.5569389$&$-10.513673$&$-10.5129976$\\[.2cm]
$-1/\tau$&$-10.513561$&$7.662\cdot 10^{-8}$&$-10.513561$&$-10.5135606$\\[.2cm]
$\tau+1$&$-12.070612$&$1.5569389$&$-10.513673$&$-10.5129975  $\\[.2cm]
\hline
\end{tabular}
}
\label{tab2}
\caption{\small A comparison of the numerical data for twice the 
regularized partition function of the Euclidean black hole NLSM 
\eqref{asodi1288912} with  $Z_{\rm reg}^{({\rm cont})}+Z^{({\rm disc})}$
for the case ${\tt k}=0$ and $n=3$.
Here $Z^{({\rm disc})}$ is given by eqs.\,\eqref{iaosid982981} and \eqref{oaspdo121}, while
$Z_{\rm reg}^{({\rm cont})}$ is defined by  \eqref{aoisdo091212}.
The table also illustrates modular invariance of the 
regularized partition function for ${\tt k }=0$.
Note that in order to achieve good accuracy for decreasing values of  $\Im m(\tau)$ 
one must take into account an increasing number of terms in the sum over ${\tt u}$ and ${\tt w}$
for $Z^{({\rm cont})}$ as well as ${\tt a}$ and ${\tt b}$ in eq.\,\eqref{parfun2}.
This significantly increases the computer time.
\label{tab01}}
\end{table}
\begin{table}
\begin{center}
\scalebox{1}{
\begin{tabular}{|c|c|c|c|c|}
\hline
 & & & & \\[-0.4cm]
$\tau$ & $Z_{\rm reg}^{({\rm cont})}$ & $Z^{({\rm disc})}$ & 
$Z_{\rm reg}^{({\rm cont})}+Z^{({\rm disc})}$ 
& $2\,Z_{{\rm \scriptscriptstyle EBH}}^{({\rm reg})}$ \\[.1cm]
\hline
&&&& \\[-0.3cm]
$0.9\,\ri$&$-3.1430392$&$0.0233941$&$-3.1196452$&$-3.1196450 $\\[.2cm]
\hline
&&&& \\[-0.3cm]
$0.2+0.9\,\ri$&$-3.0646040$&$0.0099983$&$-3.0546057$&$-3.0546064 $\\[.2cm]
\hline
&&&& \\[-0.3cm]
$0.66\,\ri$&$-3.7836669$&$0.1033699$&$-3.6802970$&$-3.6802972 $\\[.2cm]
\hline
&&&& \\[-0.3cm]
$0.2 + 0.66\,\ri$ & $-3.5074556$ & $0.0418838$ & $-3.4655718$ & $ -3.4655717 $\\[.2cm]
\hline
&&&& \\[-0.3cm]
$0.50\,\ri$&$-5.1054421$&$0.3209649$&$-4.7844771$&$-4.7844724 $\\[.2cm]
\hline
&&&& \\[-0.3cm]
$0.33\,\ri$&$-11.2855973$&$1.6391928$&$-9.6464045$& $-9.646289 $\\[.2cm]
\hline
&&&& \\[-0.3cm]
$0.25\,\ri$&$-26.5761236$&$5.4010183$&$-21.1751053$&$ -21.171536 $\\[.2cm]
\hline
\end{tabular}
}
\caption{\small
The last column contains numerical data for $2\,Z_{{\rm \scriptscriptstyle EBH}}^{({\rm reg})}$
\eqref{asodi1288912} 
with the parameters set to be ${\tt k}=-0.1$ and $n=3$. 
This is compared to 
$Z_{\rm reg}^{({\rm cont})}+Z^{({\rm disc})}$,
where $Z^{({\rm disc})}$ was computed using eqs.\,\eqref{iaosid982981},\,\eqref{oaspdo121} and
$Z_{\rm reg}^{({\rm cont})}$ via  \eqref{aoisdo091212}.
\label{tab02}}
\end{center}
\end{table}

The following comments concerning some statements 
appearing in the literature are in order here.
It was proposed in the works \cite{Maldacena:2000kv,Hanany:2002ev}
that $Z_{{\rm \scriptscriptstyle EBH}}$ \eqref{parfun2} could be represented as
\be\label{pasod190291}
2\,Z_{{\rm \scriptscriptstyle EBH}}\big|_{{\tt k}=0}=\sum_{{\tt u},{\tt w}=-\infty}^\infty\int_{0}^{\infty}
\rd s\, 2\rho(s)\
 {\rm ch}_{\bar{p},s}(\bar{{\tt q}})\,{\rm ch}_{p,s}({\tt q})\ +\ Z^{({\rm disc})}
\qquad  (\,{\rm from\ refs.\,}[9,10]\,)\ .
\ee
The character entering into this formula is given by \eqref{iasodi1092091}, while 
for the density of states
\be\label{iasoid10928}
\rho(s)=
\frac{2}{\pi}\, \log (1/\epsilon)+\frac{1}{2\pi\ri}\ 
\partial_s\log\bigg[
\, \frac{\Gamma(\frac{1}{2}+p-{\ri s})\,\Gamma(\frac{1}{2}+{\bar p}-{\ri s})}
{\Gamma(\frac{1}{2}+p+{\ri s})\,\Gamma(\frac{1}{2}+{\bar p}+{\ri s})}\,\bigg]\ .
\ee
A numerical check of the consistency of \eqref{pasod190291} with \eqref{parfun2} rules the conjecture out.
Also the formula for the contribution of the discrete spectrum
to the partition function $Z_{{\rm \scriptscriptstyle EBH}}|_{{\tt k}=0}$
is given in ref.\cite{Ribault:2003ss}.
It appears to be identical with $\frac{1}{2}\,Z^{({\rm disc})}$ 
from \eqref{oaspdo121} specialized to ${\tt k}=0$.
However, eqs.\,(2.5) and (2.10) from ref.\cite{Ribault:2003ss}
do not quite  correctly take into account the contribution of the states to
$Z_{{\rm\scriptscriptstyle EBH}}^{(\rm disc)}$
with $\mathfrak{j}=-\frac{n+1}{2},\,-\frac{1}{2}$
corresponding to the boundary of the interval in the set $\mathfrak{J}({\tt v},{\tt u})$ \eqref{iaosid982981}.
Finally, the highly non-trivial formula  \eqref{asiodio1902} is in full  agreement with the
 original observation of ref.\cite{Ikhlef:2011ay}.
However, let's emphasize that in order to state
that the Euclidean black hole NLSM governs the critical behaviour of the
${\cal Z}_2$ invariant inhomogeneous six-vertex model, this relation is insufficient.
Among others, 
the numerical study of the finite size corrections to the CFT Hamiltonian
 performed in \cite{Bazhanov:2019xvyA}, which
 are controlled by irrelevant perturbations,
show that the extended conformal symmetry algebra is the 
$\overline{W}_\infty\otimes W_\infty$\,-\,algebra with $c<2$. 
\bigskip

In view of eqs.\,\eqref{aoisdo091212A} and \eqref{aoisdo091212}
one arrives at a conjecture for the equilibrium density matrix of the Euclidean black hole NLSM.
Namely,
being restricted to the level subspaces of the irreps of the 
$\overline{W}_\infty\otimes W_\infty$\,-\,algebra belonging to 
${\cal H}_{{\rm\scriptscriptstyle EBH}}^{({\rm cont})}$,
it is given by a formula similar to eq.\,\eqref{aoisd98923}:
\be\label{aoisd98923AAA}
\hat{\rho}_{{\rm\scriptscriptstyle EBH}}\,
\big|_{\overline{{\cal W}}_{{\bar{p}},s}^{(\bar{\tt L})}\otimes{\cal W}_{p,s}^{({\tt L})}}=
\bigg[\,\frac{2}{\pi}\,\log(1/\epsilon)+\frac{\tilde{\rho}_{\bar{p},p}^{(\bar{\tt L},{\tt L})}(s)}
{{\rm par}_2(\bar{\tt L})\,{\rm par}_2({\tt L})}\,
\bigg]\ 
\bar{{\tt q}}^{-\frac{1}{12}+\frac{s^2}{n}+\frac{\bar{p}^2}{n+2}+\bar{{\tt L}}}\ 
{\tt q}^{-\frac{1}{12}+\frac{s^2}{n}+\frac{p^2}{n+2}+{\tt L}}\ \hat{{\bf 1}}
\ee
with $\tilde{\rho}_{\bar{p},p}^{(\bar{\tt L},{\tt L})}(s)$ being  defined via eqs.\,\eqref{aasas312321A}-\eqref{Zdef1b}.
It is important to keep in mind that the irreps appearing in the decomposition 
of ${\cal H}_{{\rm\scriptscriptstyle EBH}}^{({\rm cont})}$
\eqref{iaosasdidoi192009AA} are those of the 
$\overline{W}_\infty\otimes W_\infty$\,-\,algebra with $c=2+\frac{6}{n}>2$.
The above formula   is expected to be applicable to
the case of twisted boundary conditions \eqref{iasod129102} with generic ${\tt k}$.
For the model with periodic boundary conditions, the density matrix is obtained
via  a taking of the limit ${\tt k}\to 0$. 
Some care is needed for the irreps with $p=\frac{1}{2}\,{\tt u}+\frac{1}{2}\,(n+2)\,{\tt k}$,
$\bar{p}=\frac{1}{2}\,{\tt u}-\frac{1}{2}\,(n+2)\,{\tt k}$ and
 ${\tt u}$ odd,  as  
$\tilde{\rho}_{\bar{p},p}^{(\bar{\tt L},{\tt L})}(s)$ could contain simple poles at 
$s=\pm\frac{\ri}{2}\,(n+2)\,{\tt k}$,  which approach the real axis for vanishing ${\tt k}$.
This gives rise to contact terms as in  eq.\,\eqref{aposdoio12099102}.
Finally  $\hat{\rho}_{{\rm\scriptscriptstyle EBH}}$, being restricted to the discrete component 
${\cal H}^{({\rm disc})}_{{\rm\scriptscriptstyle EBH}}$ of the space of states of the
Euclidean black hole NLSM
coincides with the usual
thermal density matrix
\be\label{oaisodi9812891}
\hat{\rho}_{{\rm\scriptscriptstyle EBH}}\big|_{{\cal H}^{({\rm disc})}_{{\rm\scriptscriptstyle EBH}}}=
{\tt q}^{L_0-\frac{c}{24}}\ \bar{{\tt q}}^{\bar{L}_0-\frac{c}{24}}\ .
\ee

\section{Conclusion}
In this work we apply the results obtained for
the ${\cal Z}_2$ invariant  integrable spin chain to the  
 study of two NLSMs. These are of interest since their target space geometries
mimic that of a Lorentzian black hole and its Euclidean version.
\bigskip

The space of states occurring in the scaling limit of the low energy states of the spin chain contains both a
discrete ${\cal H}^{({\rm disc})}$ and continuous  ${\cal H}^{({\rm cont})}$ 
component. For the latter, the spectrum of conformal dimensions 
forms a continuous distribution which is characterized by a density of states.
We conjecture that  the pseudo-Hilbert space of the Lorentzian black hole NLSM
coincides with a subspace of the ${\cal C}$ even sector of ${\cal H}^{({\rm cont})}$  
in the case of periodic boundary conditions for the spin chain. 
In turn, from the density of states
restricted to this subspace
we construct an equilibrium density matrix for the NLSM.
An important point is that  ${\cal H}^{({\rm disc})}$ is excluded from the identification.
This was motivated through the study of the Hermitian structures for the spin chain, which suggests
that the states from ${\cal H}^{({\rm disc})}$  and ${\cal H}^{({\rm cont})}$ 
can not be interpreted simultaneously as normalizable states within a single CFT.
\bigskip

Contrary to the Lorentzian  black hole NLSM,  the Hilbert space of the
Euclidean one contains a discrete component made up of normalizable states, whose wavefunction(als)
are localized in the vicinity of the tip of the target manifold.
Remarkably, their contribution to the CFT partition function coincides with one half of
the contribution of the states from ${\cal H}^{({\rm disc})}$ to the  partition function of the spin chain. 
Using the full density of states for ${\cal H}^{({\rm cont})}$ 
 an equilibrium density matrix for the  Euclidean black hole NLSM is proposed,
which reproduces
the modular invariant partition
function originally obtained by Maldacena, Ooguri and Son in ref.\cite{Maldacena:2000kv}.
\bigskip

\section*{Acknowledgments}
The authors thank H.~Saleur, V.~Schomerus and A.~B.~Zamolodchikov for stimulating discussions.

\medskip
\noindent
VB acknowledges the support of the Australian Research Council grant DP190103144.

\medskip
\noindent
The research of GK is funded by the Deutsche Forschungsgemeinschaft (DFG, German Research
Foundation) under Germany's Excellence Strategy -- EXC 2121 ``Quantum Universe'' -- 390833306.

\medskip
\noindent
The work of SL is supported by the 
Rutgers New High
Energy Theory Center.

\pagebreak

\end{document}